\documentclass[aps,preprint,showpacs,nofootinbib,amsmath]{revtex4}
\usepackage{graphicx}

\newcommand*{\tcaps}{\setlength{\baselineskip}{3ex}}
\newcommand*{\fs}[1]{#1\!\!\!/}

\begin{document}

\title{\boldmath
Some aspects of $\Theta^+$ parity determination in the reaction
$\gamma\, N\to \Theta^+\, \bar{K}\to N \, K \bar{K} $}


 \author{A.\,I.~Titov,$^{ab}$  H.~Ejiri,$^{cd}$
  H.~Haberzettl,$^{ef}$ and  K.~Nakayama$^{eg}$}
 \affiliation{
 $^a$Advanced Photon Research Center, Japan Atomic Energy Research Institute,
 Kizu, Kyoto,
 619-0215, Japan\\
 $^b$Bogoliubov Laboratory of Theoretical Physics, JINR,
  Dubna 141980, Russia\\
 $^c$Natural Science, International Christian University,
Osawa, Mitaka, Tokyo, 181-8585, Japan \\
 $^d$Research Center for Nuclear Physics, Osaka University,
 Ibaraki, Osaka 567-0047, Japan\\
 $^e$Institute f\"ur Kernphysik (Theorie), Forschungzentrum
J\"ulich, D-52425 J\"ulich, Germany\\
$^f$Center for Nuclear Studies, Department of Physics, The George
Washington University, Washington, DC 20052, USA\\
$^g$Department of Physics and Astronomy, University of Georgia,
Athens, GA 30602, USA
}

\begin{abstract}
We analyze the problem of how to determine the parity of the $\Theta^+$ pentaquark in the
reaction $\gamma N\to K\Theta\to NK\bar{K}$, where $N=n,p$. Our model calculations
indicate that the contribution of the non-resonant background of the reaction $\gamma
N\to NK\bar{K}$ cannot be neglected, and that suggestions to determine the parity based
solely on the initial-stage process $\gamma N\to K\Theta$ cannot be implemented cleanly.
We discuss the various mechanisms that contribute to the background, and we identify some
spin observables which are sensitive mostly to the $\Theta^+$ parity rather than to the
details of the production mechanism.
\end{abstract}


\maketitle


\section{Introduction}

The first evidence for the $\Theta^+$ pentaquark discovered by the LEPS collaboration at
SPring-8~\cite{Nakano03} was subsequently confirmed in other
experiments~\cite{DIANA,CLAS1,CLAS2,SAPHIR03,Asratyan,NOMAD}. None of these experiments
could determine the spin and the parity of the $\Theta^+$. Some proposals to do this in
photoproduction processes using single and double polarization observables were discussed
in Refs.~\cite{NT03,Zhao03,ZhaoKhal04}. The difficulty in determining the spin and parity
of $\Theta^+$ in the reaction $\gamma N\to \Theta^+ K^-$ is due to the way the pentaquark
state is produced. The models of the $\Theta^+$ photoproduction from the nucleon based on
the effective Lagrangian approach have been developed in
Refs.~\cite{Hosaka03,OKL031,NT03,LiuKo031,LiuKo032,LiuKo033,Zhao03,ZhaoKhal04,Oh-2,CloseZhao}.
As has been pointed out, there are great ambiguities in calculating the (spin)
unpolarized and polarized observables. In the effective Lagrangian formalism the problems
are summarized as follows:

 (1) Dependence on the coupling operator for the $\Theta^+NK$ interaction, i.e., whether
one chooses pseudoscalar (PS) or pseudovector (PV) couplings. In the case of PV coupling,
gauge invariance requires a Kroll-Ruderman-type contact term even for undressed
particles which affects both unpolarized and polarized observables. For dressed
particles, in a tree-level description, contact currents also are required for PS
coupling.

(2) Ambiguity due to the choice of the coupling constants. At the simplest level, five
unknown coupling constants and their phases enter the formalism: $g_{\Theta NK}$ in the
$\Theta^+ NK $ interaction; the vector and tensor couplings $g_{\Theta NK^*}$ and
$\kappa^*$, respectively, in the $\Theta^+ NK^* $ interaction, and the tensor coupling
$\kappa_\Theta$ in the electromagnetic $\gamma \Theta^+$ interaction. To fix the absolute
value of $g_{\Theta NK}$, one can use the relation between $g_{\Theta NK}$ and the
$\Theta$ decay width $\Gamma_\Theta$. This provides, however,  only an upper limit for
$|g_{\Theta NK}|$ because all the experiments give only upper limits of the decay width
(about 25\,MeV) which are comparable with the experimental resolution.

(3) Dependence on the choice of the phenomenological form factors: (i) form factors
suppress the individual channels in  different ways, and (ii) form factors generate
(modify) the contact terms for the PS (PV) coupling schemes which affect the theoretical
predictions.

A possible solution to these problems is to use more complicated ``model-independent"
(triple) spin observables, discussed by Ejiri~\cite{Ejiri},  Rekalo and
Tomassi-Gustafsson~\cite{RT-G04}, and Nakayama and Love~\cite{NL04}. These spin
observables involve the linear polarization of the incoming photon, and the polarizations
of the target nucleon and the outgoing $\Theta^+$. Using basic principles, such as the
invariance of the transition amplitude under rotation, parity inversion and, in
particular, the reflection symmetry with respect to the scattering plane, one can arrive
at unambiguous predictions which depend only on the $\Theta^+$ parity in the reaction
$\gamma N\to \Theta^+ \bar{K}$. The key aspect of the model-independent predictions is
that in the final state the total internal parity of outgoing particles are different for
positive and negative $\Theta^+$ parity. We skip the discussions of the practical
implementation of using the triple spin observables since experimental observations of
the spin orientation of the strongly decaying $\Theta^+$ is extremely difficult. Instead,
we focus on the basic aspects of this idea. For simplicity, in the following we limit our
discussion for determining the $\Theta^+$ parity to {isoscalar} spin-1/2 $\Theta^+$. In
fact, most theories predict the $J^P$ of $\Theta^+$ to be $1/2^+$ or $1/2^-$.

There are two difficulties in applying  the ``model-independent" formalism for the parity
determination. First, the final state in the photoproduction experiment is the three-body
state $NK\bar{K}$ (and not the two body $\Theta^+\bar{K}$)-state. The spin observables in
the initial and final channels are deduced by their parities irrespective of the
intermediate $\Theta^+$ parity. It is difficult, therefore, to find the pertinent
``model-independent" observables for this case. Second, the contribution of the
non-resonant background of the reaction $\gamma N\to NK\bar{K}$ cannot be neglected. The
observed ratio of the resonance peak  to the non-resonant continuum reported in the
photoproduction experiments is about $1.5-2$. This means that the difference between the
resonant and non-resonant amplitudes is a factor of two and therefore the non-resonant
background may modify the spin-observables considerably.

\begin{figure}[t]
\parbox{.45\textwidth}{\centering
  \includegraphics[width=.42\textwidth]{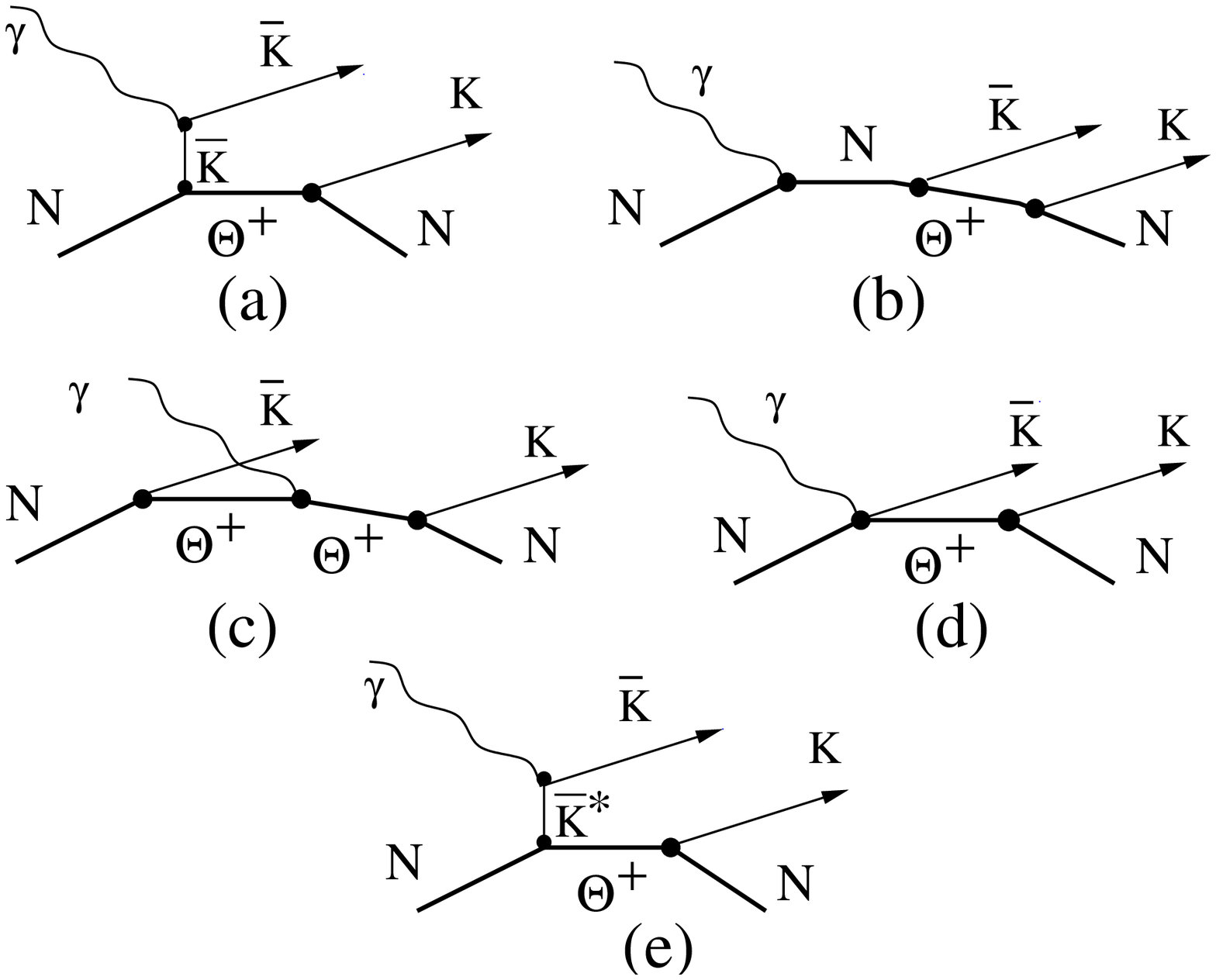}
 \caption{\label{fig:1}\tcaps%
 Tree-level diagrams for the reaction
 $\gamma N\to \Theta^+\bar{K}\to NK\bar{K}$.}
}
\hfill
\parbox{.5\textwidth}{\centering
\includegraphics[width=.5\textwidth]{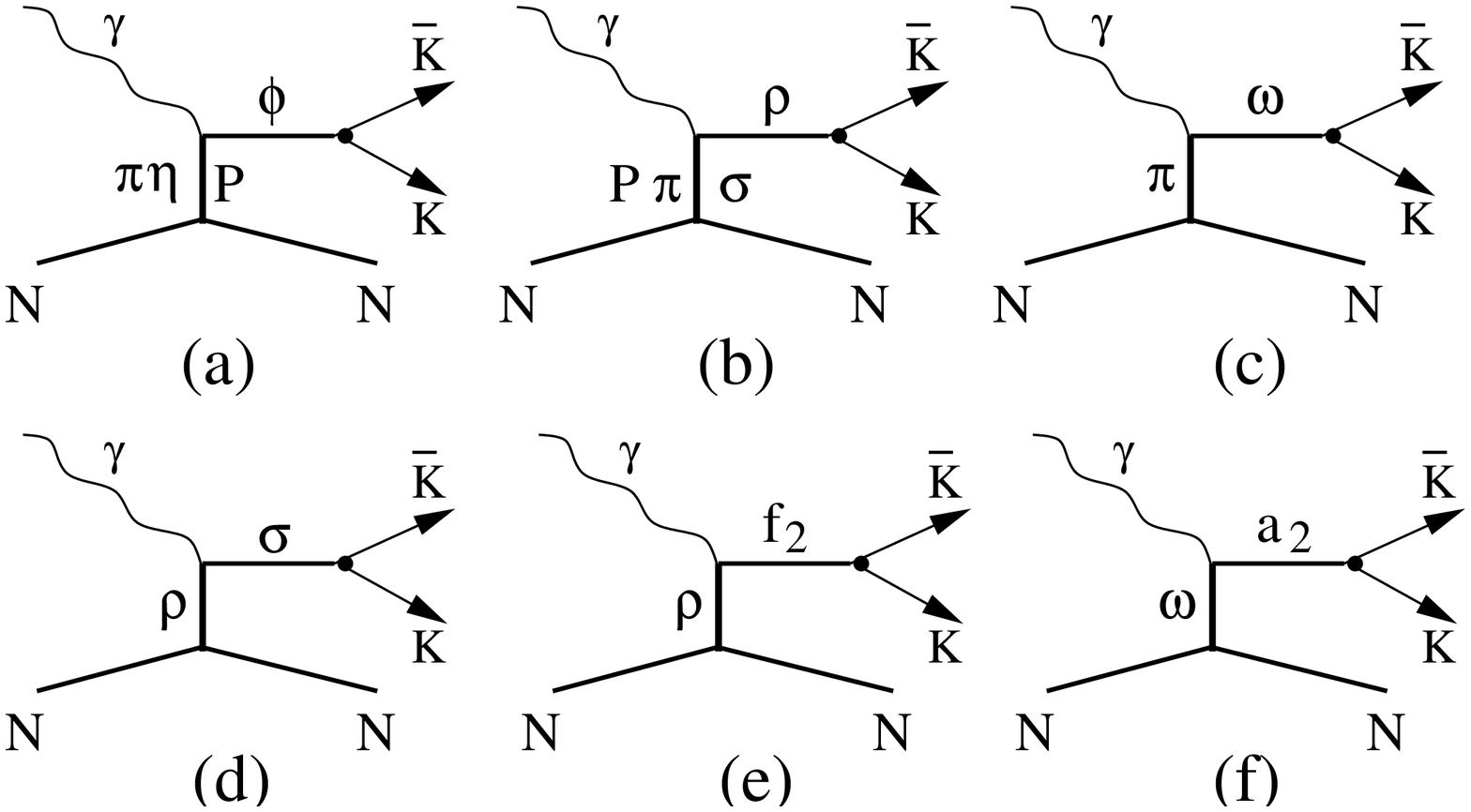}
 \caption{\label{fig:2}\tcaps%
 Diagrams for the background process for the $\gamma N\to MN\to NK\bar{K}$
 reaction, where $M$ denotes the mesons $\rho$, $\omega$, $\phi$, $\sigma$, $f_2$, and $a_2$.}
 }
\end{figure}

The aim of  the present paper  is to discuss these important aspects. We show that strict
predictions  for the $\gamma N\to \Theta^+\bar{K}$ reactions lose their definiteness in
the case of the $\gamma N\to NK\bar{K}$ processes, where $\Theta^+$ decays strongly into
$NK$; they become model-dependent. Nevertheless, we try to identify the kinematic regions
where this dependence is weak and a clear difference is expected for different $\Theta^+$
parities. In the following discussion, the term ``resonant" is applied to processes which
proceed through the virtual $\Theta^+$-state, whereas the term ``non-resonant" is used
for all other processes. The latter may have intermediate non-exotic resonant states as
well. The resonant amplitude consists of $s$-, $u$-, $t$-channel terms and the contact
($c$) term defined by the $\Theta^+NK$ interaction, as depicted in Fig.~\ref{fig:1}a-d.
We also have a $t$-channel $K^*$ exchange as shown in Fig.~\ref{fig:1}e. We found that
the main contribution to the ``non-resonant" background comes from the virtual
vector-meson photoproduction $\gamma N\to VN\to NK\bar{K}$, depicted in
Fig.~\ref{fig:2}a-c. We also have the excitation of the virtual scalar ($\sigma$) and
tensor ($f_2, a_2$) mesons shown in Fig.~\ref{fig:2}d-f, respectively, and found that
their contribution in the near-threshold region with $E_\gamma\approx 2$ GeV is
negligible.

 We write $g_{\Theta NK^*}=\alpha g_{\Theta NK}$ in order to be able to pull out an
overall factor of $g^2_{\Theta NK}$ from all contributions shown in  Fig.~\ref{fig:1}. In
view of the proportionality $g_{\Theta NK}^2\propto \Gamma_\Theta$,  the dependence of
the amplitudes on $\Gamma_\Theta$ then disappears if we consider the observables at the
resonance position. The total amplitude depends on the relative-strength parameter
$\alpha$ which will be fixed by experimental data. The dominance of the $K^*$ exchange
channel~\cite{Oh-2,CloseZhao} allows us to reduce the number of relevant input parameters
at a given coupling scheme to four: the sign and absolute value of $\alpha= g_{\Theta
NK^*}/g_{\Theta NK}$ and the sign and absolute value of $\kappa^*$.

 We will analyze two reactions, $\gamma p\to pK^0 \bar{K}^0$ and $\gamma n\to nK^+ K^-$,
and we shall refer to them as the $\gamma p$ and the $\gamma n$ reactions, respectively.

In Sec.~II, we describe our  formalism. In Sec.~III, we discuss the non-resonant
background. The procedure to fix the  parameters for the resonant amplitude is discussed
in Sec.~IV. The results of our numerical calculations for unpolarized and spin
observables are presented in Sec.~V. Our summary is given in Sec.~VI. In Appendix~A, we
show an explicit form of the transition operators for the resonance amplitude. In
Appendix~B, we discuss the Pomeron exchange amplitude, and in Appendix~C, we provide the
parameters of the background amplitude.

\section{Formalism}

\subsection{Kinematics and cross sections}

 \begin{figure}[b] \centering
   \includegraphics[width=100mm]{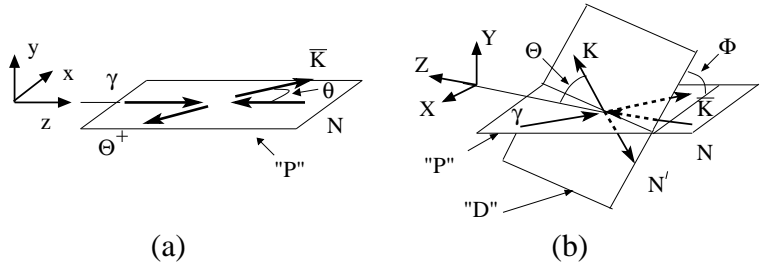}
 \caption{\label{fig:28}\tcaps%
 Schematic description  of the $\Theta^+$ production
 in  (a) the $\gamma N \to \bar{K} \Theta^+$ reaction in the center of mass system
and in (b) the reaction  $\gamma N\to NK\bar{K}$ in the $\Theta^+$ rest frame.
 The notations ``P" and ``D" correspond
 to the production and decay planes, respectively.}
 \end{figure}

The scattering amplitude $T$ of the $\gamma N \rightarrow NK\bar{K}$ reaction is related
to the $S$-matrix by
\begin{equation}
S^{}_{fi} = \delta^{}_{fi} - i(2\pi)^4 \delta^{(4)}(k + p - q -\bar{q}
- p') {\cal T}^{}_{fi}~,
\label{S:conv}
\end{equation}
where $k$, $p$, $q$, $\bar q$, and $p'$ denote the four-momenta of the incoming photon,
the initial nucleon, the outgoing $K$ and $\bar{K}$ mesons, and the recoil nucleon,
respectively. The standard Mandelstam variables for the virtual $\Theta^+$
photoproduction are defined by $t= (\bar{q}-k)^2$, $s \equiv W^2 = (p+k)^2$.  The
$\bar{K}$ meson production angle $\theta$ in the center-of-mass system (cms) is given by
$\cos\theta = \mathbf{k} \cdot \mathbf{\bar{q}} / |\mathbf{k}| |\mathbf{\bar{q}}|$. The
$\Theta^+$ decay distribution is described by the polar ($\Theta$) and azimuthal ($\Phi$)
angles of the outgoing kaon, with solid-angle element $d\Omega_Z= d\cos\Theta\, d\Phi$.
In the center-of-mass system, the quantization axis ($\bf z$) is chosen along the photon
beam momentum, and the $\bf y$-axis is perpendicular to the production plane, i.e., $\bf
y=k\times \bar q/|k\times \bar q|$. The $\Theta^+$ decay distribution is analyzed in the
$\Theta^+$ rest frame, where the quantization axis {$\bf Z$} is chosen along the incoming
(target) nucleon and $\bf{Y}=\bf{y}$. For completeness, the production and decay planes
with the corresponding coordinate systems are depicted in Fig.~\ref{fig:28}. We use the
convention of Bjorken and Drell~\cite{BD} to define the $\gamma$ matrices; the Dirac
spinors are normalized as $\bar u(p)\gamma_\alpha u(p)=2p_\alpha$. The invariant
amplitude {$T_{fi}$ is related to ${\cal T}_{fi}$ by}
\begin{equation}
 {\cal T}^{}_{fi} = \frac{T_{fi}} {\sqrt{(2\pi)^{15}\,
 2|\mathbf{k}|\,
 2E_{K}(\mathbf{q})\, 2E_{\bar{K}}(\mathbf{\bar{q}}')\,
 2E_{N}(\mathbf{p})\, 2E_{N}(\mathbf{p}')} }~,
 \label{T:conv}
\end{equation}
where $E^{}_i(\mathbf{p}) = \sqrt{M^2_i + \mathbf{p}^2}$, with $M^{}_i$ denoting the mass of
particle $i$. The differential cross section is related to the invariant amplitude by
\begin{eqnarray}
 \frac{d\sigma_{fi}}{d\Omega d\Omega_Z dM_{\Theta}}=
 \frac{p_F\sqrt{\lambda(s,M_{\Theta}^2,M_{\bar{K}}^2)}}
 {32(2\pi)^5W(W^2-M_N^2)}\, \frac14\sum_{m_i,m_f,\lambda_\gamma}
 |T_{fi}|^2~,
 \label{cs}
\end{eqnarray}
where $\lambda(x,y,z)$ is the standard triangle kinematics function,
$p_F=\sqrt{\lambda(M_{\Theta}^2,M_N^2,M_K^2)}/2M_{\Theta}$ is the $\Theta^+$ decay
momentum, $M_{\Theta}$ is the invariant mass of the outgoing nucleon and $K$-meson, $m_i$
and $m_f$ are the nucleon spin projections in the initial and the final states,
respectively, and $\lambda_\gamma$ is the incoming  photon helicity.

In the following, we consider the observables at the resonance position where the
invariant mass of the outgoing nucleon and $K$ meson is equal to the $\Theta^+$ mass,
$M_{\Theta}=M_0=1540$ MeV. In this case, the invariant amplitude of the resonant part is
expressed through the $\Theta^+$ photoproduction (${ A}$) and the $\Theta^+$ decay ($D$)
amplitudes according to
\begin{eqnarray}
 T^{\pm}_{m_f;m_i,\lambda_\gamma}
 =\sum_{m_{\Theta}}\,{
 A}^{\pm}_{m_{\Theta};m_i,\lambda_\gamma}\frac{1}{M_\Theta\Gamma_\Theta}
 D^{\pm}_{m_{\Theta};m_f}~,
 \label{TAD}
\end{eqnarray}
where  plus (minus) corresponds to the positive (negative) $\Theta^+$ parity
($J^P=\frac{1}{2}^{\pm}$), $m_{\Theta}$ is the $\Theta^+$ spin projection. In the
$\Theta^+$ rest frame, the decay amplitudes (p- and s-waves for the positive and
negative $\Theta^+$ parity) read
\begin{equation}
\begin{split}
 \frac{1}{M_\Theta\Gamma_\Theta}D^{+}_{m;m_f}
 &=D^0\left(2m\,\delta_{m,  m_f}\cos\Theta + \delta_{-m, m_f}\,\text{e}^{2im\Phi}\,\sin\Theta\right)~,
 \\
 \frac{1}{M_\Theta\Gamma_\Theta} D^{-}_{m; m_f}&=-D^0\,\delta_{m, m_f}~,
 \qquad\qquad
  D^{0}=\sqrt{\frac{4\pi}{p_F\Gamma_{\Theta}}}~,
 \end{split}
 \label{Dpm}
\end{equation}
where $\Gamma_{\Theta}$ is the total width of the $\Theta^+$ decay. After integrating
over the decay angles ($d\Omega_Z$) in Eq.~(\ref{cs}), one obtains
\begin{eqnarray}
 \frac{d\sigma^R_{fi}}{d\Omega\, dM_{\Theta}}\Bigg|_{M_\Theta=M_0}=
 \frac{1}{\pi\Gamma_{\Theta}}\frac{d\sigma^{\Theta^+}_{fi}}{d\Omega}~,
 \label{csR}
\end{eqnarray}
where
\begin{eqnarray}
 \frac{d\sigma^{\Theta^+}_{fi}}{d\Omega}=
 \frac{\sqrt{\lambda(s,M_{\Theta}^2,M_{\bar{K}}^2)}}{64\pi^2
  W^2(W^2-M_N^2)}\,
 \frac14\sum_{m_i,m_f,\lambda_\gamma}
 |{ A}_{m_f;m_i,\lambda_\gamma}|^2
 \label{csT}
\end{eqnarray}
is the cross section of the $\Theta^+$ photoproduction in the $\gamma N\to
\Theta^+\bar{K}$ reaction, with $m_f=m_{\Theta}$. By using the linear relation $g_{\Theta
NK^*}=\alpha g_{\Theta NK}$, one finds that ${d\sigma^R}/{d\Omega\, dM_{\Theta}}$ does
not depend on the $\Theta^+$ decay width at the resonance position, whereas
${d\sigma^{\Theta^+}}/{d\Omega}$ does.

 \subsection{Effective Lagrangians for the  Resonant Amplitudes}

 As mentioned before, we describe the basic resonance process by considering the
photoproduction of $\Theta^+$, with a subsequent decay of $\Theta^+$ into a nucleon and a
kaon, as shown in  Figs.~\ref{fig:1}a-d. We neglect, therefore, the contributions
resulting from the photon interacting with the final decay vertex (see Ref.~\cite{NT03}
for the corresponding three additional graphs). In view of the chosen kinematics, where
the invariant mass of the final $KN$ pair is at the resonance position, this is a good
approximation since in the neglected graphs the $\Theta^+$ is far off-shell and the
graphs of Figs.~\ref{fig:1}a-d dominate the resonance contribution. From a formal point
of view, then, we lose gauge invariance of the process since this necessarily requires
also the contributions arising from the electromagnetic interaction with the decay
vertex. However, following Ref.~\cite{hhgauge} for the initial photoproduction process,
we will construct an overall conserved current by an appropriate choice of the contact
term of Fig.~\ref{fig:1}d. In view of the dominance of the resonance graphs we do take
into account, numerically we expect very little difference between our present
current-conserving results and those of a full gauge-invariant calculation.

The effective Lagrangians which define the amplitudes shown in Figs.~\ref{fig:1}a-d  are
discussed in
Refs.~\cite{Hosaka03,OKL031,NT03,LiuKo031,LiuKo032,LiuKo033,Zhao03,ZhaoKhal04,Oh-2}.
Note, different papers employ different phase conventions. Therefore, for easy reference,
we list here the explicit forms of the effective Lagrangians used in the present work. We
have\footnote{%
Throughout this paper, isospin operators will be suppressed in all
the Lagrangians and matrix elements for simplicity. They can be
easily accounted for in the corresponding coupling constants.}
\begin{subequations}
\label{Leff}
\begin{eqnarray}
 {\cal L}_{\gamma KK}^{} &=& i e\,
 (K^- \partial^\mu K^+
 - K^+\partial^\mu K^-)A_\mu~,
 \label{AKK}\\
{\cal L}^{}_{\gamma\Theta\Theta}
  &=& - e \,\bar \Theta \left(\gamma_\mu
  -\frac{\kappa_\Theta}{2M_\Theta}\sigma_{\mu\nu}\partial^\nu\right) A^\mu\Theta~,
  \label{gTT}\\
{\cal L}^{}_{\gamma NN}
  &=& - e \,\bar N \left(e_N\gamma_\mu -
  \frac{\kappa_N}{2M_N}\sigma_{\mu\nu}\partial^\nu\right) A^\mu N~,
  \label{gNN}\\
 {\cal L}^{\pm[\text{pv}]}_{\Theta NK}
  &=& \mp \frac{g_{\Theta NK}}{M_\Theta \pm M_N}
  \bar \Theta \Gamma^{\pm}_\mu (\partial^\mu K) N  + \text{h.c.}~,
  \label{TNK}\\
 {\cal L}^{[\rm pv]}_{\gamma\Theta NK}
  &=& -i \frac{eg_{\Theta NK}}{M_\Theta \pm M_N}
  \bar \Theta \Gamma^{\pm}_\mu A^\mu K N + \text{h.c.}~,
  \label{gTNK}\\
  {\cal L}^{\pm[\text{ps}]}_{\Theta NK}
  &=& -i {g_{\Theta NK}}
  \bar \Theta \Gamma^{\pm} N  + \text{h.c.}~,
  \label{TNKps}
  \end{eqnarray}
  \end{subequations}
where $A,\,\Theta$, $K$, and $N$ are the photon, $\Theta^+$, kaon,
and the nucleon fields, respectively, $\Gamma^\pm_\mu\equiv
\Gamma^{\pm} \gamma_\mu$
(with $\Gamma^+=\gamma_5$ and $\Gamma^-=1$ for positive and negative parity, respectively),
$e_p=1$, $e_n=0$,  and $\kappa_{N}$ is the nucleon  anomalous
magnetic moment ($\kappa_p=1.79$ and $\kappa_n=-1.91$).
Equation~(\ref{gTNK}) describes the  contact (Kroll-Ruderman)
interaction in the pseudo-vector coupling scheme (see
Fig.~\ref{fig:1}d), which is absent in case of the pseudo-scalar
coupling [Eq.~(\ref{TNKps})]. In addition, we consider the $K^*$
exchange process shown in Fig.~\ref{fig:1}e which is described by
the two effective Lagrangians
\begin{subequations}
\begin{eqnarray}
 {\cal L}_{\gamma K K^*}^{} &=& \frac{e g_{\gamma KK^*}}{M_{K^*}}
  \epsilon^{\alpha\beta\mu\nu}
  \partial_\alpha A_\beta
  \partial_\mu \bar{K}^{*}_\nu K  +  \text{h.c.}~,
 \label{gammaKK*}\\
 {\cal L}_{\Theta NK^{*}}^{\pm} &=& - g_{\Theta NK^*}
 \,\bar \Theta \,\Gamma^{\mp}\left(\gamma_\mu  -
 \frac{\kappa^*}
 {M_\Theta+M_N}\sigma_{\mu\nu}\partial^\nu \right) \bar{K}^{*\mu} N
 +  \text{h.c.}
 \label{TNK*}
\end{eqnarray}
\end{subequations}
 In calculating the invariant amplitudes we dress the vertices by form factors. In the
present tree-level approach, with our chosen kinematics, only the lines connecting the
electromagnetic vertex to the initial $\Theta^+KN$ vertex correspond to off-shell
hadrons. We describe the product of both the electromagnetic and the hadronic form-factor
contributions along these off-shell lines by the covariant phenomenological function
\begin{eqnarray}
 F(M,p^2)=\frac{\Lambda^4}{\Lambda^4 + (p^2-M^2)^2}~,
 \label{FF}
\end{eqnarray}
where $p$ is the corresponding off-shell four-momentum, $M$ is the mass, and $\Lambda$ is
the cutoff parameter. We conserve the electromagnetic current of the entire amplitude by
making the initial photoproduction process gauge invariant. To this end, we apply the
gauge-invariance prescription by Haberzettl~\cite{hhgauge}, in the modification by
Davidson and Workman~\cite{DavWork} (which renders the required additional contact terms
free of kinematical singularities), to construct a contact term for the initial process
$\gamma N \to \Theta^+ \bar{K}$. For pseudovector coupling, the inclusion of form factors
not only modifies the usual Kroll-Ruderman term, but also requires additional contact
terms contributing to the amplitudes. We emphasize that contributions of the latter type
are necessary even for pure pseudoscalar coupling.

The resonance amplitudes obtained for the $\gamma n$ and $\gamma p$ reactions read
\begin{subequations}
\label{res_ampl}
\begin{eqnarray}
 &&  {A}^{\pm}_{fi}(\gamma n)=\bar u_{\Theta}(p_\Theta)
 \left[
  {{{\cal M}^s}}^{\pm}_\mu +
  {{{\cal M}^t}}^{\pm}_\mu +
  {{{\cal M}^u}}^{\pm}_\mu +
  {{{\cal M}^c}}^{\pm}_\mu +
  {{{\cal M}^t}}^{\pm}_\mu(K^*)
  \right]
 u_p(p)\varepsilon^\mu,\\
&& {\cal A}^{\pm}_{fi}(\gamma p)=\bar u_{\Theta}(p_\Theta)
  \left[
 {{{\cal M}^s}}^{\pm}_\mu +
  {{{\cal M}^u}}^{\pm}_\mu +
  {{{\cal M}^c}}^{\pm}_\mu +
  {{{\cal M}^t}}^{\pm}_\mu(K^*)
  \right]
 u_p(p)\varepsilon^\mu~.
\end{eqnarray}
\end{subequations}
The explicit forms of the transition operators  ${\cal M}^{i}_\mu$ for the $\gamma n\to
\Theta^+ K^-\to n K^-K^+$ and $\gamma p\to \Theta^+\bar{K}^0\to p K^0\bar{K}^0$ reactions
are shown in Appendix A. The choice of the strength parameters $g_i$ in the effective
Lagrangians and the cutoff parameters $\Lambda=\Lambda_B$ will be discussed in Sec.~IV.

\section{Non-resonant background}

The non-resonant background for the reaction $\gamma n\to n
K^+K^-$ has been discussed qualitatively by Nakayama and
Tsushima~\cite{NT03}. Together with the virtual vector-meson
contribution $V\to K^+K^-$ ($V=\omega,\rho$, and $\phi$ mesons),
they have included the excitation of the virtual $\Sigma$
hyperons. The coupling constants in the corresponding effective
Lagrangians were fixed using the known decay widths and SU(3)
symmetry, and the particles were taken as undressed \emph{but}
with physical masses. The present analysis with dressing form
factors shows that at $E_\gamma\approx 2$ GeV, the  main
contribution comes from the virtual vector-meson photoproduction.
For completeness, we also explore the contribution from the scalar
($\sigma$) and tensor ($a_2,f_2$)  mesons. As we shall show in
Sec.~III.E, the latter is found to be negligibly small in the
$E_\gamma \approx 2$\,GeV region.

\subsection{\boldmath Vector meson contributions:  $\rho,\omega,\phi$}

\begin{figure}[b]\centering
 \parbox{.35\textwidth}{\includegraphics[width=.35\textwidth]{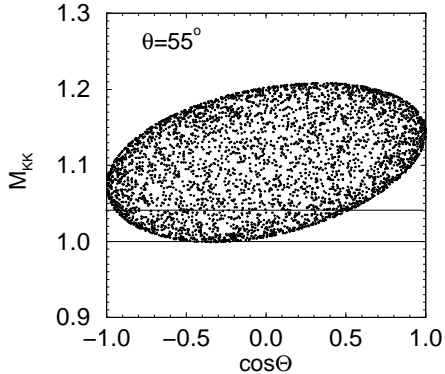}}
 \hfill
 \parbox{.6\textwidth}{%
 \caption{\label{fig:3}\tcaps%
 The phase-space diagram for the $\Theta^+$ photoproduction
 at $E_\gamma=2$ GeV: the invariant mass  $M_{K\bar{K}}$
 \textit{versus} $\cos\Theta$ at a fixed $\bar{K}$ production angle
 ($\theta=55^\circ$). The cutoff area of the phase phase
 is placed between the two solid lines.}}
 \end{figure}

Naively,  one can expect the dominance of the intermediate $\rho$-meson channel in the
background contribution because the cross section for the real $\rho$-meson
photoproduction is about an order of magnitude larger than that for the $\omega$ meson,
and it is about two orders of magnitude larger than that for the $\phi$-meson
photoproduction. However, at $E_\gamma\approx 2$ GeV, the $K\bar{K}$ invariant mass is
distributed in the region $1\,\text{GeV}\lesssim M_{K\bar{K}}\lesssim 1.2$\,GeV which
straddles the $\phi$ mass. Fig.~\ref{fig:3} shows an example of the phase-space diagram
of the $K\bar{K}$ invariant mass versus the cosine of the $K$ decay angle at a fixed
$\bar{K}$production angle of $\theta(\text{cm})=55^\circ$. One can see that the narrow
mass distribution of the $\phi$ is within the sampled kinematic region which therefore
makes the $\phi$-meson contribution significant.

In this study we consider the contribution from $\rho,\,\omega$,
and $\phi$ mesons. The low-energy $\rho$- and $\omega$-meson
photoproductions are described within an effective meson-exchange
model~\cite{TLTS99,TitovLee02,FS96}. Thus, the $\rho$-meson
photoproduction is dominated by the $t$-channel scalar ($\sigma$)
and pseudoscalar ($\pi$) meson exchanges. The $\omega$-meson
photoproduction is mostly defined by the $\pi$ exchange. In
Ref.~\cite{Oh04} the $\sigma$ exchange in the $\rho$
photoproduction is reexamined on the basis of un-correlated
two-pion and tensor $f_2$-meson exchanges. The main problem of
this approach is the requirement of a large coupling constant for
the $f_2 \rho\gamma$ interaction. This results in a  branching
ratio of $f_2\to\rho\gamma$ decay which seems to be unreasonably
large. Another ambiguity is related to the unknown $f_2NN$
coupling and the form factors for the off-shell $f_2$ meson. Since
the quality of the description of the experimental data using
either the $\sigma$ exchange or the $f_2$ exchange is comparable
to each other \cite{Oh04}, and to avoid introducing too many
unknown parameters, we employ the $\sigma$-exchange model in this
work; this is quite reasonable for the present purposes. When the
photon energy increases, we have to add the Pomeron exchange as
well. But at $E_\gamma\approx 2$ GeV, it is important only for the
$\rho$ channel where the Pomeron exchange gives about 30\% of its
contribution. In the $\omega$ channel, the Pomeron contribution is
suppressed by the factor $(\gamma_\rho/\gamma_\omega)^2\approx
6.33/72.71\approx 1/11.5$, where $\gamma_\rho$ and $\gamma_\omega$
are the $\rho$ and $\omega$ decay couplings, respectively.
Therefore, in the $\omega$ photoproduction, we limit ourselves to
the $\pi$ exchange process only. The effective Lagrangians
responsible for the meson-exchange channels read
\begin{subequations}
\label{LeffV}
\begin{eqnarray}
 {\cal L}^{}_{\pi NN}
  &=& \frac{g_{\pi NN}}{M_N}
  \bar N \gamma_5\gamma_\mu \partial^\mu \pi N~,
  \label{NNpi}\\
 {\cal L}^{}_{\sigma NN}
  &=&  g_{\sigma NN} \bar NN\sigma~,
 \label{NNsigma}\\
 {\cal L}_{\rho\sigma\gamma}^{} &=&
 \frac{e g_{\rho\sigma\gamma}}{2M_{\rho}}
 (\partial^\nu\rho^\mu - \partial^\mu\rho^\nu)
 (\partial_\nu A_\mu - \partial_\mu A_\nu)\sigma~,
 \label{gammarhosigma}\\
  {\cal L}_{V\pi\gamma}^{} &=& \frac{e g_{V\pi\gamma}}{M_{V}}
  \epsilon^{\alpha\beta\mu\nu}
  \partial_\alpha A_\beta
  \partial_\mu V_\nu \pi~,
 \label{gammarhopi}\\
{\cal L}_{V K K}^{} &=&
  ig_{V K K}
 (\bar{K}\partial^\mu K -  K\partial^\mu\bar{K} )V_\mu~,
 \label{VKK}
\end{eqnarray}
\end{subequations}
where $V$ stands for the vector meson.
The amplitudes for the $\gamma N\to NV\to NK\bar{K}$ reaction may be
expressed as
\begin{eqnarray}
 T_{fi}=\sum\limits_{\lambda_V}
 A^V_{m_f\lambda_V;m_i\lambda_\gamma} \frac{1}
 {M^2_V -M_{K\bar{K}}^2 + i\Gamma_VM_V} D^V_{\lambda_V}\, F_V(M^2_{K\bar{K}})~,
 \end{eqnarray}
 where $A^V$ and $D^V$ are the vector meson photoproduction ($\gamma N \to V N$) and
 decay ($V \to K\bar{K}$) amplitudes, respectively, $M_{K\bar{K}}$ is the $K\bar
 K$ invariant mass, $\Gamma_V$ is the total decay width of the vector meson,
 $\lambda_\gamma$ and $\lambda_V$ are the helicities of the photon and
 vector meson, respectively, and $F_V$ is the form factor of the virtual vector meson.

 The photoproduction amplitudes may be expressed in a standard form
 \begin{eqnarray}
 A^{V}_{m_f\lambda_V;m_i\lambda_\gamma}=\bar u_f{\cal M}^{}_{\mu\nu}u_i\,
 \varepsilon^\mu_{\lambda_\gamma}{\varepsilon^*}^\nu_{\lambda_V}~.
\end{eqnarray}
 In the case of the scalar (S: $\sigma$) and pseudoscalar (PS: $\pi$)
 meson exchange, the transition operators ${\cal M}^{}_{\mu\nu}$ read
\begin{eqnarray}
 {\cal M}_{\mu\nu}^{\text{S}}&=&
  \frac{eg_{\rho\sigma\gamma} g_{\sigma NN} }{M_\rho}
 \frac{ g_{\mu\nu}(k\cdot Q_V) - {Q_V}_\mu k_\nu}{\bar t-M^2_\sigma}
F_\sigma(\bar t)~,
 \label{Tbg-s}\\
  {\cal M}_{\mu\nu}^{\text{PS}}&=&
  i\frac{eg_{\rho\pi\gamma}g_{\pi NN}}{M_\rho}\gamma_5
  \frac{\varepsilon^{\mu\nu\alpha\beta} k_\alpha {Q_V}_\beta}
  {\bar t - M^2_\pi}F_\pi(\bar t)~,
 \label{Tbg-pi}
\end{eqnarray}
respectively, where $Q_V=q+\bar q$,  and $F_M$ is the product of
the two form factors of the virtual exchanged mesons in the $MNN$
and $\gamma VM$ vertices. The explicit forms of $F_M(\bar{t})$
are given in Appendix~C. Note that the four-momentum transfers to
the $K\bar{K}$ pair, $\bar t$ is different from  $t$ in $\Theta^+$
photoproduction.

The decay amplitude,
\begin{eqnarray}
 D^V_{\lambda_V}= g_{VKK}( q-\bar
q)_\mu\varepsilon^\mu_{\lambda_V}~,
\label{D1}
\end{eqnarray}
has a simple form in the $V$ rest frame, with the quantization axis
$\mathbf{z}$ parallel to the beam momentum (Gottfried-Jackson
system), i.e.,
\begin{eqnarray}
  D^V_{\lambda_V}=2k_f g_{VKK} \sqrt{\frac{4\pi}{3}}
  Y_{1\lambda_V}(\widehat{\mathbf{q}})~,
  \qquad
  k_f=\frac{M_{K\bar{K}}}{2}\sqrt{1-4\frac{M_K^2}{M^2_{K\bar{K}}}}~,
  \label{D2}
\end{eqnarray}
where $\widehat{\mathbf{q}}$ is the solid angle of the direction of flight of the $K$ meson in
the $K\bar{K}$ rest frame, i.e., $ \widehat{\mathbf{q}}\equiv\Omega_K$. The $\phi$-meson
photoproduction is defined by the Pomeron and pseudoscalar ($\pi,\eta$) meson exchanges,
as described in Ref.~\cite{Titov03}. For an easy reference, we provide the transition
operator ${\cal M}^P_{\mu\nu}$ for the Pomeron exchange amplitude and the parameters
which define the $\phi$-, $\rho$-, and $\omega$-meson photoproduction in Appendices~B and
C, respectively.

\begin{figure}[t] \centering
   \includegraphics[width=.3\textwidth]{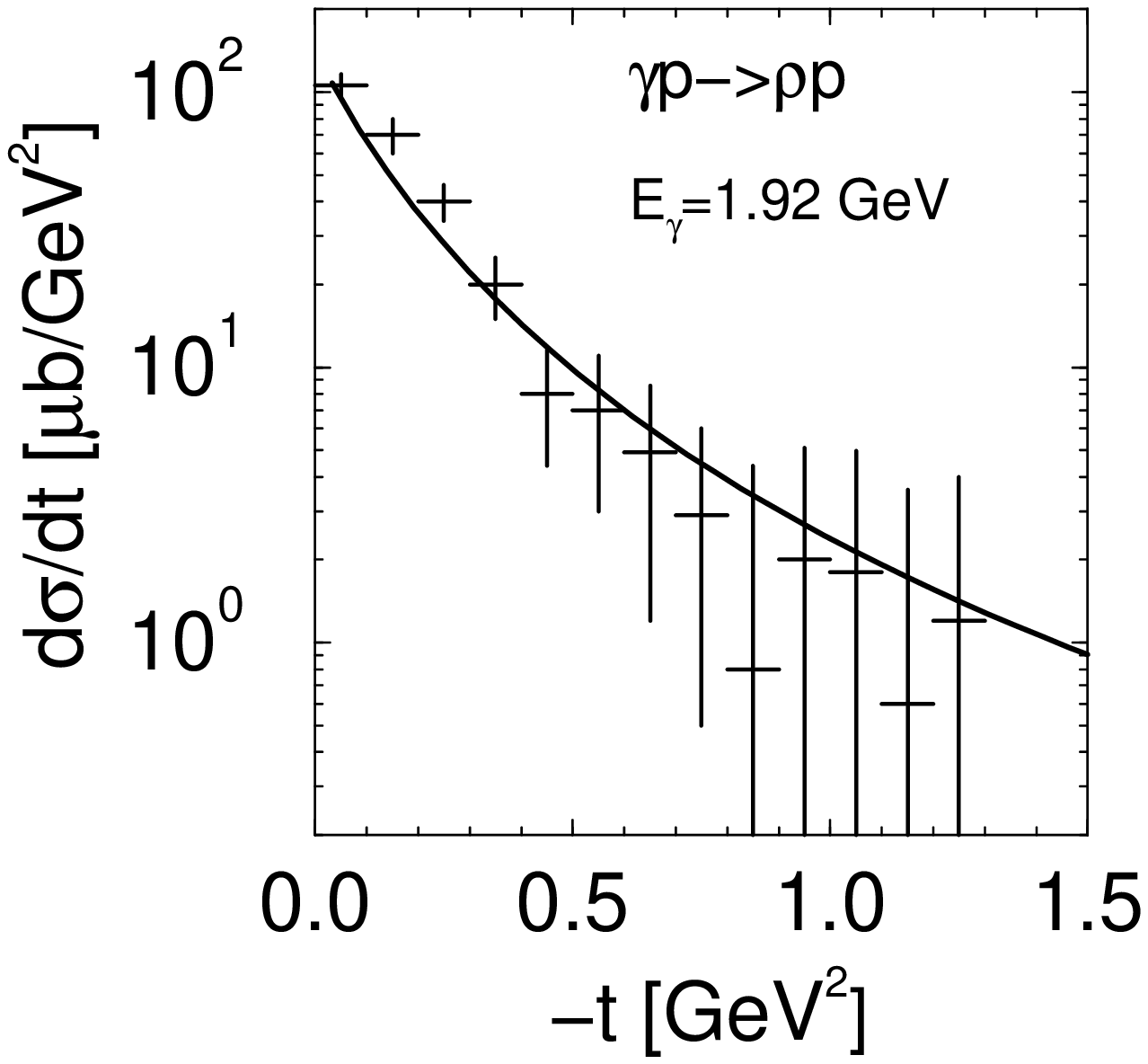}\hfill
   \includegraphics[width=.3\textwidth]{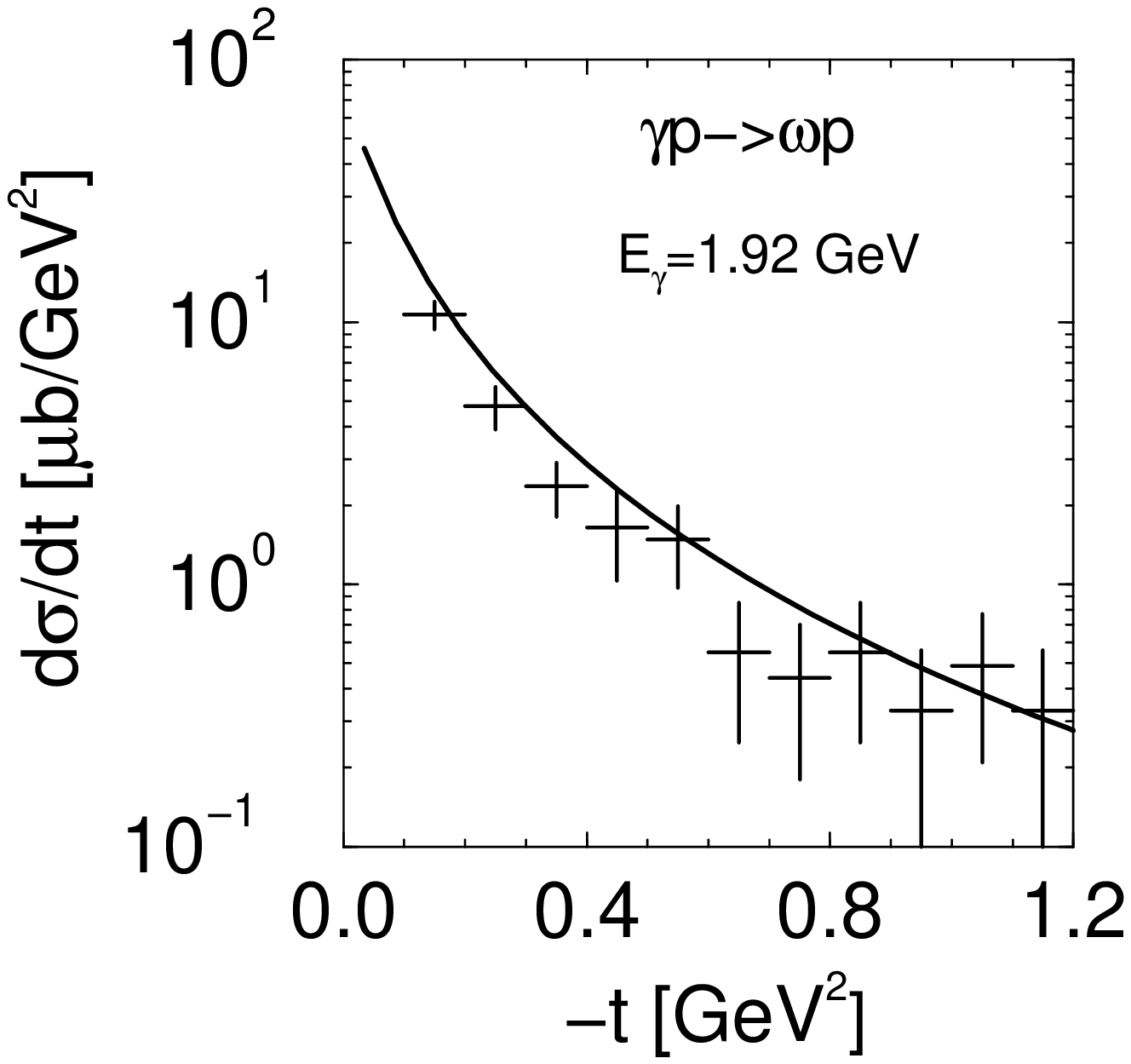}\hfill
   \includegraphics[width=.3\textwidth]{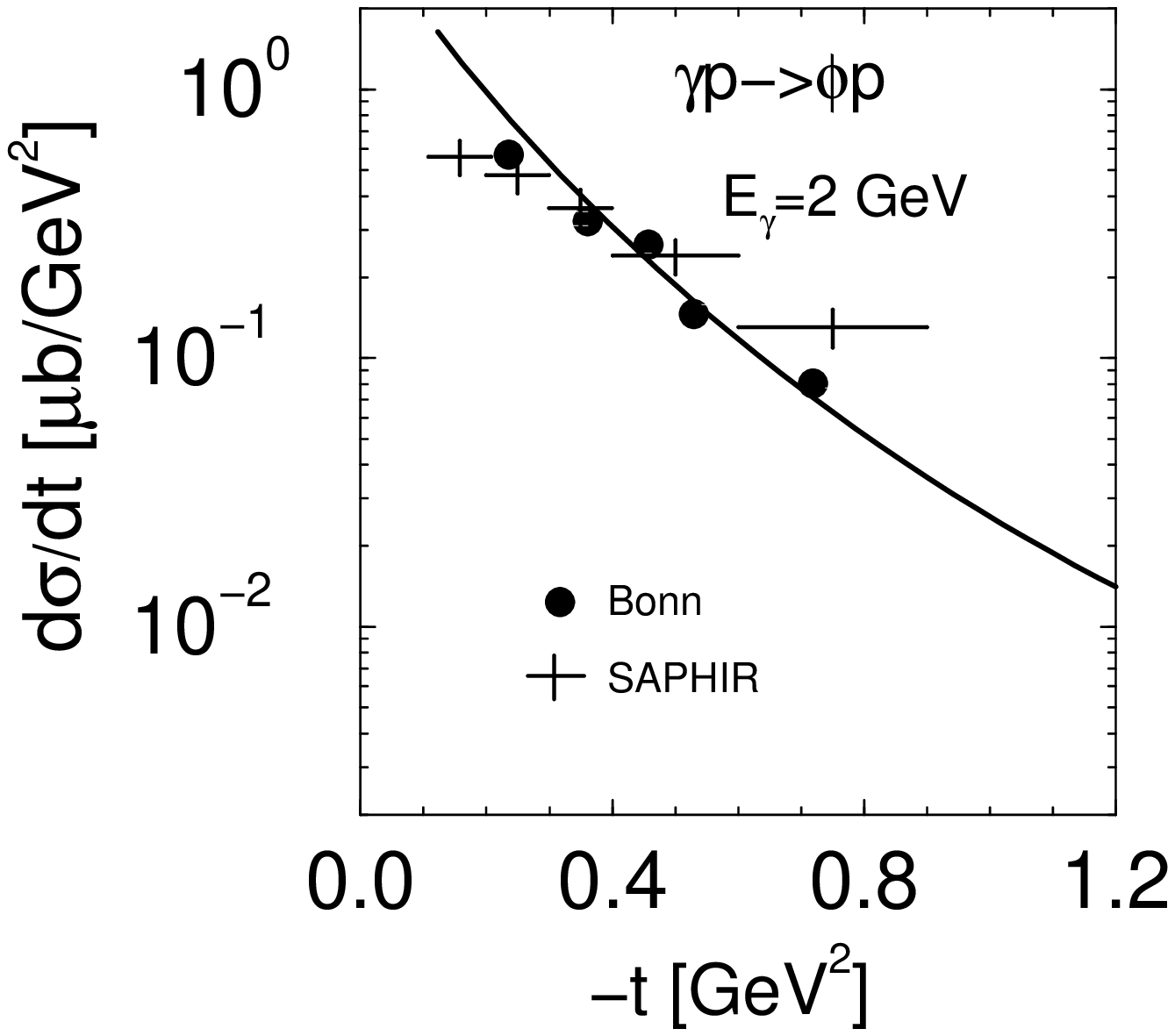}
\caption{\label{fig:4}\tcaps%
Cross sections for the $\rho$-, $\omega$-, and $\phi$-meson
 photoproduction. The experimental points are taken from
 Refs.~\protect\cite{SAPHIR,BONN,SAPHIR2}.}
\end{figure}

Figures~\ref{fig:4}a-c show the cross sections of the vector-meson photoproduction in the
reactions $\gamma p\to Vp$ ($V=\rho,\omega,\phi$) at $E_\gamma \approx 2$ GeV together with
the available data. One can see a quite reasonable agreement between the experimental
data and the calculation. This encourages us to use the same model for the description of
the non-resonant background in the $\Theta^+$ photoproduction.

\subsection{\boldmath Scalar meson contribution: $\sigma$}

The $\gamma \rho\sigma$ interaction responsible for the $\rho$-meson contribution
naturally leads to  the virtual $\sigma$-meson photoproduction and its subsequent decay
into the $K\bar{K}$ pair as  shown in Fig.~\ref{fig:2}d. The corresponding effective
Lagrangians for this transition read
\begin{subequations}
 \label{S1}
\begin{align}
 {\cal L}_{\sigma KK}&=M_\sigma g_{\sigma KK}\sigma K\bar{K}~,
\\
  {\cal L}_{\rho NN}&=-g_{\rho NN}\left(\bar N\gamma_\mu \rho^\mu N
 -\frac{\kappa_\rho}{2M_N}
 \bar N\sigma_{\mu\nu}\partial^\nu\rho^\mu N\right)~.
\end{align}
\end{subequations}
The $\sigma K\bar{K}$ coupling may be estimated from SU(3) symmetry as $g_{\sigma KK}=-
g_{\sigma \pi\pi}/2$.

 The amplitude for the $\gamma N\to\sigma N\to NK\bar{K}$ process has the form
\begin{equation}
  A^{}_{m_f;m_i\lambda_\gamma}=\bar u_f{\cal M}^{\rho}_{\mu} u_i
 \varepsilon^\mu_{\lambda_\gamma}~,
 \label{S2}
\end{equation}
where
\begin{equation}
  {\cal M}_{\mu}^{\rho}=
 \frac{ eg_{\rho\sigma\gamma} g_{\rho NN}}{M_\rho (\bar t- M_\rho^2)   }
 \left[{(1+\kappa_\rho)\hat R_\mu -{\kappa_\rho} \frac{R'_\mu}{M_N}}\right]
  F_\rho (\bar t)~,
  \label{S22}
\end{equation}
with
\begin{subequations}
\begin{align}
\hat R_\mu&=(k\cdot Q)\gamma_\mu - \fs k Q_\mu~,
\\
R'_\mu &=(k\cdot Q)p'_\mu  -  (k\cdot p') Q_\mu~,
\\
Q &= p'-p~.
\end{align}
\end{subequations}
The $\sigma \to K\bar{K}$ decay amplitude is a constant
\begin{eqnarray}
D_\sigma=-M_\sigma g_{\sigma KK}\sqrt{4\pi}Y_{00}(\widehat{\mathbf{q}})~.
\label{S3}
\end{eqnarray}
For the vector and tensor coupling constants and the cutoff parameter in $F_\rho$, we
take the corresponding values from the Bonn model~\cite{BonnMod} (see Appendix~C).

\subsection{\boldmath Tensor meson contributions: $a_2$, $f_2$}

The tensor $a_2(1320)$ and $f_2(1270)$ mesons have finite branching ratios into the
$K\bar{K}$ decay channel and, therefore, one can expect their non-negligible contributions
to the non-resonant background in the $\Theta^+$ photoproduction~\cite{Dzierba04}. The
corresponding branching ratios are $(4.9\pm 0.8)\times10^{-2}$ and $(4.6\pm
0.5)\times10^{-2}$ for the $a_2(1320)$ and $f_2(1270)$ mesons, respectively~\cite{PDG}.

Similarly to the scalar meson, the tensor mesons ($a_2$ and $f_2$) appear in the
``non-resonant" background in two ways. Firstly, $a_2$ and $f_2$ give a contribution to
the photoproduction of the $\omega$ and $\rho$ mesons , respectively. Secondly, they can
be produced directly by the incoming photon with subsequent exchange by $\omega$ and
$\rho$ mesons, as it is depicted in Fig.~\ref{fig:2}e,f. The first case results only in
some renormalization of the coupling constants in the $\omega$ and $\rho$ photoproduction
amplitudes, and it does not modify the shape of the $K\bar{K}$ invariant mass
distribution. But since the coupling constants in the $\rho$ and $\omega$ photoproduction
processes are fixed by the corresponding  experimental data anyway, we may assume that
such a renormalization effect is taken into account in those effective strength
parameters. The second contribution may change the $K\bar{K}$ invariant distribution
qualitatively  when $M_{K\bar{K}}\approx M_{f_2(a_2)}$. This is realized at higher energies
with $E_\gamma\geq 2.3$\,GeV. At $E_\gamma\approx 2$\,GeV, their contribution is not very
strong; nevertheless, for completeness, we include these processes in our consideration.
We assume that production of the $a_2$ and $f_2$ mesons is generated by the
$a_2\gamma\omega$ and $f_2\gamma\rho$ interactions, respectively.

The interaction of the tensor and the two vector fields is described by the
gauge-invariant interaction
\begin{equation}
 \label{T1}
 \begin{split}
 {\cal L}_{t_2V_1V_2}&= \frac{g_{t_2V_1V_2}}{M_{t_2}}(L^{\alpha\beta} +
 L^{\beta\alpha})\xi_{\alpha\beta}~,
 \\
 L^{\alpha\beta}&=(\partial^\alpha V_1^\mu -\partial^\mu V_1^\alpha)
 (\partial^\beta {V_2}_\mu -\partial_\mu V_2^\beta)~,
 \end{split}
\end{equation}
where $V_{1,2}$ and $\xi$ are the vector and tensor meson fields,
respectively, and $t_2=a_2, f_2$.

The $t_2 K\bar{K}$ interaction  is described by the
effective Lagrangian
\begin{eqnarray}
 {\cal L}_{t_2 K K}=\frac{g_{t_2 KK}}{M_{t_2}}
 \left(\partial^\beta\bar{K} \partial^\alpha K + \partial^\beta K\partial^\alpha \bar{K}
  \right)  \xi_{\alpha\beta}~.
\label{T2}
\end{eqnarray}
The coupling constant  $g_{t_2 KK}$ is related to the
  decay width  $\Gamma_{t_2\to K\bar{K}}$ width that subsumes
  transitions into both $K^0\bar{K}^0$ and $K^+K^-$ channels
according to
\begin{eqnarray}
 {g^2_{t_2 KK}}=
 \frac{15\pi\Gamma_{t_2\to K\bar{K}}M_{t_2}^4}{2p_{t_2}^5}~,
 \label{T3}
\end{eqnarray}
where $p_{t_2}=M_{t_2}\sqrt{1/4-M_K^2/M^2_{t_2}}$ is the $t_2$ decay momentum. Here we
assume that $\Gamma_{t_2\to K^+ K^-}\approx \Gamma_{t_2\to K^0\bar{K}^0}\approx 0.5
\Gamma_{t_2\to K\bar{K}}$. Using the known widths  $\Gamma_{a_2\to K\bar{K}}$ and
$\Gamma_{f_2\to K\bar{K}}$ from~\cite{PDG} we get
\begin{eqnarray}
 {g_{a_2 K K}}=4.9~,\qquad {g_{f_2 K K}}=7.4~.
 \label{T4}
\end{eqnarray}

The amplitude of the $\gamma N\to Nt_2\to N K\bar{K}$ transition is expressed as
\begin{eqnarray}
 T_{fi}=\sum\limits_{\lambda_\gamma\sigma}
 A^{t_2}_{m_f\sigma;m_i\lambda_\gamma} \frac{1}
 {M^2_{t_2} -M_{K\bar{K}}^2 + i\Gamma_{t_2} M_{t_2}}
  D^{t_2}_{\sigma}\, F_{t_2}(M^2_{K\bar{K}})~,
\label{T5}
 \end{eqnarray}
where $A^{t_2}$ is the tensor meson photoproduction ($\gamma N \to
t_2 N$) amplitude,   $\Gamma_{t_2}$ is the total decay width of
the tensor meson,  $\sigma$ is the spin projection of the tensor
meson, and $F_{t_2}$ is the form factor of the virtual tensor
meson. The $t_2\to K\bar{K}$ decay amplitude reads
\begin{eqnarray}
 D^{t_2}_{\sigma} =-\frac{2g_{t_2KK}k_f^2}{M_{t_2}}
 \sqrt{\frac{8\pi}{15}}Y_{2\sigma}(\widehat{\mathbf{q}})~,
 \label{T6}
\end{eqnarray}
 The $a_2$-meson  photoproduction amplitude is given by
\begin{eqnarray}
 A^{a_2}_{m_f\sigma;m_i\lambda_\gamma}=
 \frac{g_{a_2\gamma\omega} g_{\omega NN}}{M_{a_2}}
 \bar u_f H^{a_2}_{\alpha\beta\mu}\,u_i\,\varepsilon^{*\alpha\beta}\,
 \varepsilon^{\mu}_{\lambda_\gamma}\,F_\omega(\bar t)~,
\label{T7}
 \end{eqnarray}
 where $\varepsilon^{\alpha\beta}$ is the Rarita-Schwinger  spinor of the tensor meson
 \begin{eqnarray}
 \varepsilon_{\alpha\beta}(\sigma)=\sum\limits_{l_1l_2}
 \langle
 1l_1\,1l_2|2\sigma\rangle\,\varepsilon_\alpha(l_1)\varepsilon_\beta(l_2)~,
 \label{T8}
 \end{eqnarray}
and
\begin{eqnarray}
 H^{a_2}_{\alpha\beta\mu}=
 (Q_{\alpha}\gamma^\nu -Q^\nu\gamma_\alpha)
 (k_\beta g_{\mu\nu} - k_\nu g_{\mu\beta})
 +(Q_{\beta}\gamma^\nu -Q^\nu\gamma_\beta)
 (k_\alpha g_{\mu\nu} - k_\nu g_{\mu\alpha})~.
\label{T9}
 \end{eqnarray}
Similarly,  the amplitude of the $f_2$-meson  photoproduction
reads
\begin{eqnarray}
 A^{f_2}_{m_f\sigma;m_i\lambda_\gamma}=
 \frac{g_{f_2\gamma\rho} g_{\rho NN}}{M_{f_2}}
 \bar u_f H^{f_2}_{\alpha\beta\mu}\,u_i\,\varepsilon^{*\alpha\beta}\,
 \varepsilon^{\mu}_{\lambda_\gamma}\,F_\rho(\bar t)~,
\label{T10}
 \end{eqnarray}
with
 \begin{equation}
\label{T11}
 \begin{split}
 H^{f_2}_{\alpha\beta\mu}&=
 (Q_{\alpha}G^\nu -Q^\nu G_\alpha)
 (k_\beta g_{\mu\nu} - k_\nu g_{\mu\beta})
 +(Q_{\beta}G^\nu -Q^\nu G_\beta)
 (k_\alpha g_{\mu\nu} - k_\nu g_{\mu\alpha})~,\\
 G_\mu &=(1+\kappa_\rho)\gamma_\mu -\frac{\kappa_\rho}{M_N}p'_\mu~.
\end{split}
 \end{equation}

In the absence of experimental information necessary to extract the coupling constants
$g_{a_2\gamma\omega}$ and $g_{f_2\gamma\rho}$, we assume that
\begin{equation}
eg_{a_2\omega\gamma}  = 0.29~, \qquad eg_{f_2\rho\gamma}  = 0.14~.
\label{a2f2coup}
\end{equation}
These are rough estimates obtained by making use of the available data for the
decay widths for $a_2\to \omega \pi^+\pi^-$, $a_2\to \gamma\gamma$,
and $f_2\to \gamma\gamma $ in conjunction with the vector dominance model,
in addition to the Gell-Mann--Sharp--Wagner contact term \cite{GSW}.\footnote{
The value of the coupling constant $g_{a_2\omega\gamma}$ used in the present work
leads to the  $a_2\to \omega\gamma$ decay width
 \begin{equation*}
 \Gamma_{a_2\to \omega\gamma}\approx 0.29 \,\text{MeV},\qquad
  \frac{\Gamma_{a_2\to
  \omega\gamma}}{\Gamma_{tot}}\approx 2.76\times10^{-3}~,
  \label{a2-7}
 \end{equation*}
which is comparable with the ${a_2\to \pi\gamma}$ decay with the branching ratio
$(2.68\pm0.31)\times10^{-3}$. Our estimate, therefore, may be taken as an upper limit. The
value of the coupling constant $g_{f_2\omega\gamma}$ in Eq.~(\ref{a2f2coup}) results in
the branching ratio
 \begin{equation*}
 \Gamma_{f_2\to \rho\gamma}\approx 0.059 \,\text{MeV}~,
 \qquad
  \frac{\Gamma_{f_2\to
  \rho\gamma}}{\Gamma_{tot}}\approx  3.2\times10^{-4}~,
 \end{equation*}
which is about a factor 28 greater than the branching ratio for the $f_2\to\gamma\gamma$
decay, which is $(1.14\pm0.13)\times 10^{-5}$. Note that our estimate is about a factor
5.5 smaller than the similar estimate given in Ref.~\cite{Oh04} where  ${\Gamma_{f_2\to
\rho\gamma}}/{\Gamma_{f_2\to \gamma\gamma}}\approx  155$.}

\subsection{\boldmath $K\bar{K}$ invariant mass distribution}

If there is no limitation on the phase space of the outgoing kaons, then the integration
over the decay angle of the $K$ meson eliminates the interference terms among the scalar,
vector, and tensor meson exchange contributions, and the $K\bar{K}$  invariant mass
distribution may be expressed in a compact form as
\begin{align}
 \frac{64\pi^3(s-M_N^2)^2}{k_f}\,\frac{d\sigma}{d\bar t\,
 dM_{K\bar{K}}} &=
 M_\sigma^2g^2_{\sigma KK}\frac{|A_\sigma|^2}
 {(M_{K\bar{K}}^2
 -M_\sigma^2)^2+M_\sigma^2\Gamma_\sigma^2}
 \nonumber\\[2ex]
 &\quad \mbox{}
 +\frac{4k_f^2}{3}\left\vert\sum\limits_V
 \frac{g_{VKK} A^V_{fi} }
 {M_{K\bar{K}}^2 -M_V^2 + i M_V\Gamma_V}\right\vert^2
 \nonumber\\[2ex]
 &\quad \mbox{}
 +\frac{8k_f^2}{5}\left\vert\sum\limits_{t_2}\frac{k_f}{M_{t_2}}
  \frac{g_{t_2KK} A^{t_2}_{fi} }
 {M_{K\bar{K}}^2 -M_{t_2}^2 + i M_{t_2}\Gamma_{t_2}}\right\vert^2~,
\label{KK1}
 \end{align}
where the average over the spins/helicity in the initial state and the summation over
spin variable in the final states are implied. One can see that,  near threshold,  the
tensor meson contribution is suppressed by the factor $k_f^2/M_{t_2}^2\ll 1$ compared to
the vector meson contribution.

 \begin{figure}[t!] \centering
   \includegraphics[width=.35\textwidth]{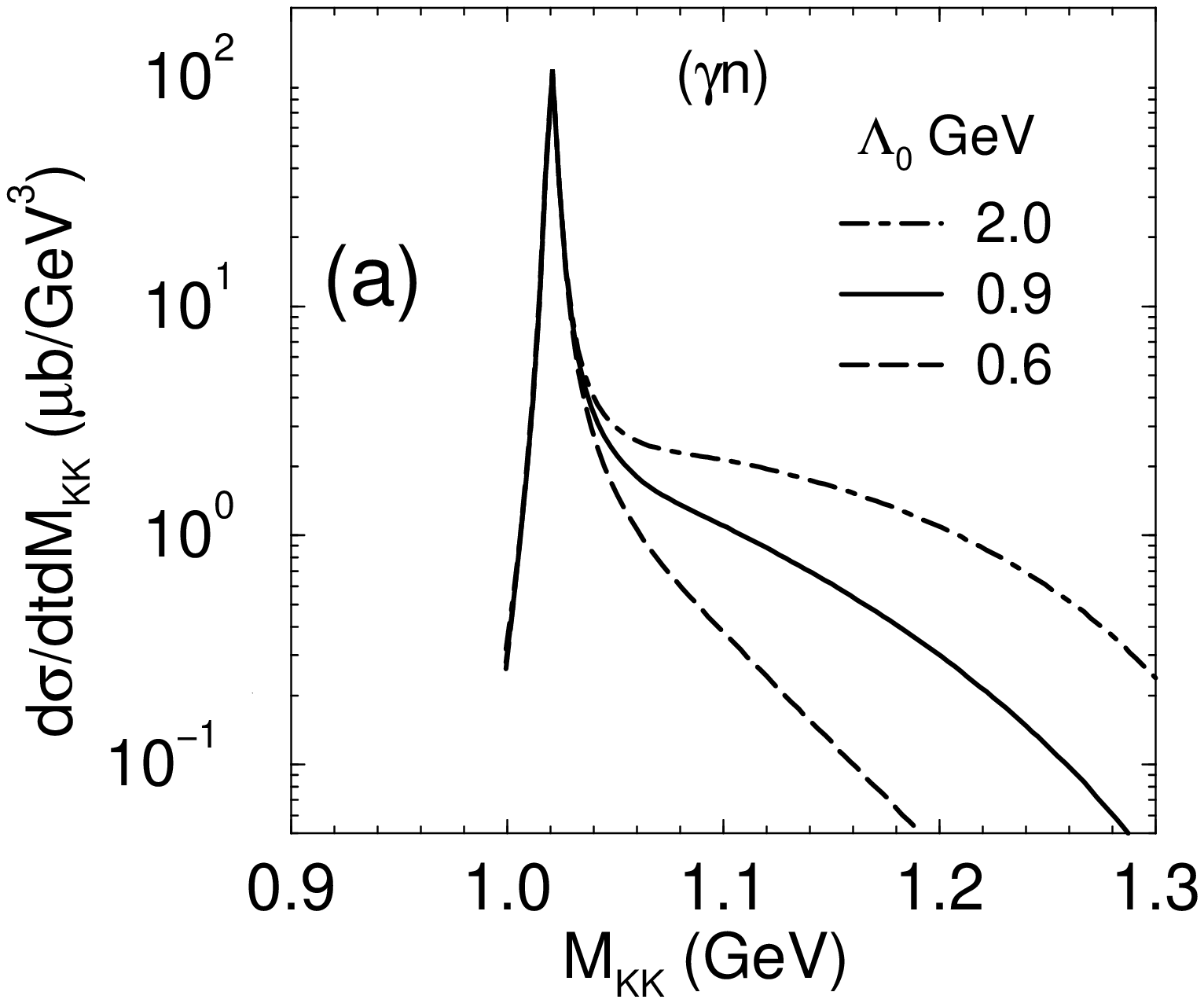}\qquad
    \includegraphics[width=.35\textwidth]{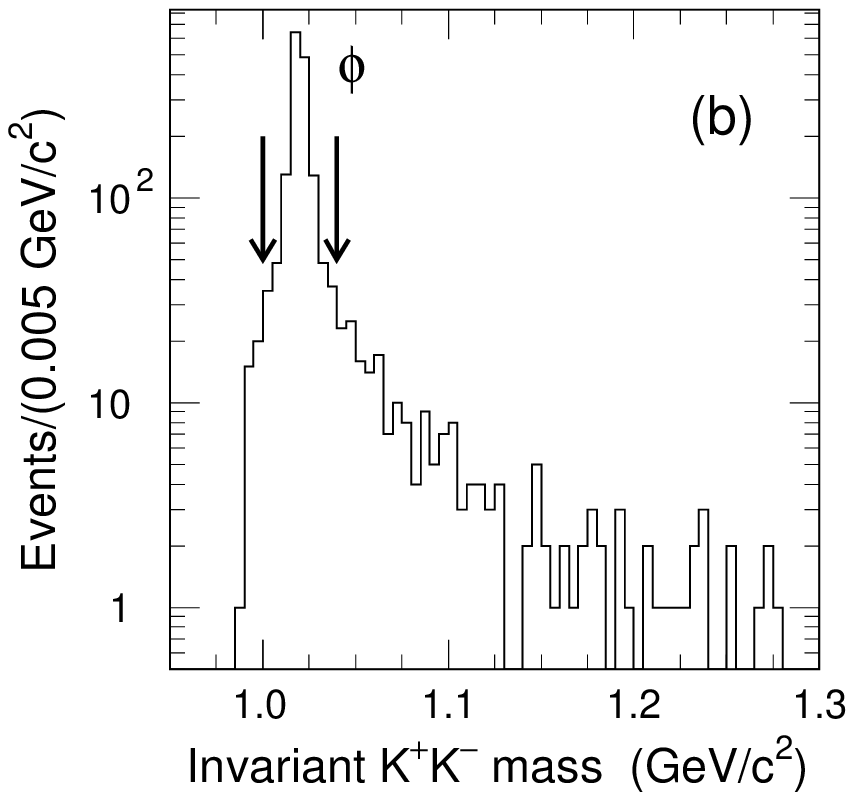}\\
 \caption{\label{fig:5}\tcaps%
(a) Dependence of the shape of the invariant mass distribution
  on the cutoff parameter $\Lambda_0$ in the $\gamma n\to K^+ K^-$ reaction.
 (b)
 $K\bar{K}$ invariant-mass distribution in
 $\gamma n\to K^+K^-$ taken from Ref.~\protect\cite{Nakano03};
 arrows indicate the $\phi$-meson cut window.}
\mbox{}\\
  \includegraphics[width=.35\textwidth]{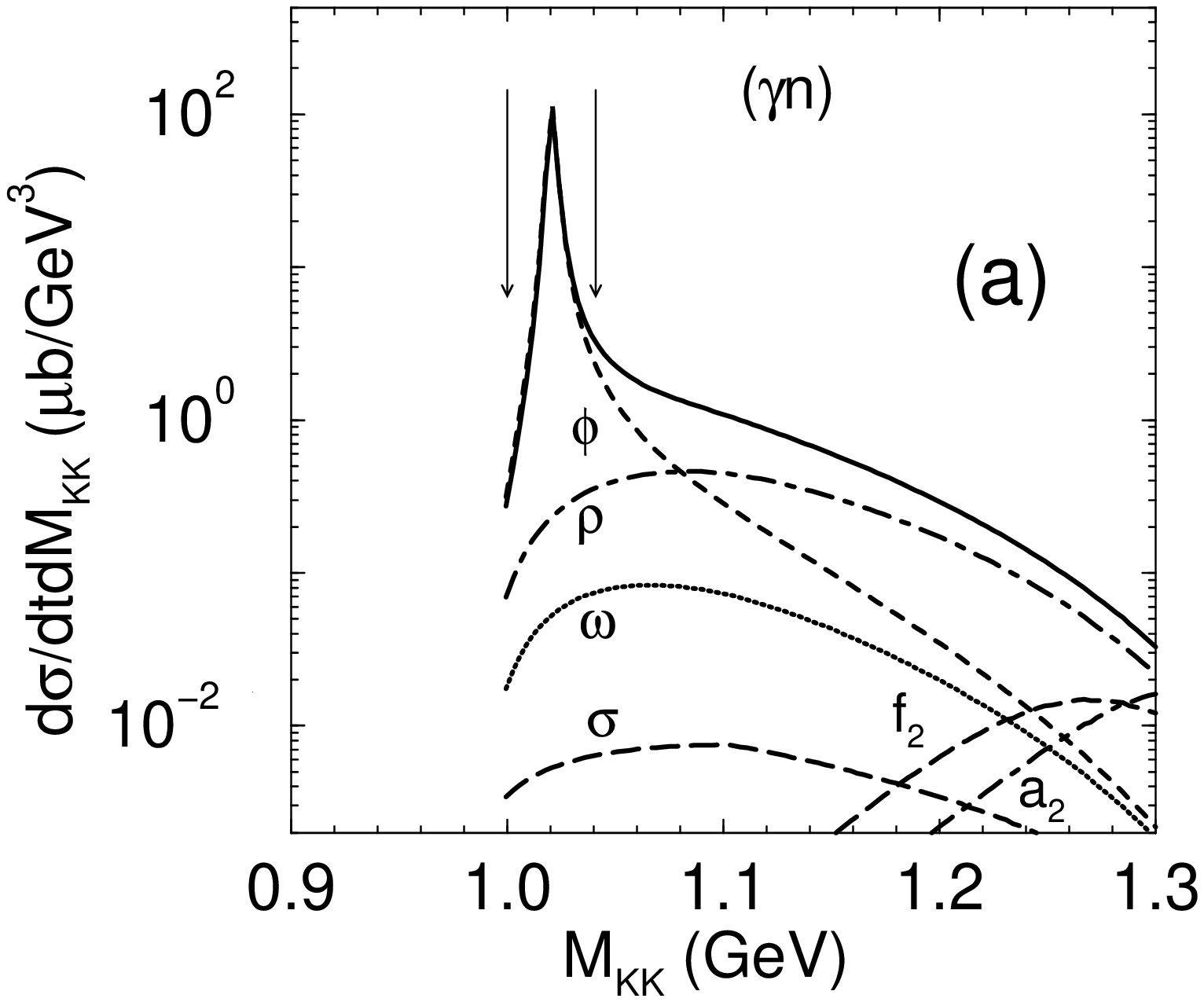}\qquad
  \includegraphics[width=.35\textwidth]{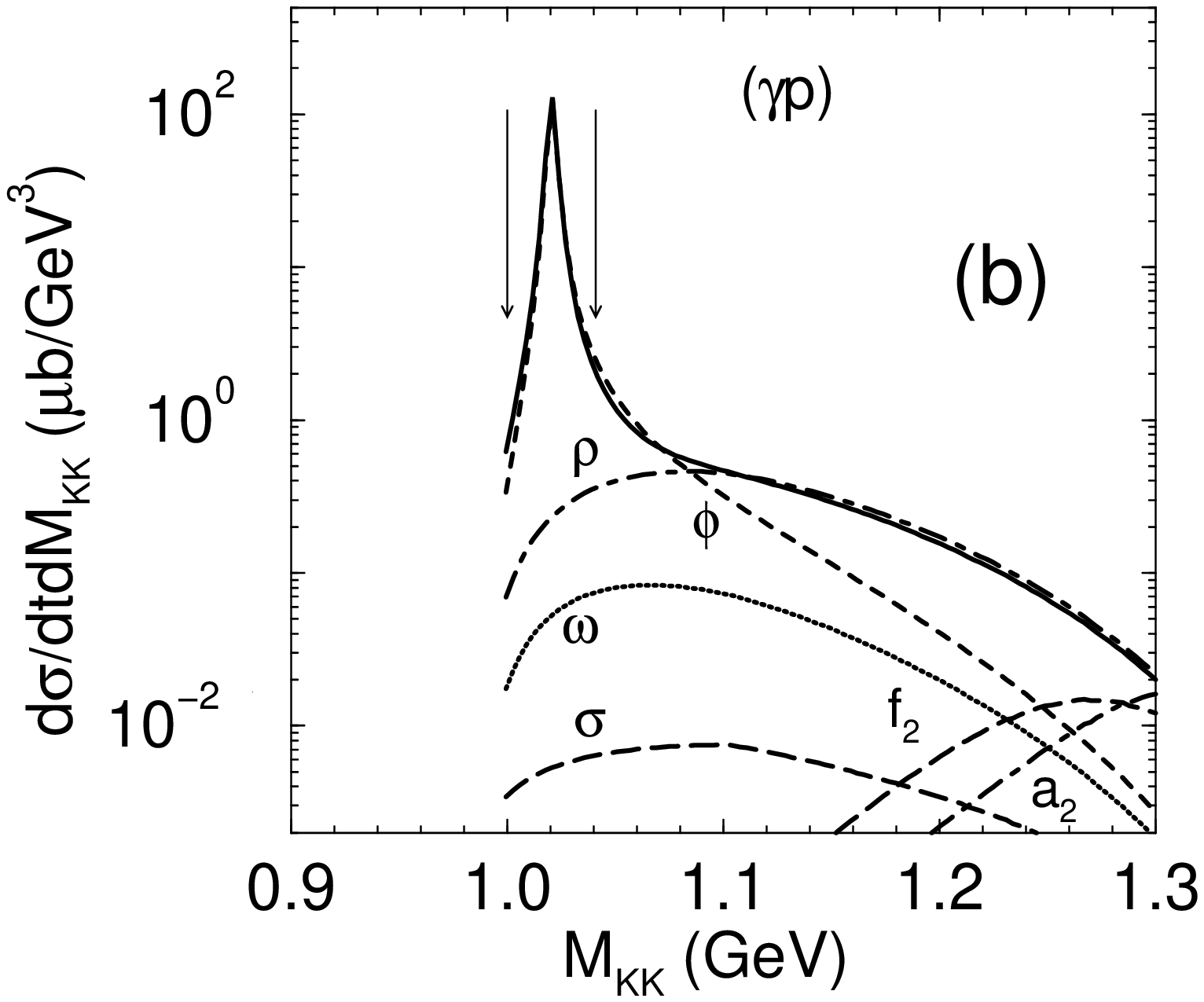}
 \caption{\label{fig:6}\tcaps%
   $K\bar{K}$ invariant-mass distribution in (a)
  $\gamma n\to K^+K^-$ and in (b)
  $\gamma p\to p K^0\bar{K}^0$. The arrows indicate the
  $\phi$-meson cut window.}
\end{figure}

The $K\bar{K}$ invariant mass distribution at a forward angle of
$K\bar{K}$ photoproduction [$\theta_{K\bar{K}}\,(\rm cm)=10^\circ$]
is shown in Figs.~\ref{fig:5} and \ref{fig:6}. The photon energy
was taken from the threshold to 2.35 GeV, in accordance with the
measurement of Ref.~\cite{Nakano03}. The  distribution has one
unknown parameter compared to the real vector meson
photoproduction. It is the cutoff parameter $\Lambda_0$ in the
form factor $F_{(V,\sigma, a_2,f_2)}=F(M_{(V,\sigma,
a_2,f_2)},M^2_{K\bar{K}})$, where $F(M,x^2)$ is defined in
Eq.~(\ref{FF}). The dependence of the $K\bar{K}$ invariant mass on
this parameter is shown in Fig.~\ref{fig:5}a. A comparison with
the measured distribution~\cite{Nakano03} shown in
Fig.~\ref{fig:5}b favors $\Lambda_0\approx 0.9$ GeV. We use this
value in our further  analysis. Figures~\ref{fig:6}a,b show the
structure of these distributions. One can see a strong
 $\phi$-meson photoproduction peak at $M_{K\bar{K}}\approx M_\phi$
 and a long tail dominated by the $\rho$-meson channel. The
 contribution from the other mesons is much smaller. The
 contribution from the $a_2$ and $f_2$ mesons becomes comparable to
 that from the vector mesons  near the threshold, $M_{K\bar
 K}\approx 1.3$ GeV (at $E_\gamma=2.2$ GeV).
 In this region of the $K\bar{K}$ invariant masses, the cross section is
 rather small compared to that around $M_{K\bar{K}}\approx M_\phi$,
 which gives the main contribution to the background of the
 $\Theta^+$ photoproduction at $E_\gamma\approx2$ GeV.

Finally, we note that the cross section for $\gamma n\to n K^+K^-$ exceeds that for
$\gamma p\to p K^0 \bar{K}^0$ by approximately a factor of two. This is due to the
distinct $\phi-\rho$ interference effect in these two reactions caused by the different
signs in the $\rho K^+K^-$ and $\rho K^0\bar{K}^0$ coupling constants. In the case of the
$\phi$-meson, the signs of the $\phi K^+K^-$ and $\phi K^0\bar{K}^0$ coupling constants
are the same.

\subsection{\boldmath Non-resonant background in $\Theta^+$ photoproduction}

Let us now examine  the background contribution to the angular distribution of the
$K\bar{K}$ pair in the final state. The $K\bar{K}$ invariant mass is taken at the
$\Theta^+$ resonance position at $M_\Theta=M_0=1.54$ GeV. The corresponding differential
cross section, $d\sigma/d\Omega\, dM_\Theta$, where $\Omega$ is the solid angle of the
$K\bar{K}$ pair, is calculated using the general expression for the differential cross
section given by Eq.~(\ref{cs}). Here, all the background channels contribute coherently
and the integration over the $\Theta^+$ decay angle $\Omega_Z$ is performed numerically.

\begin{figure}[b]\centering
\includegraphics[width=.3\textwidth]{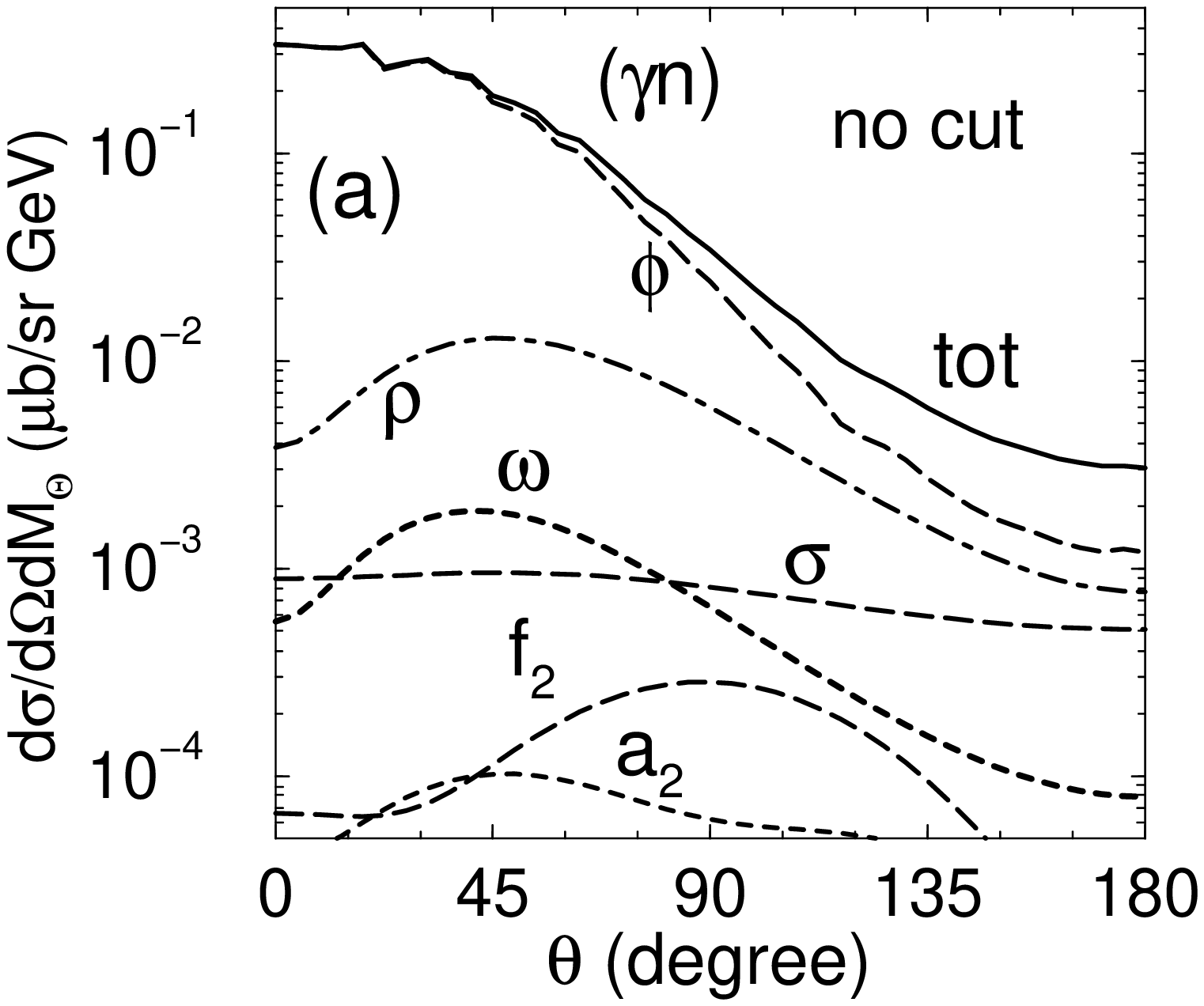}\hfill
 \includegraphics[width=.3\textwidth]{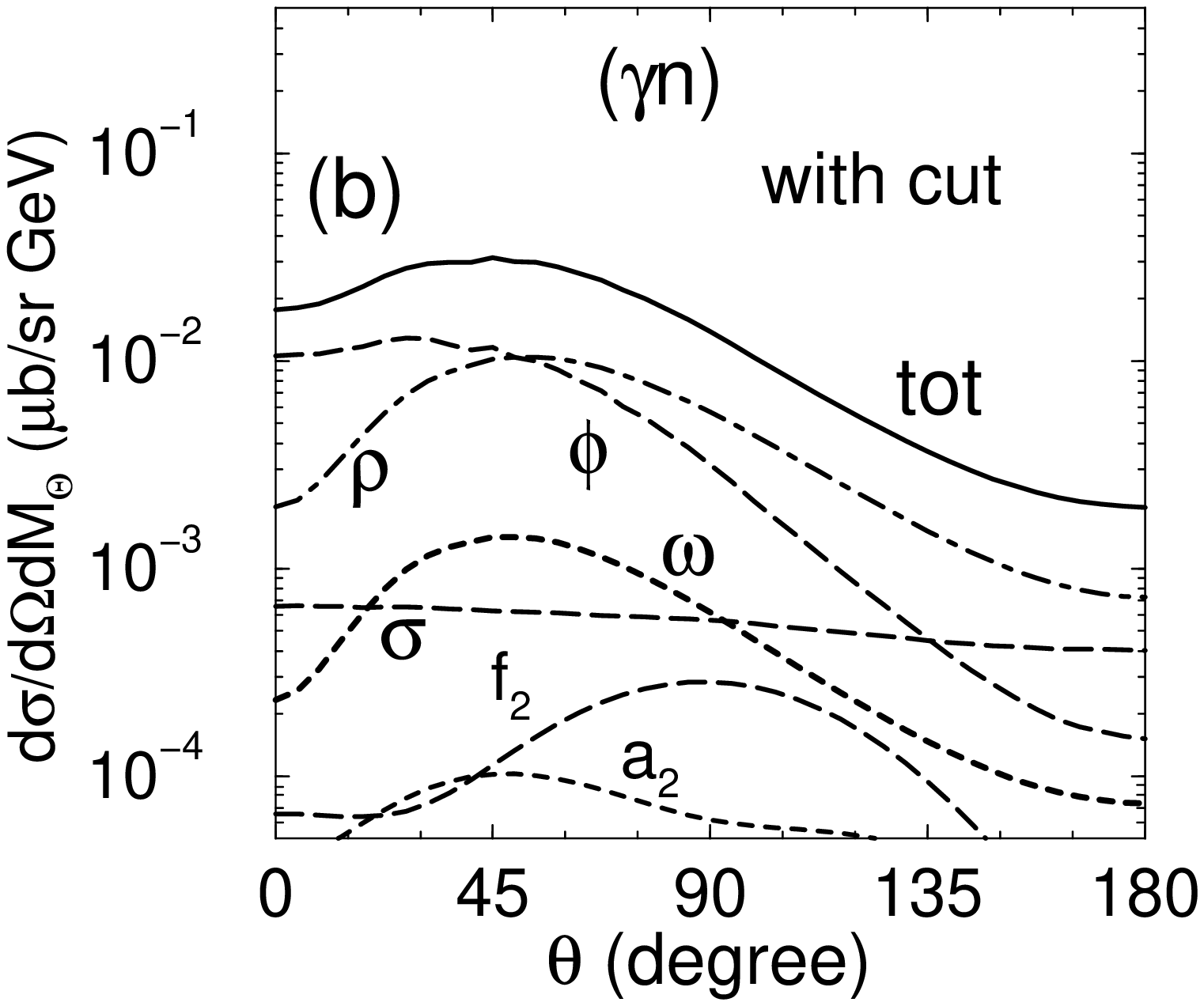}\hfill
 \includegraphics[width=.3\textwidth]{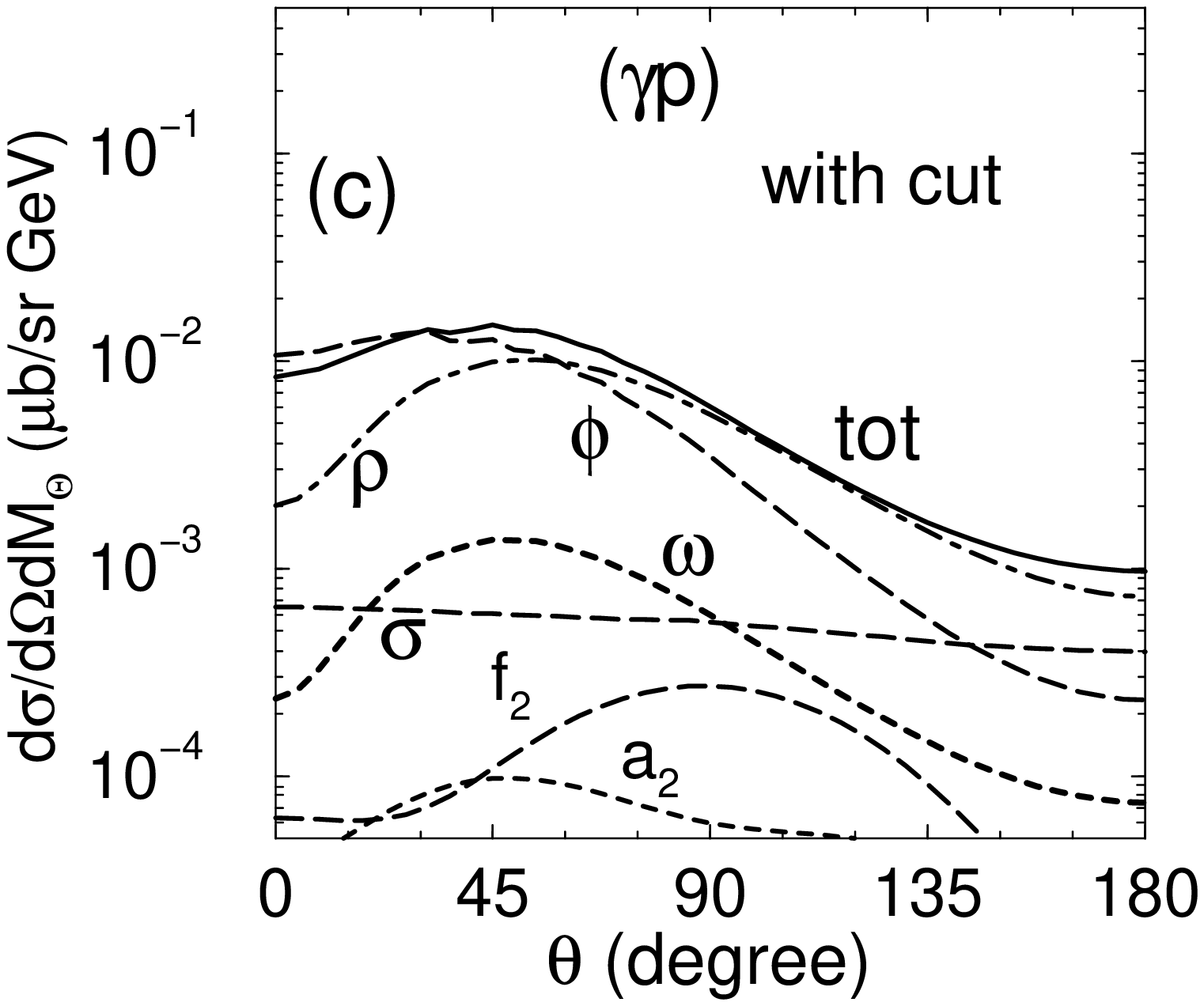}
 \caption{\label{fig:7}\tcaps%
Contribution of the background processes
 to the differential
 cross section of the $\Theta^+$ photoproduction for the (a,b)
 $\gamma n$ and (c) $\gamma p$ reactions without (a)  and
 with (b,c) the cut in the $K\bar{K}$ invariant-mass
 distribution, respectively.
 }
\end{figure}

Our result for the $\gamma n\to n K^+K^-$ reaction is shown in Fig.~\ref{fig:7}a. One can
see a rather strong contribution from the $\phi$ photoproduction. We will eliminate the
phase space with the $K\bar{K}$ invariant mass from 1.00 to 1.04 GeV following
Ref.~\cite{Nakano03} in order to reduce its contribution. The corresponding phase space,
shown schematically in Fig.~\ref{fig:3}, is about 15\% of the total phase space but it
gives about 100\% of the $\phi$-channel contribution to the background. The differential
cross section, with an $M_{K\bar{K}}$ cut for the $\gamma n$ and $\gamma p$ reactions, are
shown in Figs.~\ref{fig:7}b and \ref{fig:7}c, respectively. Now the $\phi$ and $\rho$
contributions are comparable to each other. Another interesting aspect is the enhancement
of the $\sigma$ contribution compared to the corresponding contribution in the case of
the $K\bar{K}$ invariant-mass distributions shown in Fig.~\ref{fig:5}a,b. This enhancement
is explained by the term ${\kappa_\rho}(R'\cdot\varepsilon_\gamma)/M_N$ in Eq.~(\ref{S2})
caused by the $\rho NN$ tensor coupling. The contribution from this term increases
strongly with increasing $|\mathbf{p}'|\sin\theta_{K}$. When the momentum transfer to the
$K\bar{K}$ pair is small (see Fig.~\ref{fig:6}a,b), this contribution is rather small,
whereas in the kinematic condition of Fig.~\ref{fig:7} it is large.

As can be seen from Fig.~\ref{fig:7} and from the discussion in
connection to Fig.~\ref{fig:6}a,b in Sec.~III.D, the
contributions from the tensor mesons, $a_2$ and $f_2$, at
$E_\gamma\approx 2$ GeV are found to be very small (even for large
coupling constants~\cite{Oh04} {and different choices of their
signs}). Therefore, hereafter, the tensor mesons will be omitted
in our calculations.

\section{Fixing the parameters of the resonant amplitude}

(1) The magnitude of the coupling constant, $g_{\gamma KK^*}$, is
extracted from the width of the $K^*\to \gamma K$
decay~\cite{PDG}. Its sign is fixed by SU(3) symmetry. We have
$eg_{\gamma K^{0}K^{*0}}=-0.35$ and $eg_{\gamma K^{\pm}K^{*\pm}}=
0.23$.

(2) The contribution of the $s$-channel (Fig.~\ref{fig:1}b) is  small which leads to a
rather weak dependence of the total amplitude on the tensor coupling $\kappa_\Theta$ in
the $\gamma \Theta\Theta$ interaction within a ``reasonable" range of
$0\lesssim|\kappa_\Theta| \lesssim0.5$~\cite{MagMom}. Therefore, we choose
$\kappa_\Theta=0$ for both parities.

(3) The coupling constant $g_{\Theta NK}$  for the positive and negative
$\Theta^+$ parity is found from the $\Theta^+$ decay width,
\begin{eqnarray}
\Gamma_{\Theta}=\frac{[g^{\pm}_{\Theta NK}]^2 p_F}{2\pi M_\Theta}
(\sqrt{M_N^2 + p_F^2}\mp M_N)~.
\end{eqnarray}
There are several indications that the  $\Theta^+$ decay width is most likely to be  of
the order of one MeV~\cite{SmallWidth} and that the observed width $\Gamma_{\text{exp}}$
in the $\Theta$ photoproduction is rather to be regarded as the experimental resolution
($\Delta$).
 By construction (see Introduction), the differential cross section for the $\Theta^+$
photoproduction defined by Eq.~(\ref{cs}) at the resonance position does not depend on
the $\Theta$ decay width. Instead, it depends on the ratio of the coupling constants
$g_{\Theta N K^*}$ and $g_{\Theta N K}$ (as well as, in general, also on other
parameters). In principle, this ratio may be extracted from the experimental data by
comparing the resonant and the background contributions because the cross section due to
the background is known in the present case. However, a proper comparison requires to
account for the ``experimental resolution" in our calculation.
The smearing of the $\Theta^+$ invariant mass distribution results in a
suppression of the resonant cross section at the resonance position by a factor of
 \begin{eqnarray}
 d^2\approx \frac{\Gamma_\Theta}{\Delta}\approx\frac{1}{10}~,
 \end{eqnarray}
where we use an averaged value of $\Delta\approx 20$ MeV and $\Gamma_\Theta\approx 2$ MeV. We
include the effect of this smearing by multiplying all the resonance amplitudes by this
factor $d$. In our calculation we will use the ratio of the resonance plus background to
background processes from the experiment to fix the model parameters. This ratio is
different in different experiments and varies from 2~\cite{CLAS1} to 7~\cite{SAPHIR03}.
We will use a value of 3.4 which corresponds to that found in the LEPS
experiment.

\begin{figure}[t] \centering
 \includegraphics[width=.3\textwidth]{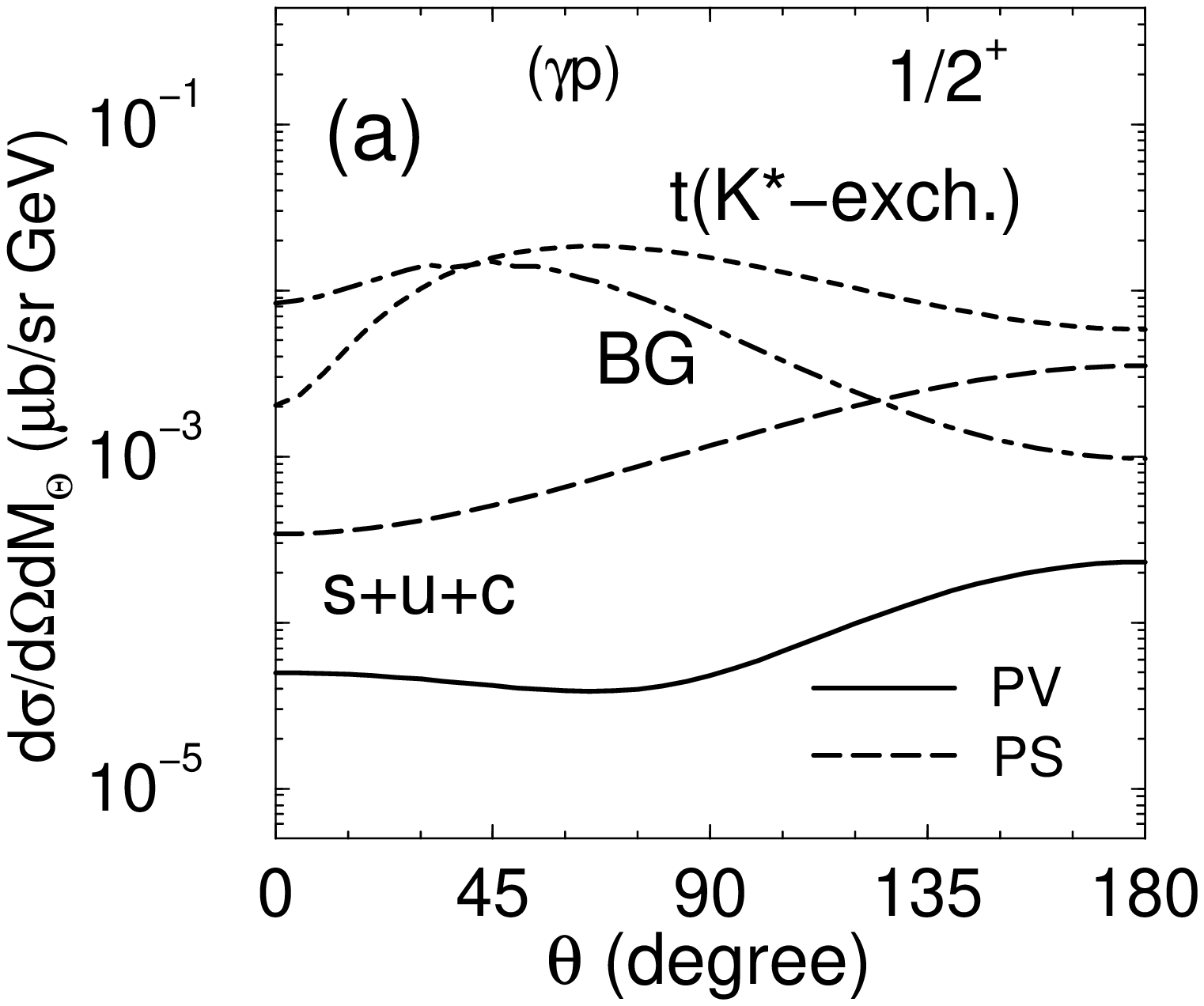}\qquad\qquad
 \includegraphics[width=.3\textwidth]{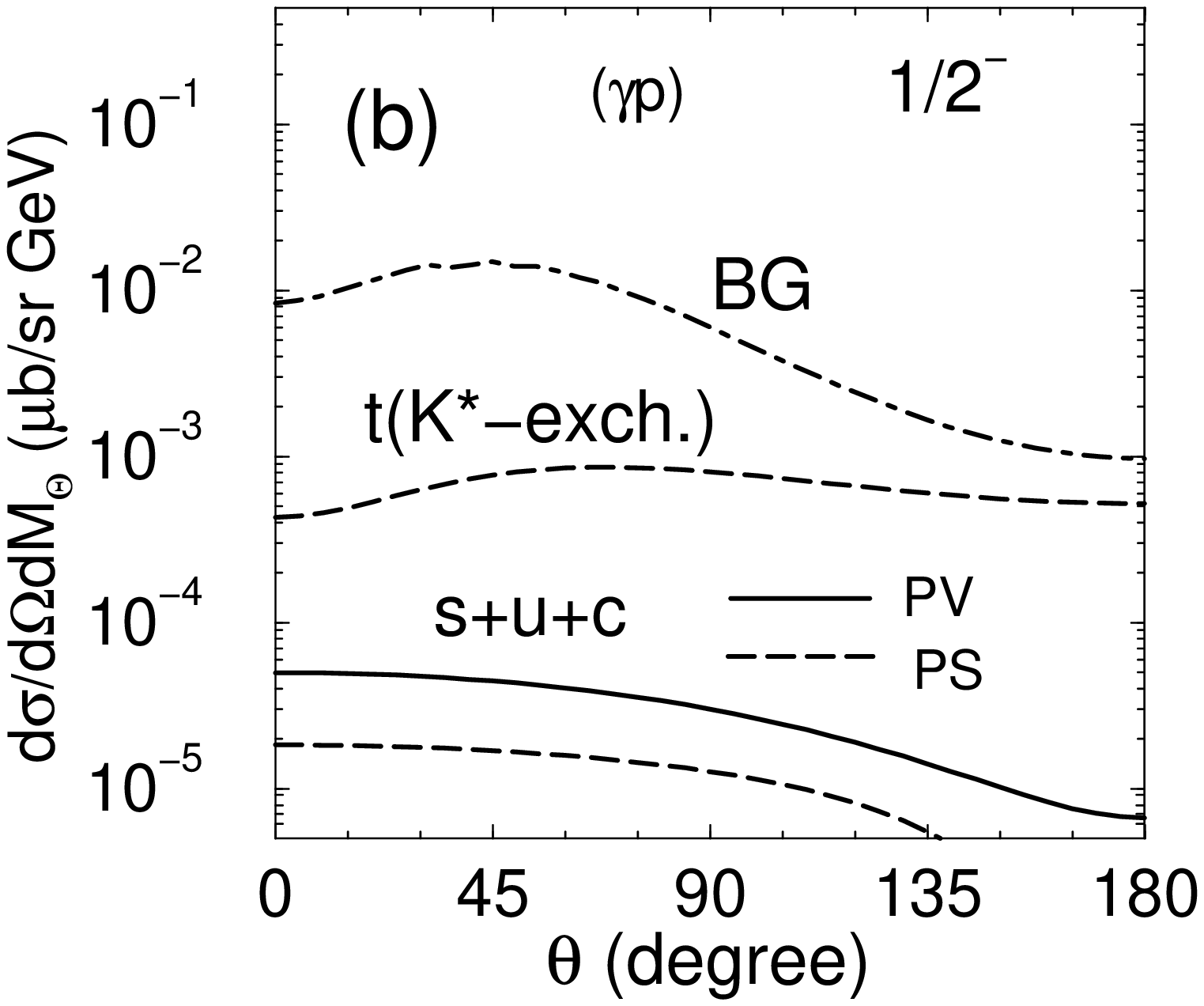}\\
~\\
 \includegraphics[width=.3\textwidth]{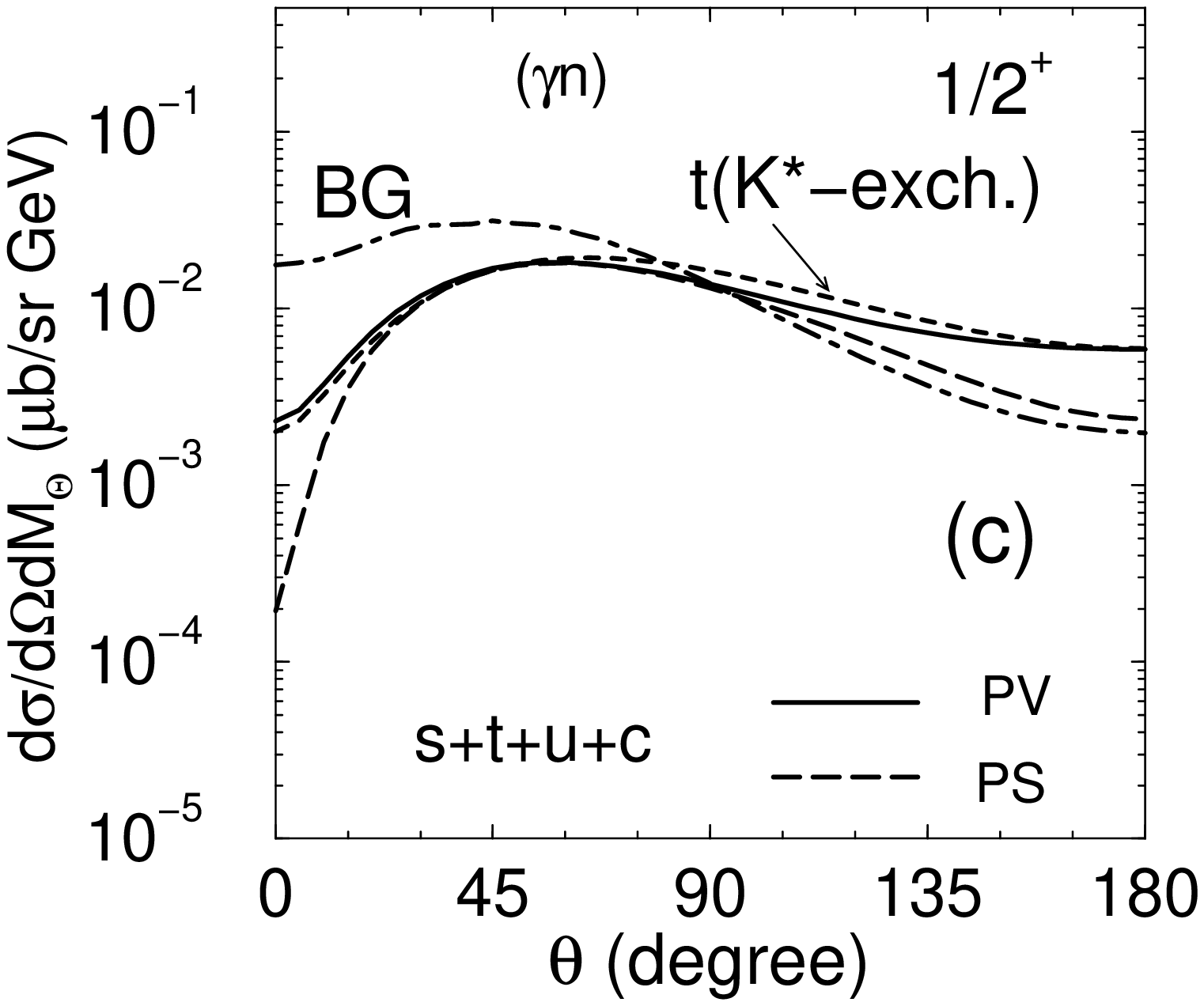}\qquad\qquad
 \includegraphics[width=.3\textwidth]{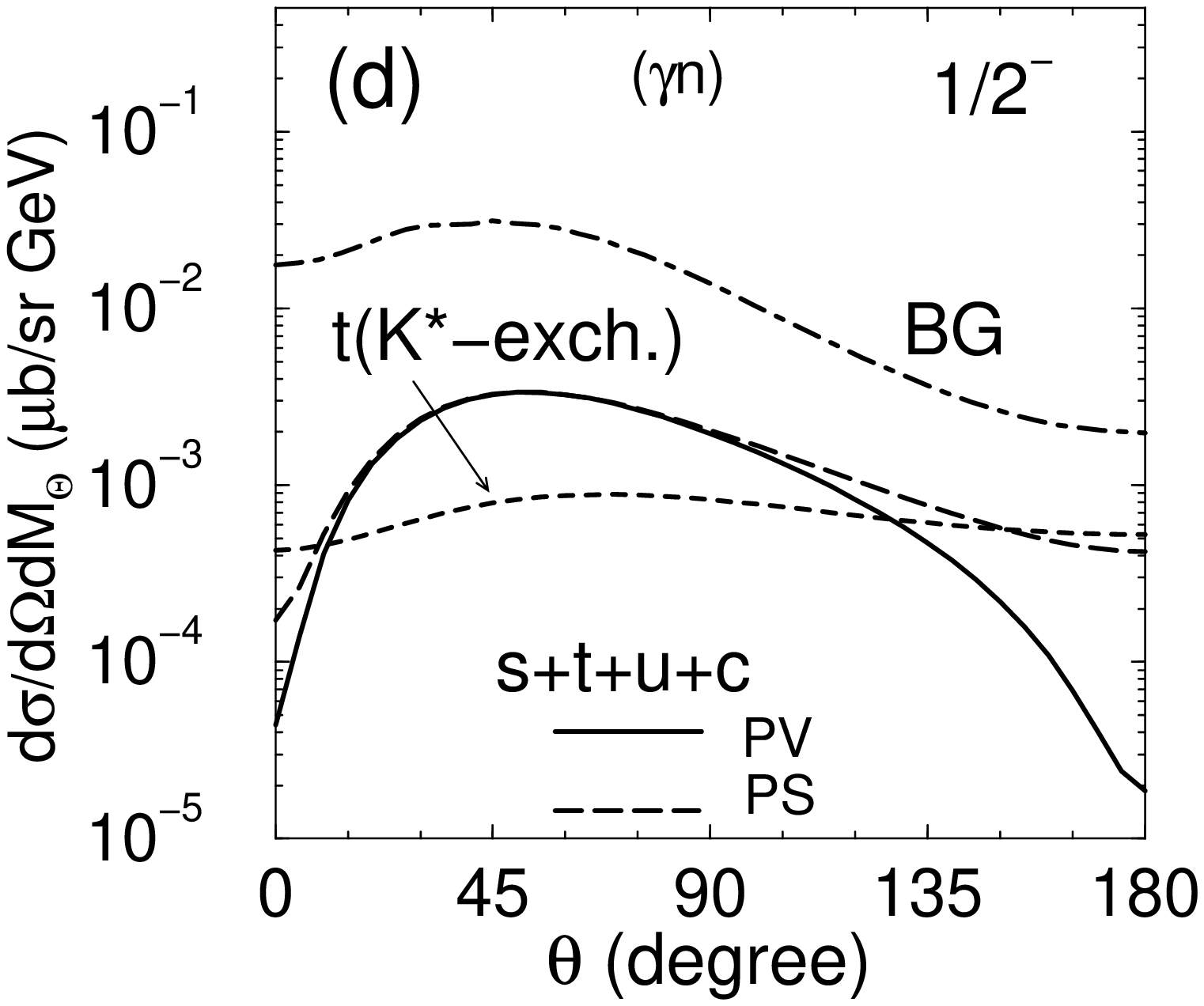}
 \caption{\label{fig:8}\tcaps%
 Contributions of the Born $s$, $t$, $u$
 and the contact $c$ terms in the  (a,b) $\gamma p$
and  (c,d) $\gamma n$ reactions,
 together with  the $t$-channel $K^*$ exchange and the background
 processes. The  respective cases for positive and negative $\pi_\Theta$
 are depicted in (a,c) and (b,d).}
\end{figure}

(4) The next important point is to fix the cutoff parameters. In our model, we have two
different cutoff parameters. One ($\Lambda_{K^*}$) is in the $t$-channel $K^*$ exchange
amplitude. Another one ($\Lambda_B$) defines the Born terms of the $s$-, $u$-, and $t$-channels and
the current-conserving contact terms $c$. Generally speaking, $\Lambda_{K^*}$
together with $g_{\Theta NK*}$ (or the ratio $\alpha=g_{\Theta NK^*}/g_{\Theta NK}$)
defines the strength of the $K^*$ exchange amplitude. Increasing  $\Lambda_{K^*}$ leads
to a decreasing $\alpha$. If we assume for the moment that the Born terms are negligible
compared to the $K^*$ exchange, then taking as a guide the quark model prediction for
$\alpha$ for the positive $\Theta^+$ parity~\cite{QMKK*},
\begin{eqnarray}
\alpha\approx\sqrt{3}~,
\end{eqnarray}
we can fix $\Lambda_{K^*}$ using the calculated background cross section and measured
ratio of the resonance contribution to the background [signal-to-noise ($S/N$)]. The
$\gamma p\to p K^0\bar{K}^0$ reaction is close to this ideal case, where the
$K^*$-exchange contribution is much larger than the Born terms. But in the case of the
$\gamma n\to n K^+\bar{K}^-$ reaction, the situation is more complicated. The
$t$-channels and the contact terms are strong and comparable to the $K^*$ exchange. These
situations for the $\gamma p$ and $\gamma n$ reactions may be understood from
Fig.~\ref{fig:8}. We show here the differential cross sections for the $\gamma N\to N
K\bar{K}$ reactions as a function of $\bar{K}$-production angle in the center-of-mass
system at the resonance position and for positive and negative $\Theta^+$ parities. We
display the individual contributions from the Born terms, the $K^*$ exchange channel and
the background processes for both the pseudoscalar and pseudovector couplings with
$\Lambda_B=1$\,GeV. The parameters for the $K^*$ exchange amplitude read
$\Lambda_K^*=1.5$, $\alpha=1$ and $\kappa^*=0$. One can see that for the $\gamma p$
reactions the $K^*$-exchange channel is dominant, whereas for the $\gamma n$ reactions
all the individual contributions become comparable to each other and the problem of
fixing the cut off parameter $\Lambda_B$ must be solved consistently. Note that the
inclusion of the $\Sigma$ and $\Lambda$ photoproduction~\cite{LambdaSigma} processes
results in a larger ambiguity in the choice of $\Lambda_B$ which varies from 0.5 to 2
(GeV) depending on the coupling scheme, method of conserving the electromagnetic current,
etc.

To choose $\Lambda_B$, we use the following strategy. We assume that the ratio $\alpha$,
as well as the cutoff $\Lambda_{K^{*}}$, must be the same in the $\gamma p$ and $\gamma
n$ reaction. Then,  fixing $\alpha$ and $\Lambda_{K^{*}}$ from the $\gamma p$ reaction
and using the signal-to-noise ratio $S/N$, we determine $\Lambda_B$ unambiguously for the
$\gamma n$ reaction and its value will depend on the type of the coupling (PV or PS), the
value and the sign of tensor coupling ($\kappa^*$), and sign of $\alpha$.

\begin{figure}[t] \centering
 \includegraphics[width=.33\textwidth]{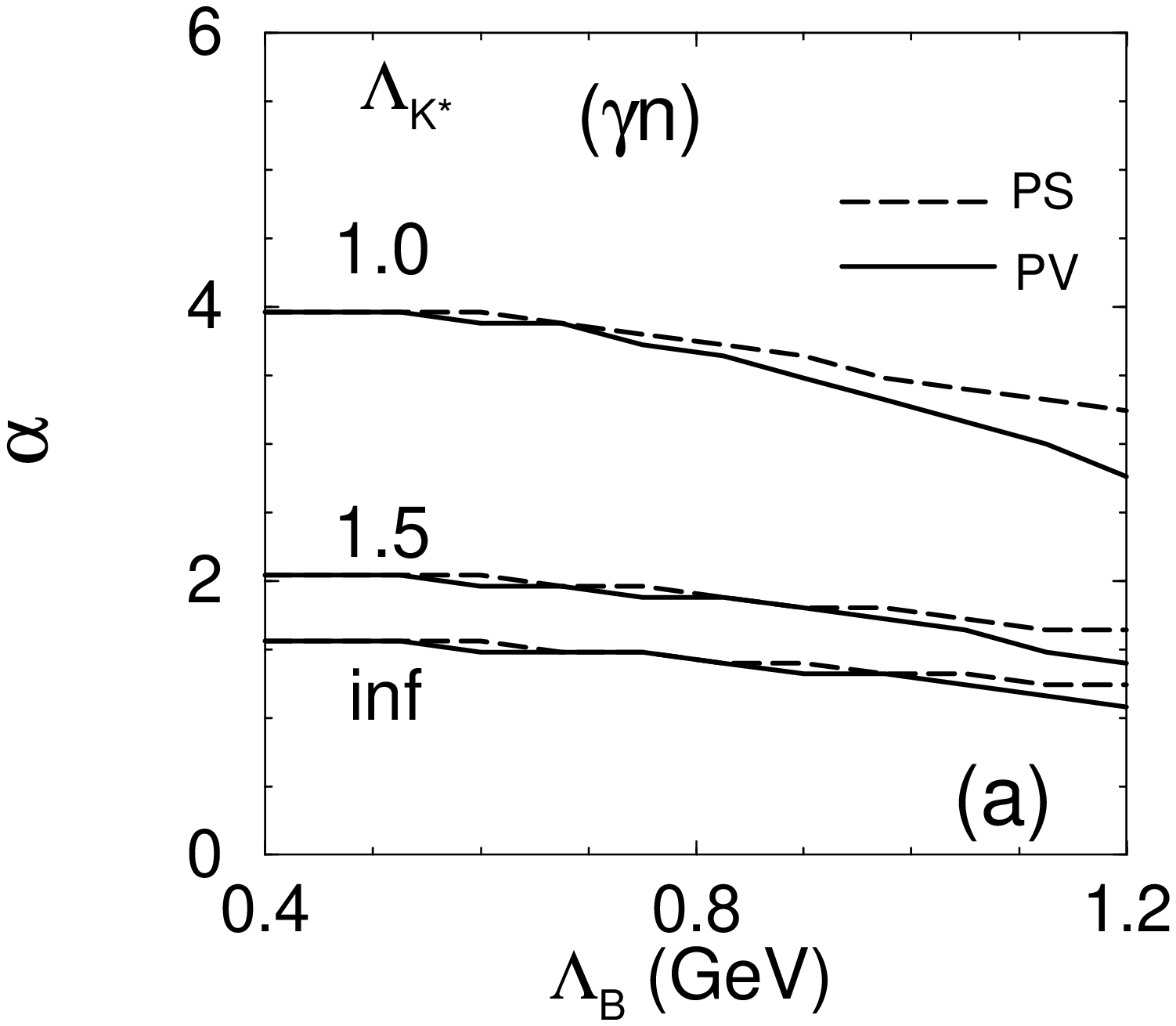}\qquad
 \includegraphics[width=.33\textwidth]{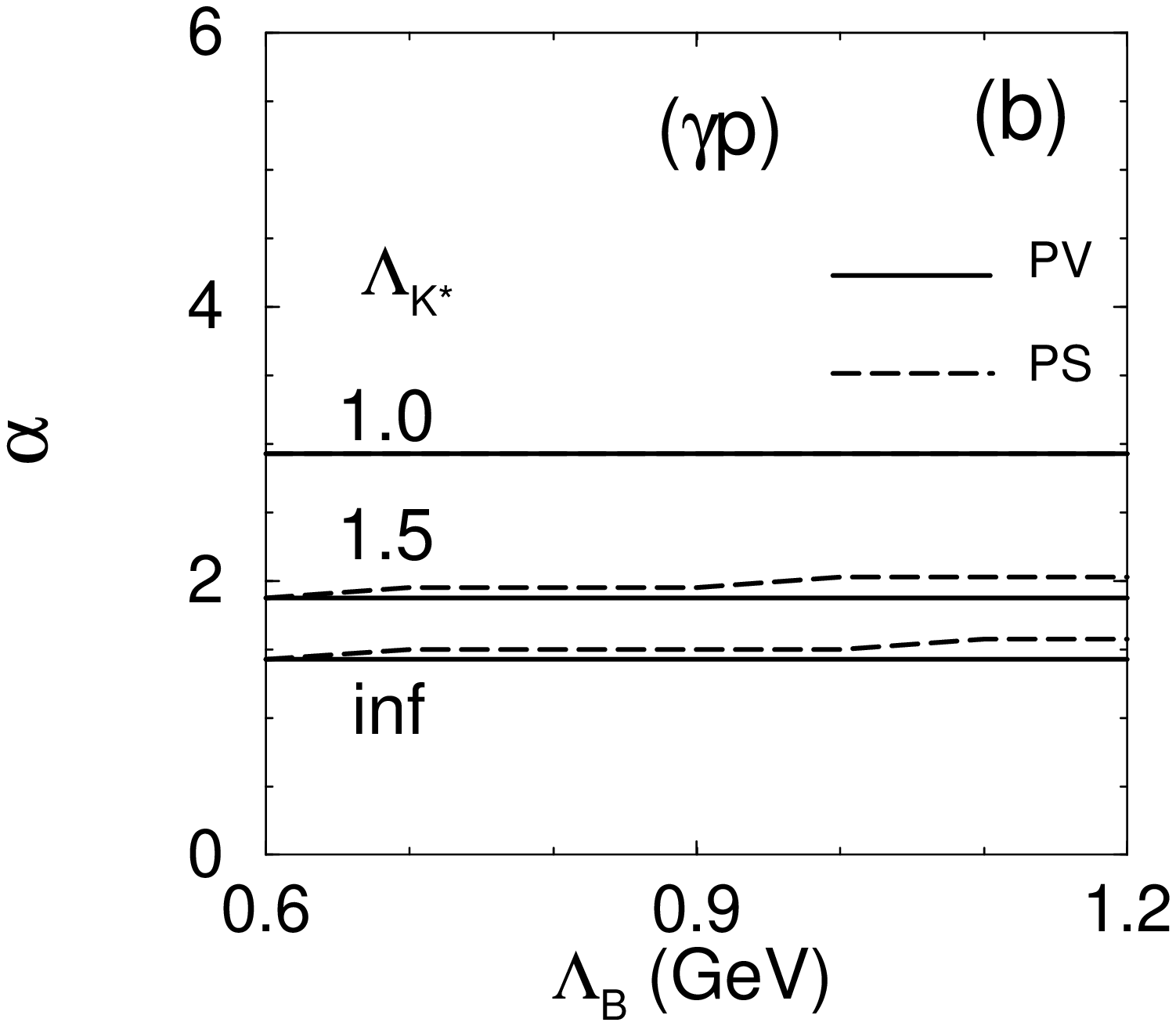}
 \caption{\label{fig:9}\tcaps%
 The scale parameter $\alpha$ as a function of
 the cutoff parameters $\Lambda_{K^*}$ and $\Lambda_B$.}
\end{figure}

\begin{figure}[b] \centering
 \includegraphics[width=.3\textwidth]{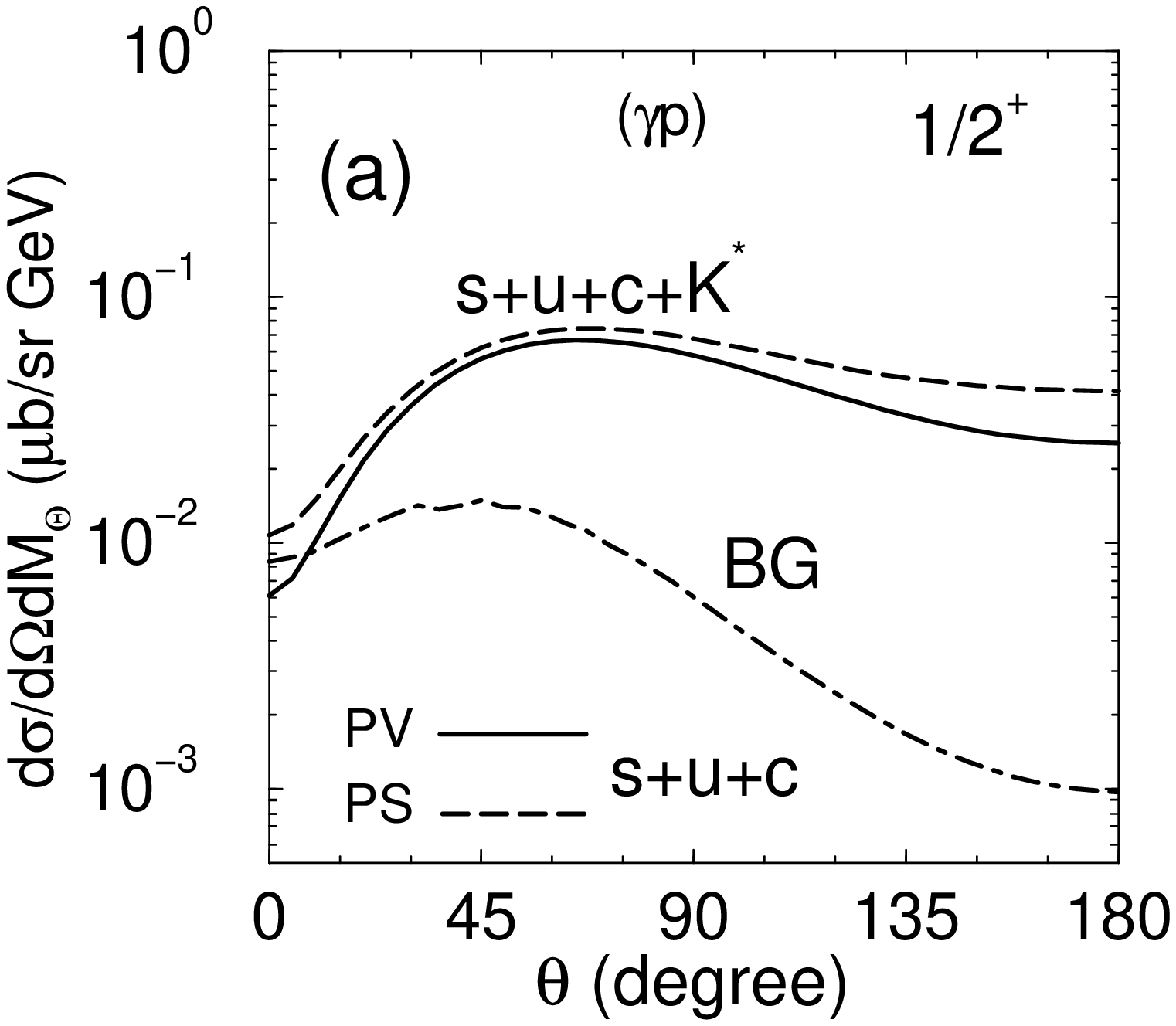}\qquad
 \includegraphics[width=.3\textwidth]{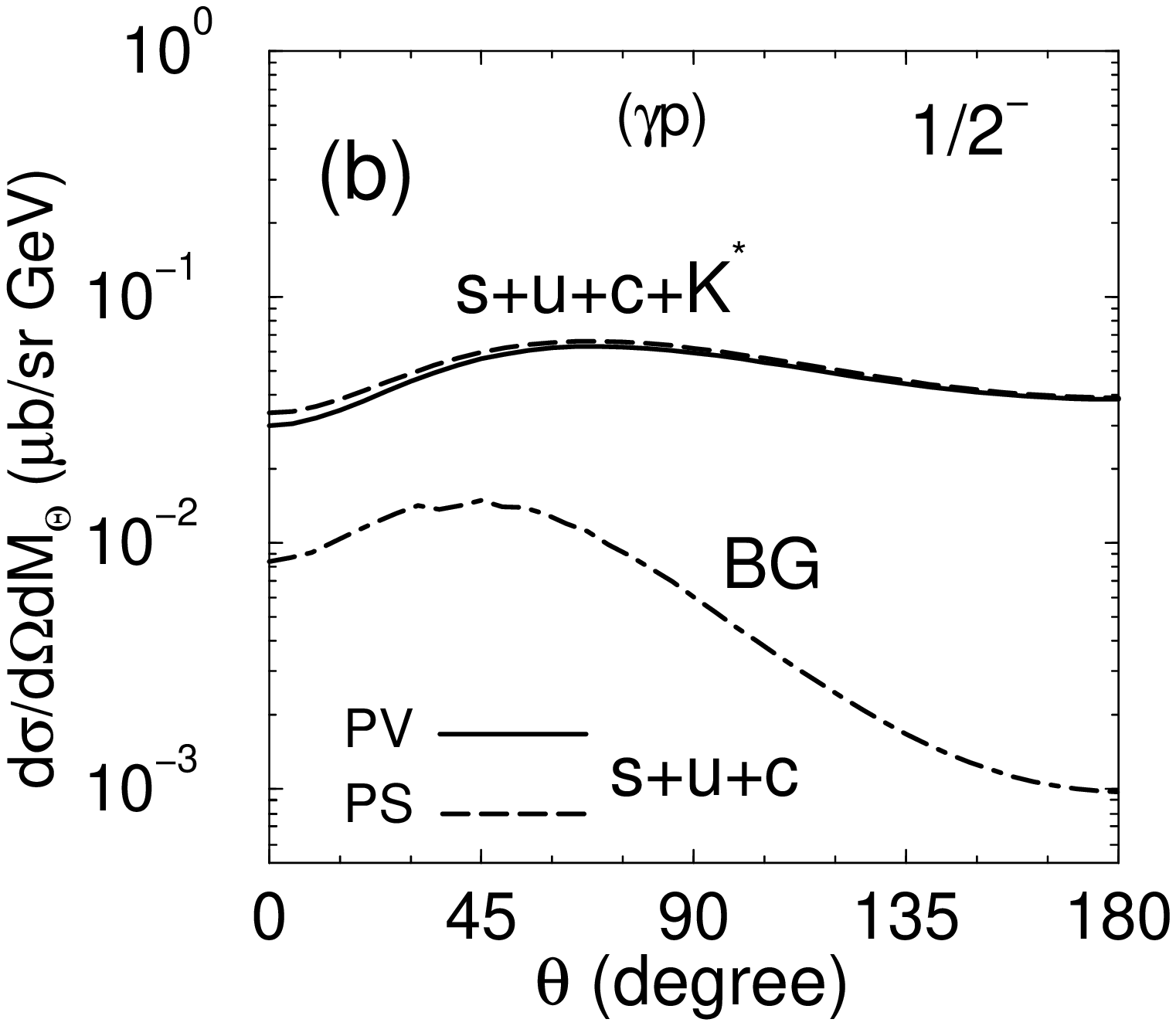}\\
 ~\\
 \includegraphics[width=.3\textwidth]{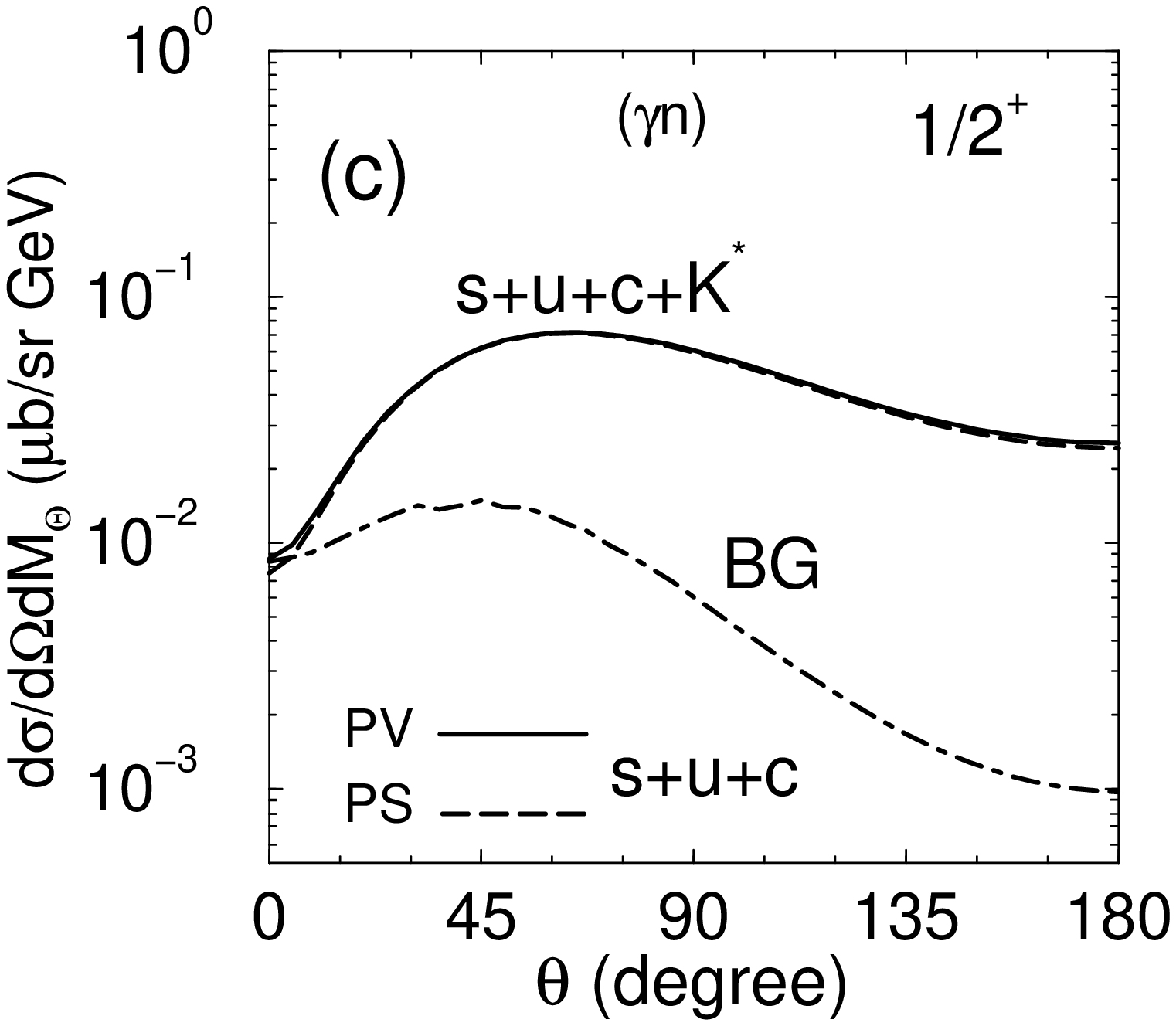}\qquad
 \includegraphics[width=.3\textwidth]{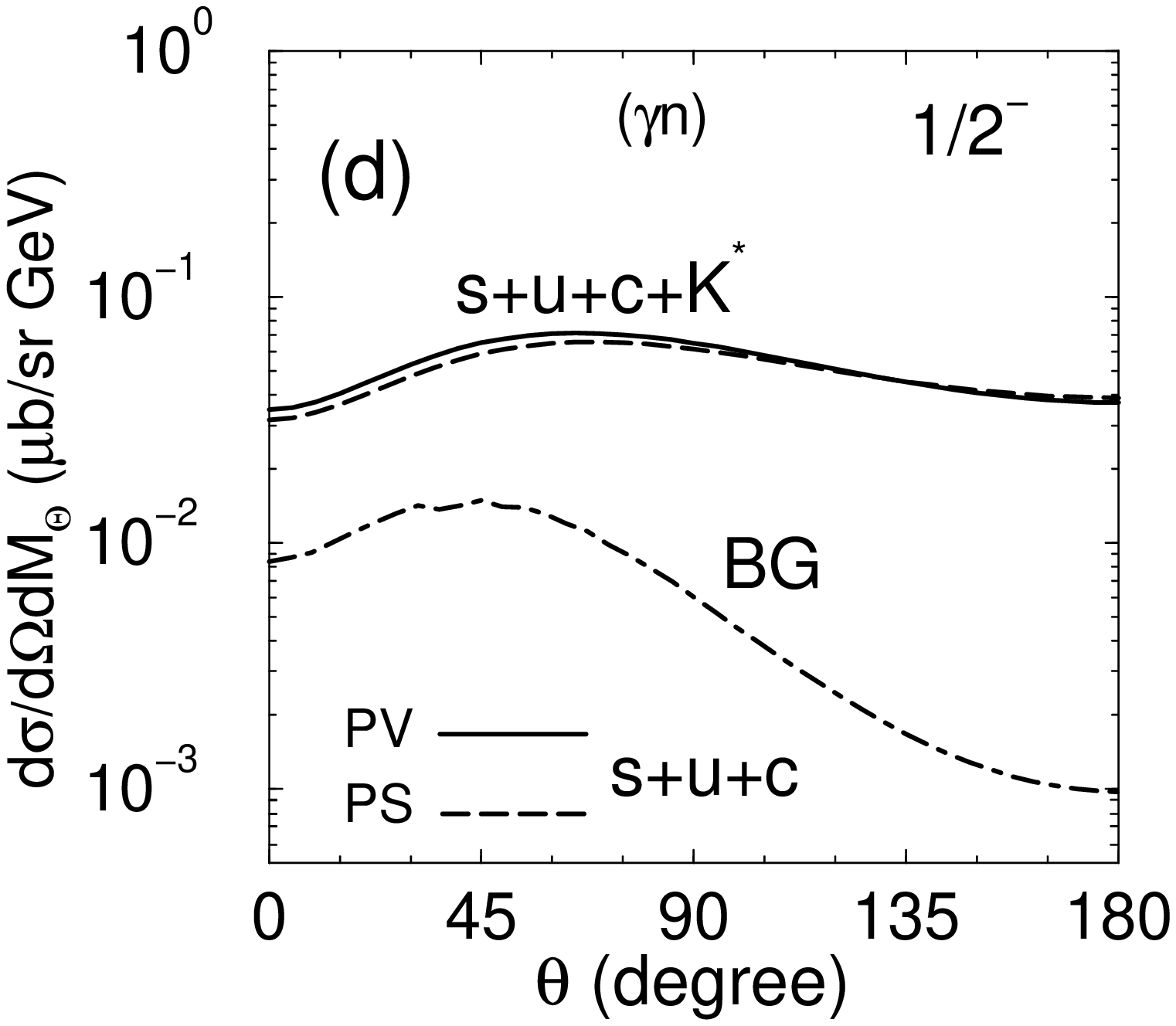}
 \caption{\label{fig:10}\tcaps%
 The cross section of the resonance
 $\Theta^+$ photoproduction and the background processes as a function of the $\bar
 K$-production angle for the (a,b) $\gamma p$ and (c,d) $\gamma n$
 reactions and for the  (a,c) positive and  (b,d) negative parities of
 $\Theta^+$.}
\end{figure}

Figure~\ref{fig:9} shows the dependence of $\alpha$ on $\Lambda_B$ at fixed
$\Lambda_{K^*}$ and $\kappa^*=0$. For the $\gamma p$ reaction, this dependence is rather
weak but for $\gamma n$, it is strong. For further analysis of the $\gamma p$ reaction,
we chose an averaged value of $\Lambda_B=1$ GeV. We will analyze observables at three
values of the tensor coupling $\kappa^*=0,\pm 0.5$. The corresponding values of $\alpha$
depending on $\kappa^*$ and the $\Theta^+$ parity are shown in Table~\ref{tab:1}. All the
calculations are done at the $\bar{K}$ production angle of $\theta=55^\circ$, where the
resonance processes are largest. Also, at this condition, the background contribution is
dominated by the known $\phi$ and $\rho$ channels and therefore it is better established.
Figures~\ref{fig:10}a-d show the result for the differential cross section for the
$\gamma N\to N K\bar{K}$ reactions, calculated by considering a coherent sum of all the
resonance processes with the scaled $K^*$ exchange amplitude by a factor of $\alpha$ and
appropriate $\Lambda_B$. One can see that in the vicinity of $\theta\approx 55^\circ$,
the unpolarized cross sections for the different coupling schemes  and different parities
are close to each other for both the $\gamma p$ and $\gamma n$ reaction. The difference
between these two reactions may appear in the angular distribution of the $\Theta^+\to N
K$ decay and in the corresponding spin observables.

 \begin{table}[t]
\parbox{.35\textwidth}{\flushleft
 \begin{tabular}{l|@{\extracolsep{1em}}ccc}
 \hline\hline
 $ 1/2^P~\setminus~\kappa^*$ &   $ 0.5$ & $0.0 $ & $-0.5$   \\ \hline
 $1/2^+$ &   1.67    & 1.875  & 2.01     \\
 $1/2^-$ &   9.38    & 8.625  & 7.88 \\
  \hline\hline
\end{tabular}
}
\hfill
\parbox{.58\textwidth}{%
  \caption{\label{tab:1}\tcaps%
  Ratio $\alpha=g_{\Theta NK^*}/g_{\Theta NK}$ for different
  $\Theta^+$ parity and  tensor coupling $\kappa^*=0,\pm 0.5$.  }
}
\end{table}

\section{Results and discussion}

\subsection{\boldmath $\Theta^+$ decay distribution}

\begin{figure}[t] \centering
 \includegraphics[width=.3\textwidth]{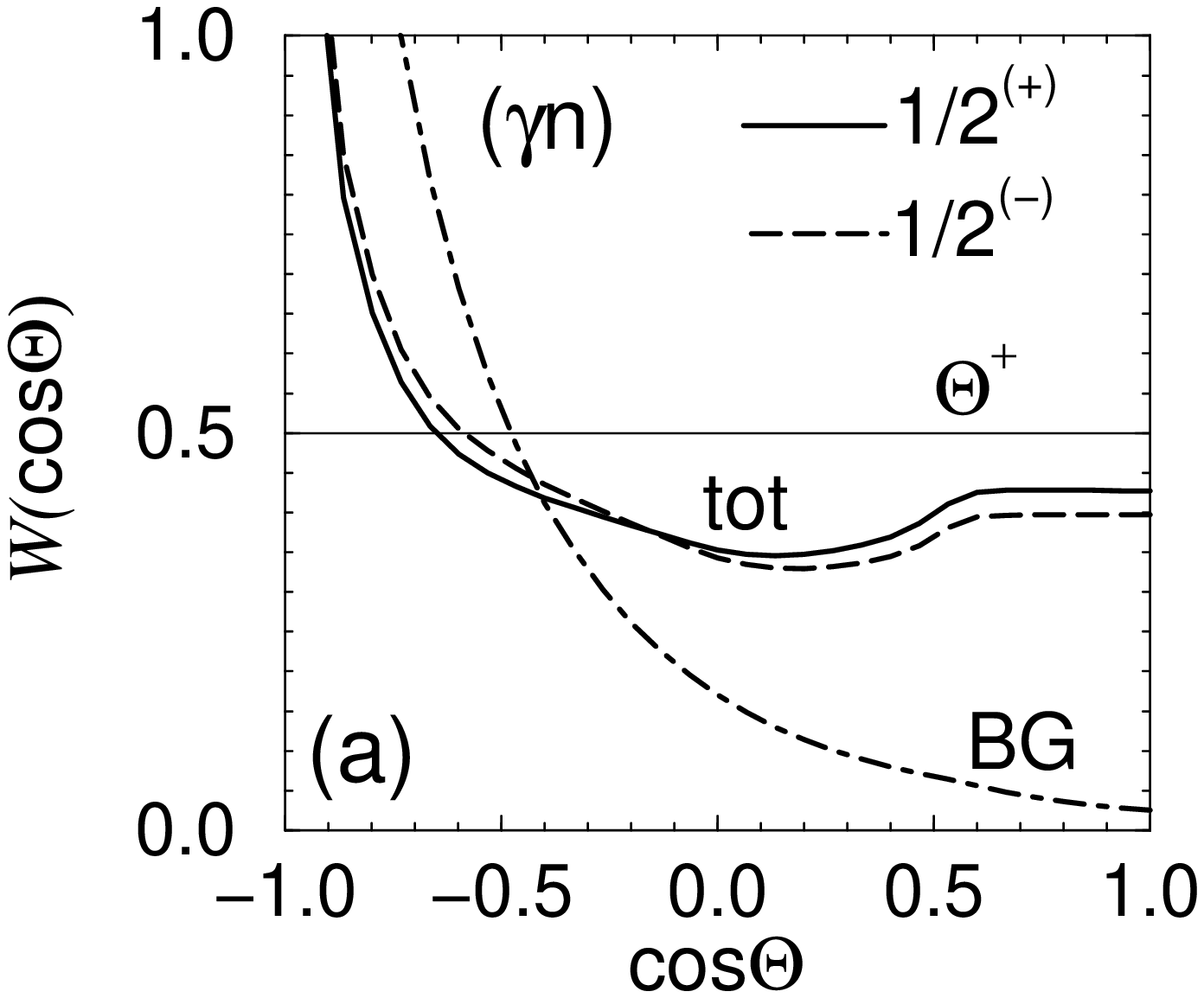}\hfill
 \includegraphics[width=.3\textwidth]{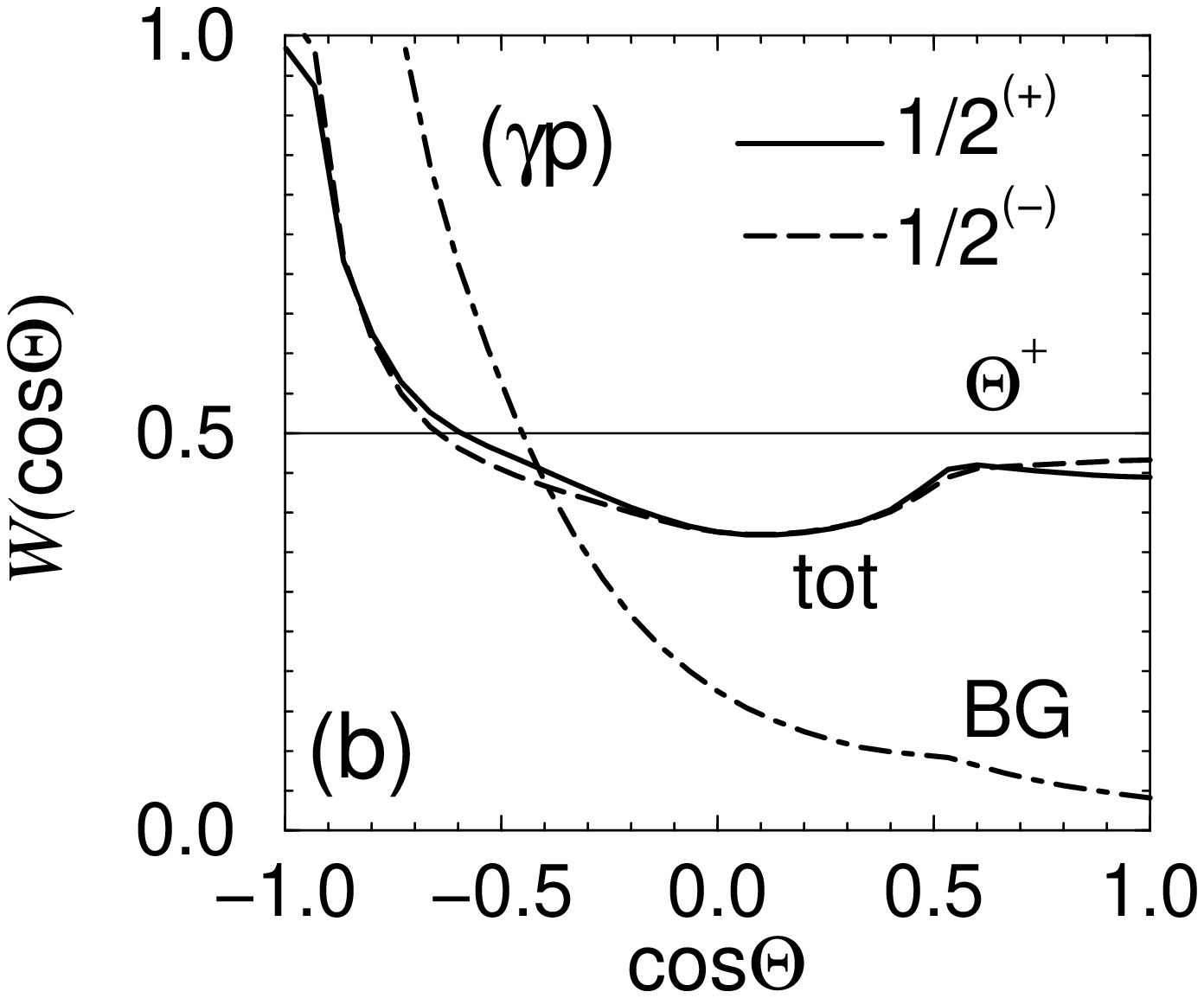}\hfill
 \includegraphics[width=.3\textwidth]{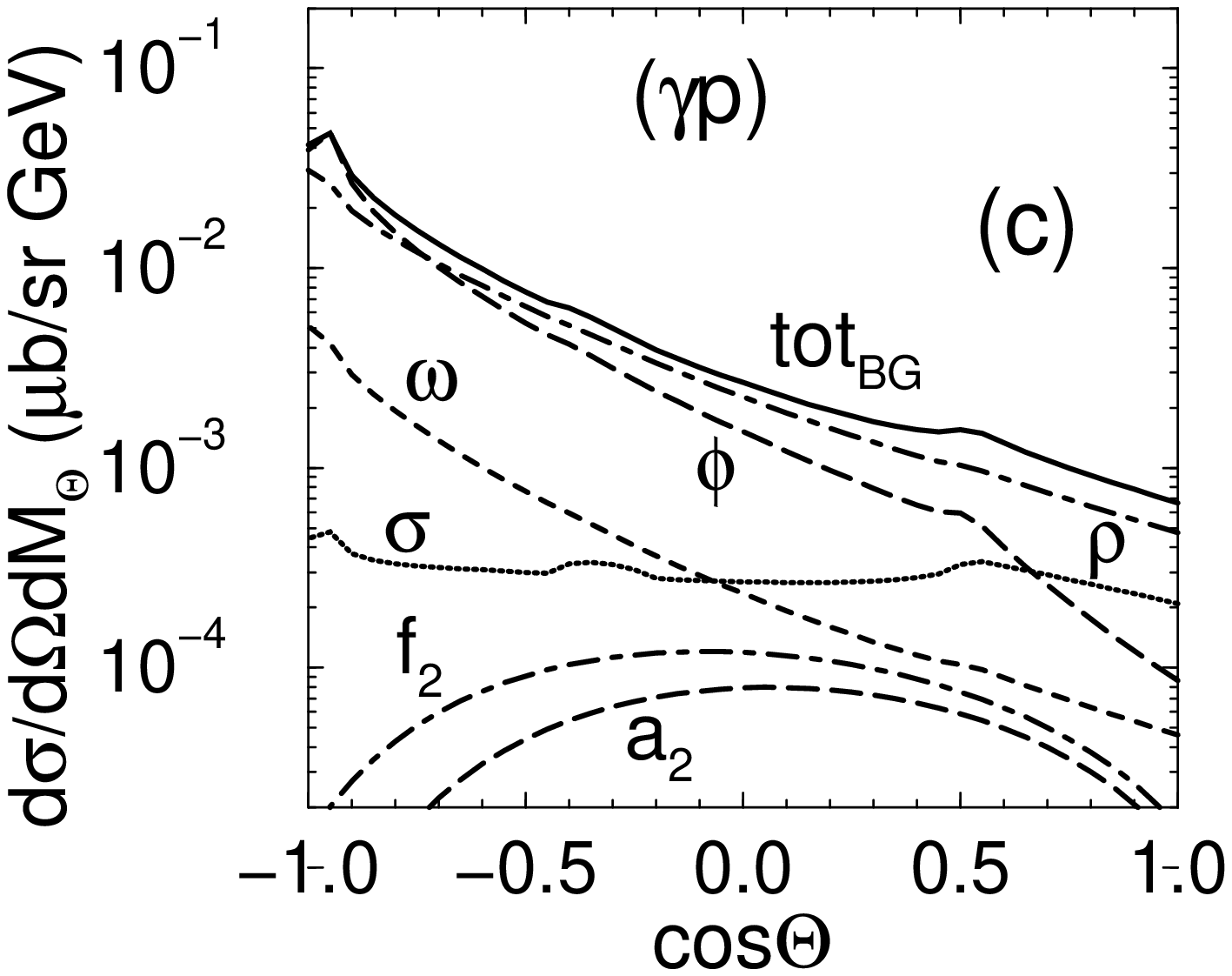}
 \caption{\label{fig:11}\tcaps%
 The decay distributions of  (a) $nK^+$ and (b) $pK^0$ in
 the reactions $\gamma \to nK^+ K^-$ and
 $\gamma \to pK^0\bar{K}^0$, respectively.
 The solid  and dashed curves  correspond to the positive and
 negative $\Theta^+$ parities, respectively. The distribution from
 the background is shown by the dot-dashed curve, whereas the
 contribution from the pure resonance channel is shown by the solid
 thin line. (c) The differential cross section of the background
 channels.}
\end{figure}

The angular distribution of the $\Theta^+\to N K$  decay is described by the decay
amplitudes $D^{\pm}$ in Eqs.~(\ref{TAD}) and (~\ref{Dpm}). Later we will discuss the
polar angle ($\Theta$) distribution integrated over the azimuthal angle ($\Phi$). The
decay amplitudes exhibit a p- or an s-wave distribution depending on the parity of
$\Theta^+$ ($\pi_{\Theta}$) being positive or negative, respectively. However, if the
spin state of the recoil nucleon is not fixed, the difference in the angular
distribution, normalized as
\begin{eqnarray}
\int W(\cos\Theta)\,  d\cos\Theta=1~,
\end{eqnarray}
disappears (for a spin-$\frac12$ $\Theta^+$). The pure resonant amplitude results in an
isotropic distribution with
\begin{eqnarray}
 W^R(\cos\Theta)=\frac{1}{2}~.
\end{eqnarray}
The interference between the resonant and the background amplitudes leads to an
anisotropy in the angular distribution. Figures~\ref{fig:11}a,b  show the angular
distribution $W(\cos\Theta)$ for the $\gamma n$ and $\gamma p$ reactions. Here we chose
the PV-coupling scheme with positive $\alpha$ and $\kappa^*=0$. The results for other
input parameters are similar to those shown in Fig.~\ref{fig:11}. The solid  and dashed
curves correspond to the positive and negative parities for $\Theta^+$, respectively. The
angular distribution due to the background is shown by the dot-dashed curve, whereas the
contribution from the pure resonance channel is shown by the solid thin line. Some
non-monotonic behavior of $W(\cos\Theta)$ around $\Theta\approx\pi/3$ is caused by the
sharp cut in the $K\bar{K}$ invariant mass with $|M_{K\bar{K}}-M_\phi|<20$ MeV as
described in Sec.~III.E. We do not see any essential difference between the calculations
corresponding to different $\pi_\Theta$.

The differential cross section due to the background channels in $\gamma p\to p
K^0\bar{K}^0$ is shown in Fig.~\ref{fig:11}c. It is peaked at $\cos\Theta\approx -1$ for
the following two reasons. First, the momentum transfers to the $K \bar{K}$ pair ($|\bar
t|$) reaches its minimum value at $\cos\Theta= -1$ and second, the polar angle of the
$K\bar{K}$ decay distribution ($\theta_K$) with respect to the photon momentum in the
$K\bar{K}$ rest frame is close to $\pi/2$ (when $\theta\approx 55^\circ$ and
$\Theta\approx\pi$). In this region the dominant background contribution arising from the
vector meson channels has a maximum because the cross section due to the $ \gamma N\to
NV\to NK\bar{K}$ process is proportional to $\sin^2\theta_K$.

In summarizing this subsection, we conclude that: (i) the decay distribution for
unpolarized $\Theta^+$ photoproduction is not sensitive to the $\Theta^+$ parity and
(ii)  the vector meson dominance of the background leads to a specific decay distribution
which can be checked experimentally.

\subsection{Spin observables}

\subsubsection{Single spin observables}

Let us consider the beam asymmetry defined as
\begin{eqnarray}
\Sigma_B=\frac{\sigma(\perp)-\sigma(\parallel)}
              {\sigma(\perp)+\sigma(\parallel)}~,
\label{BeamA}
\end{eqnarray}
where $\sigma(\perp)$ and $\sigma(\parallel)$ are the cross sections for $\Theta^+$
photoproduction with the photon beam polarized perpendicular
($\varepsilon_\gamma=\widehat{\mathbf{y}}$) or parallel
($\varepsilon_\gamma=\widehat{\mathbf{x}}$) to the production plane, respectively. We will
analyze the beam asymmetry as a function of the $\Theta^+$ polar decay angle $\Theta$.
For the resonant channel contribution, $\sigma(\perp)$ and $\sigma(\parallel)$ do not
depend on the decay angle $\Theta$ if the nucleon spin states are not fixed. Therefore,
in this case, the beam asymmetry is a constant and its value depends on the details of
the production mechanism. The interference with the background amplitude results in some
structure of the beam asymmetry. The question is whether or not the beam asymmetries for
positive and negative $\pi_\Theta$ are different from each other at a qualitative level.

\begin{figure}[p]\centering
 \includegraphics[width=.23\textwidth]{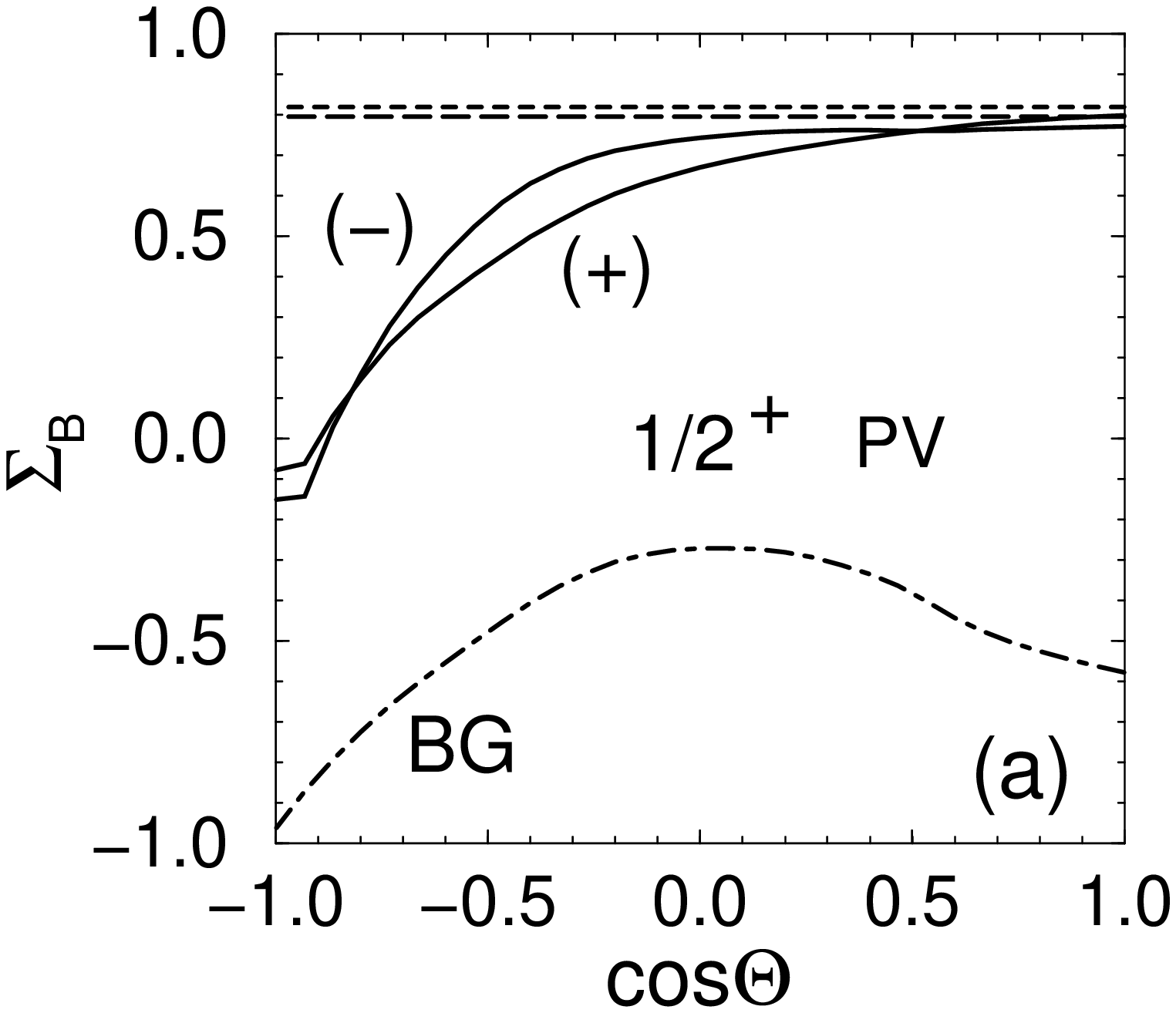}\hfill
 \includegraphics[width=.23\textwidth]{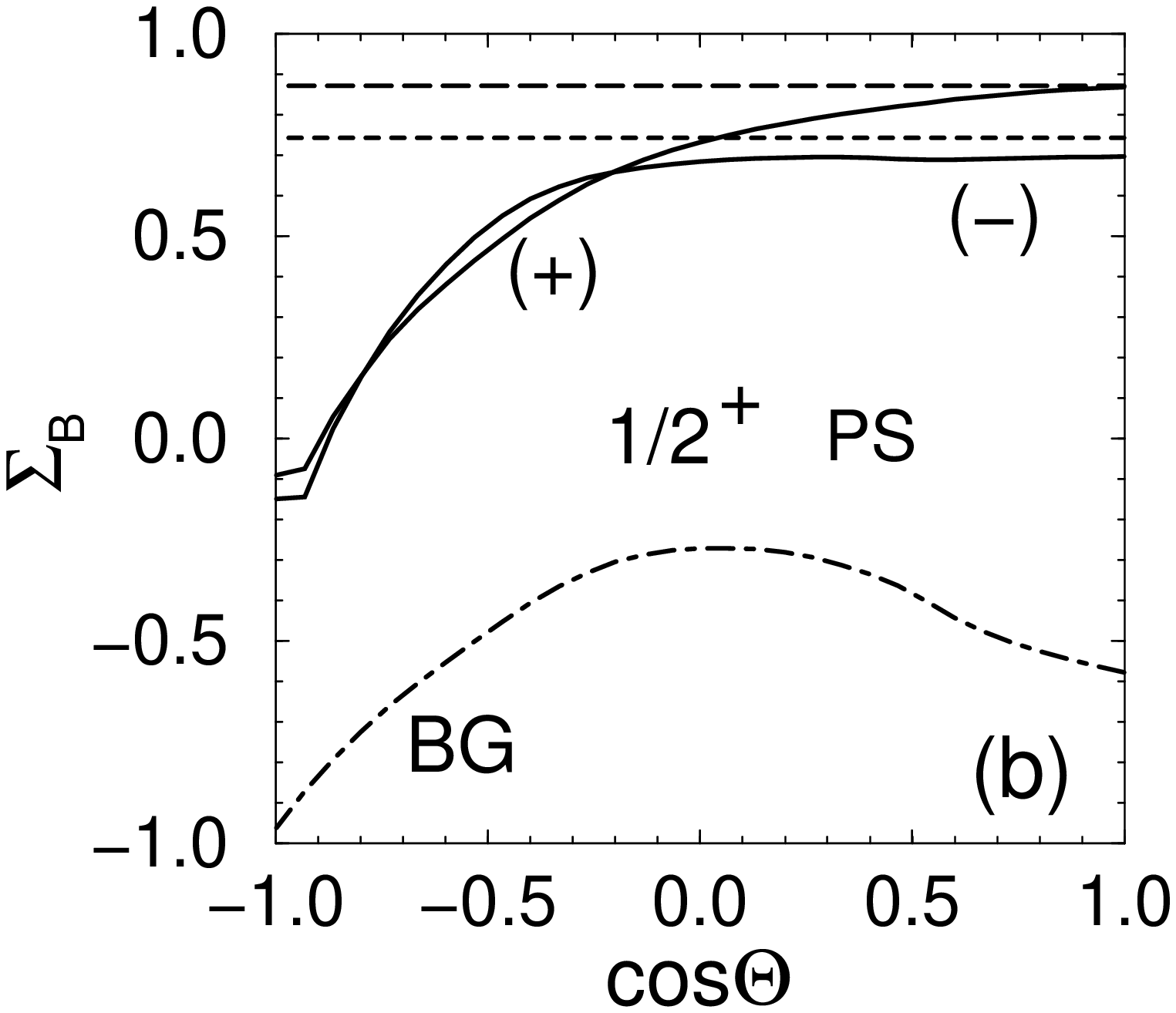}\hfill
 \includegraphics[width=.23\textwidth]{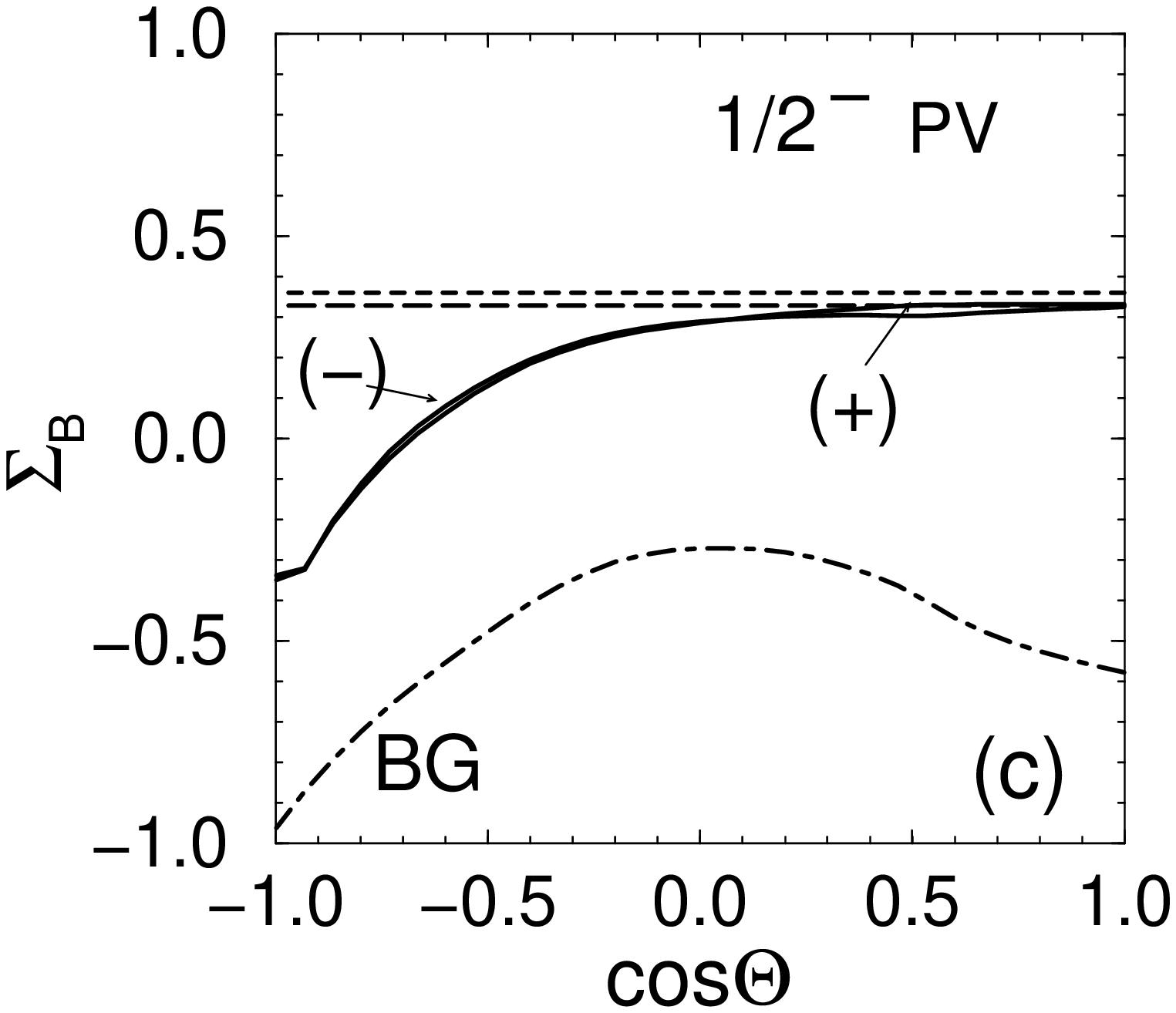}\hfill
 \includegraphics[width=.23\textwidth]{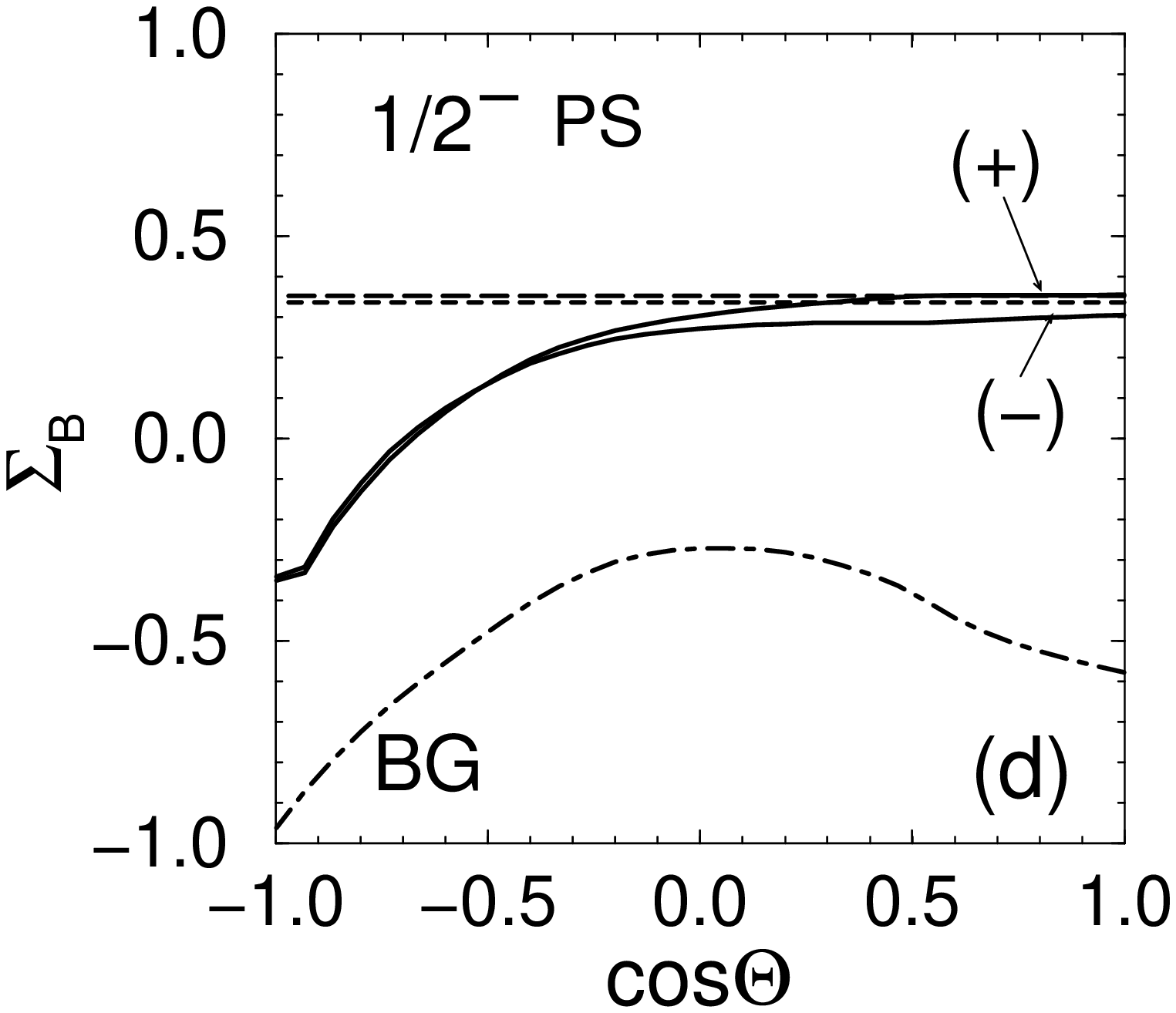}
 \caption{\label{fig:12}\tcaps%
 The beam asymmetry in
 $\gamma p\to pK^0\bar{K}^0$ as a function of the $K$ decay angle
 for $\kappa^*=0$.
 The results for $\pi_\Theta=+$ and $\pi_\Theta=-$ are shown
 in (a,b) and (c,d), respectively.
 The results for pseudovector (PV) and pseudoscalar (PS) couplings are
 displayed in (a,c) and (b,d), respectively. The symbol ($\pm$)
 corresponds to positive and negative $\alpha.$
 The asymmetries due to the
 resonant channel are shown by the long dashed ($\alpha>0$) and
 dashed ($\alpha<0$) lines, respectively. The asymmetry from the
 background is shown by the dot-dashed curves.}

\mbox{}\\

 \includegraphics[width=.23\textwidth]{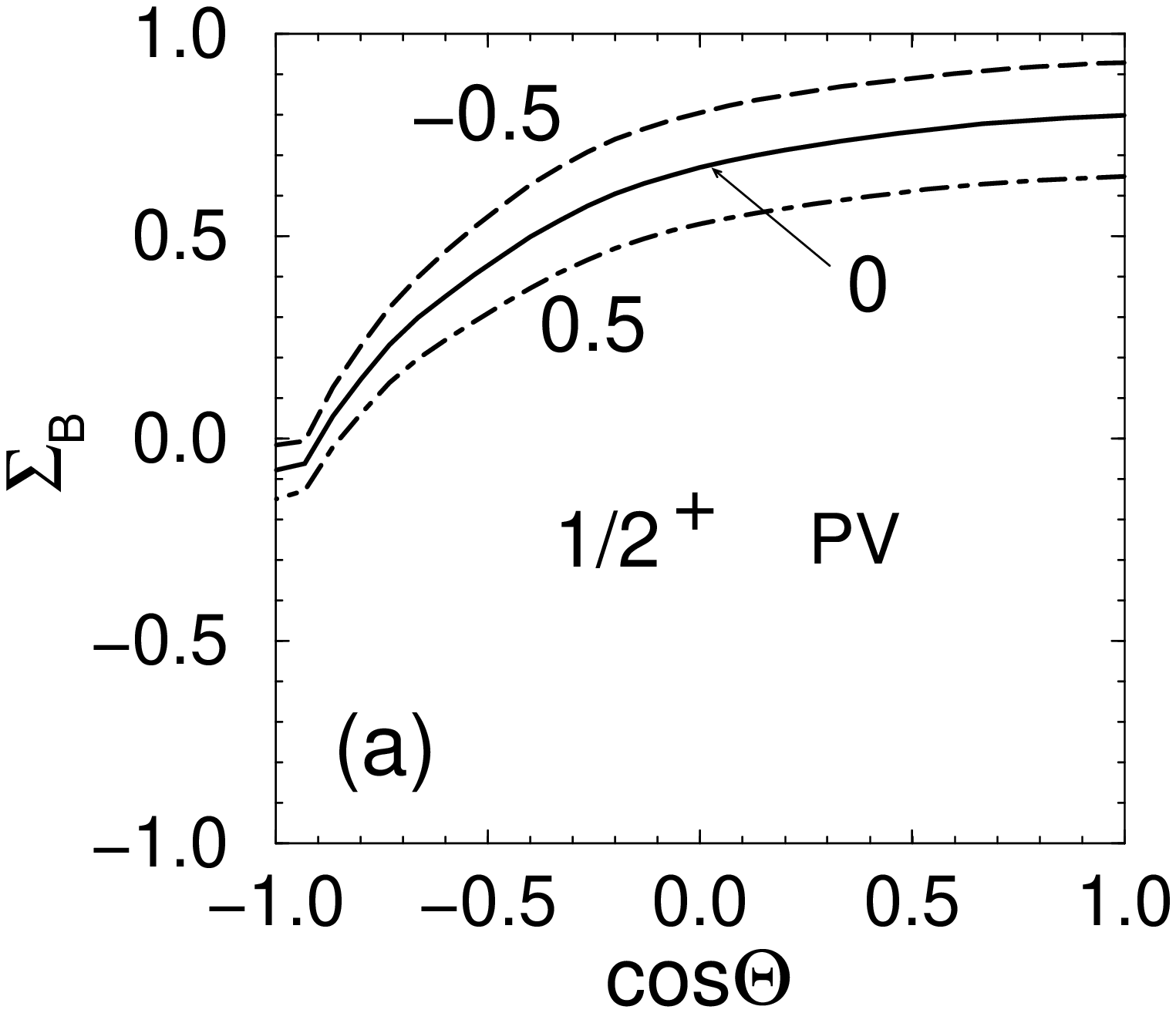}\hfill
 \includegraphics[width=.23\textwidth]{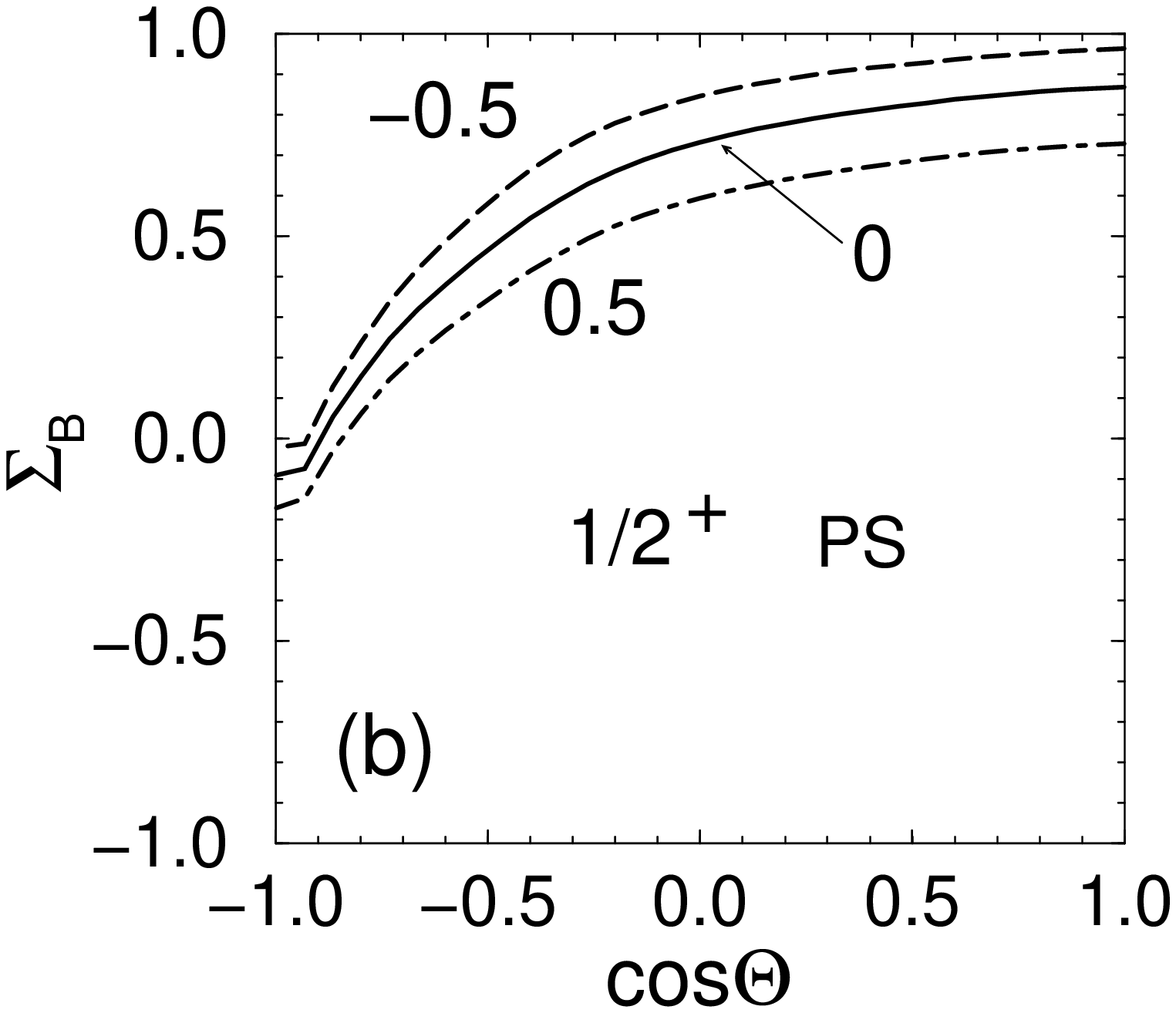}\hfill
 \includegraphics[width=.23\textwidth]{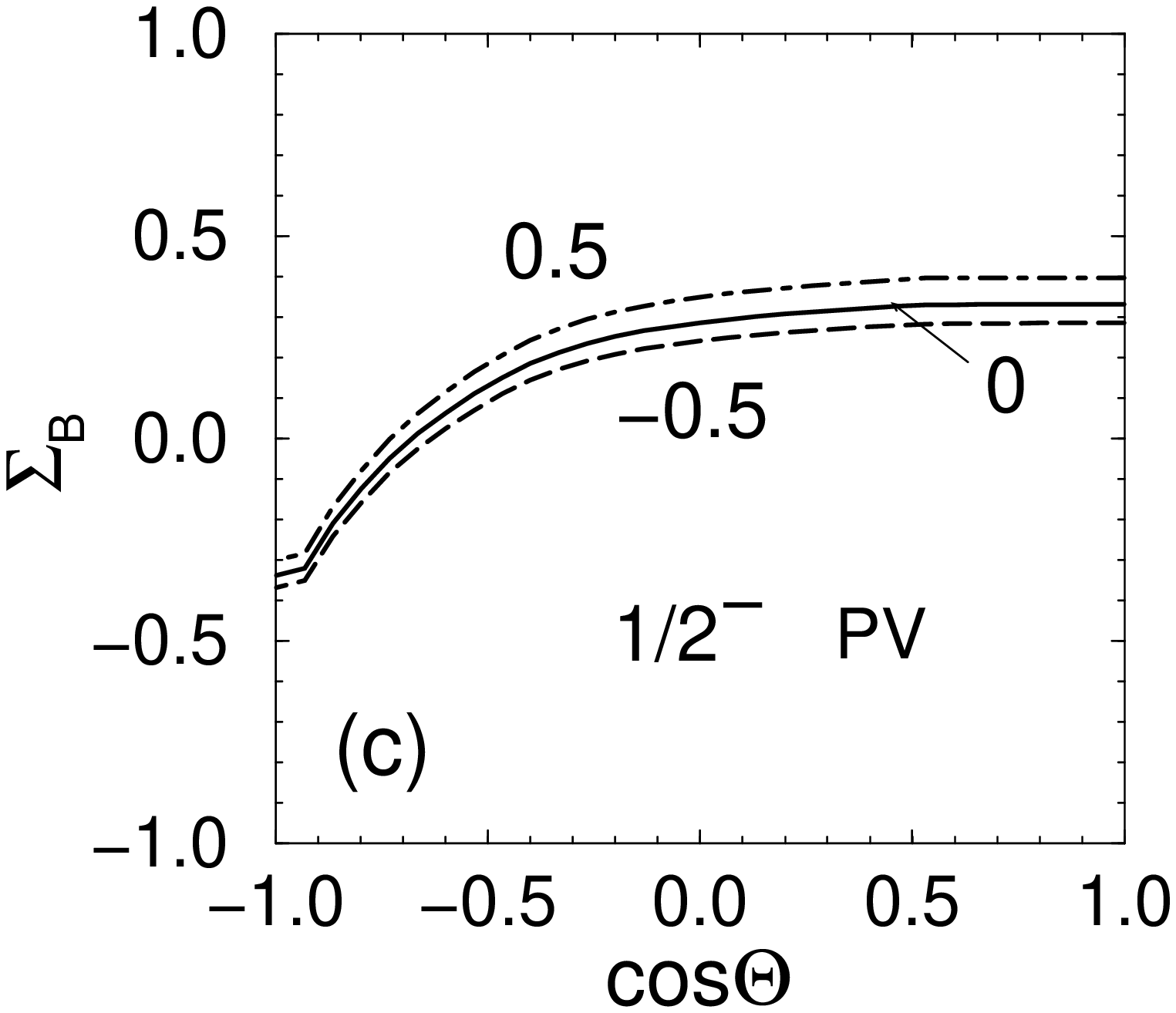}\hfill
 \includegraphics[width=.23\textwidth]{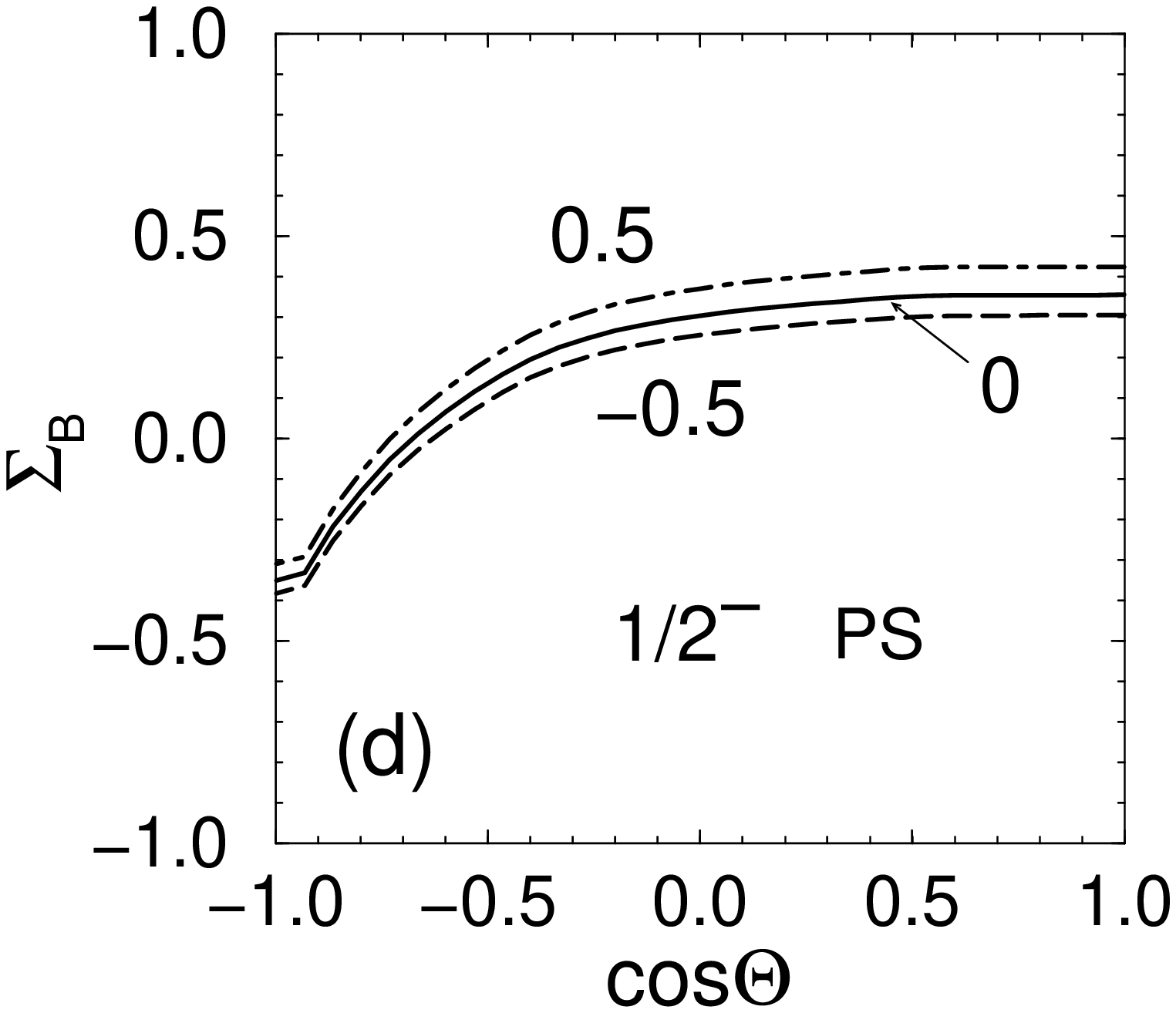}
 \caption{\label{fig:13}\tcaps%
 The beam asymmetry in
 $\gamma p\to pK^0\bar{K}^0$ as a function of the $K$ decay angle.
 The result is for $\bar\alpha>0$ and $\kappa^*=0,\, \pm0.5$.
 Other notations are the same as in Fig.~\protect\ref{fig:12}.}

\mbox{}\\

 \includegraphics[width=.23\textwidth]{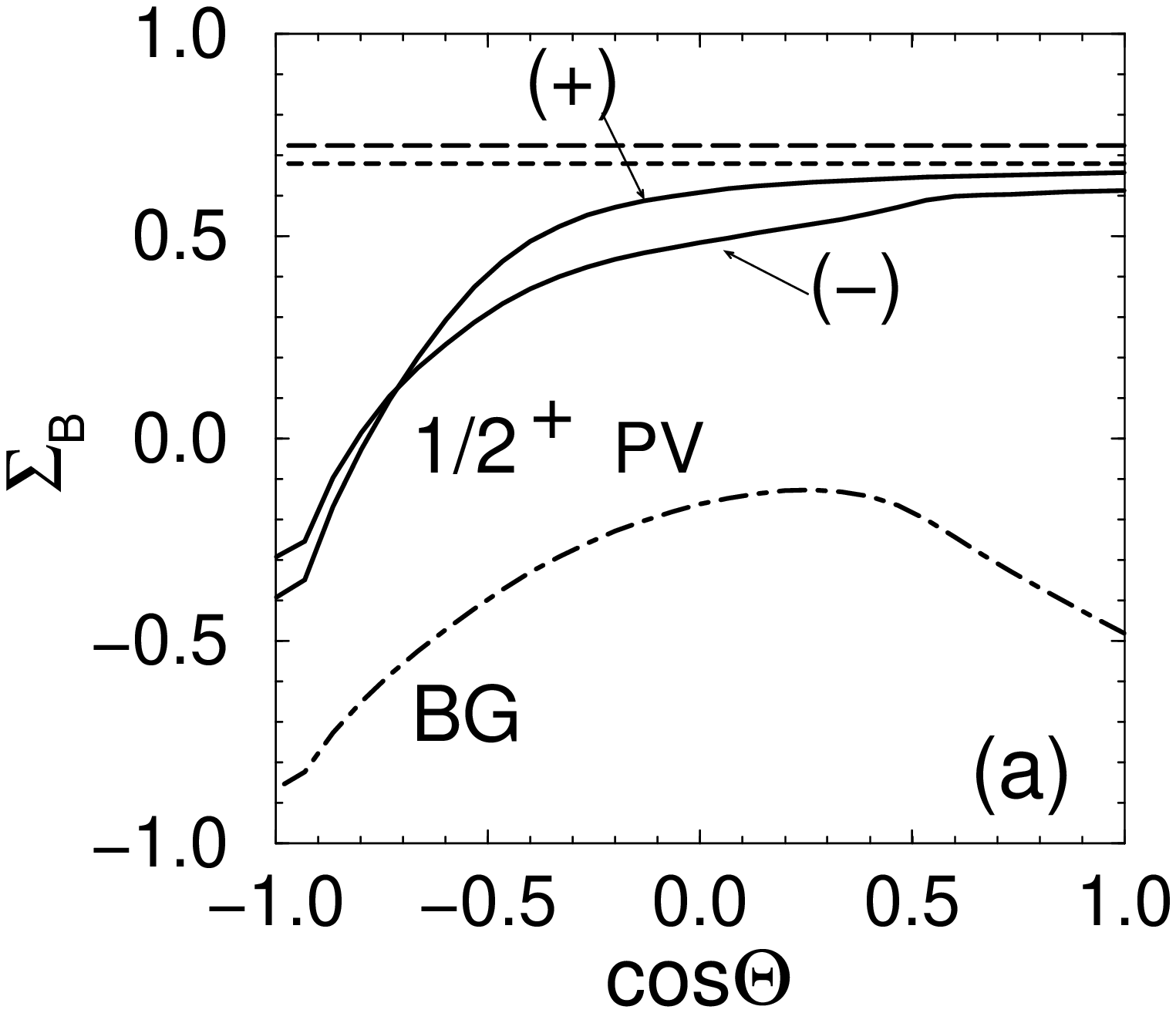}\hfill
 \includegraphics[width=.23\textwidth]{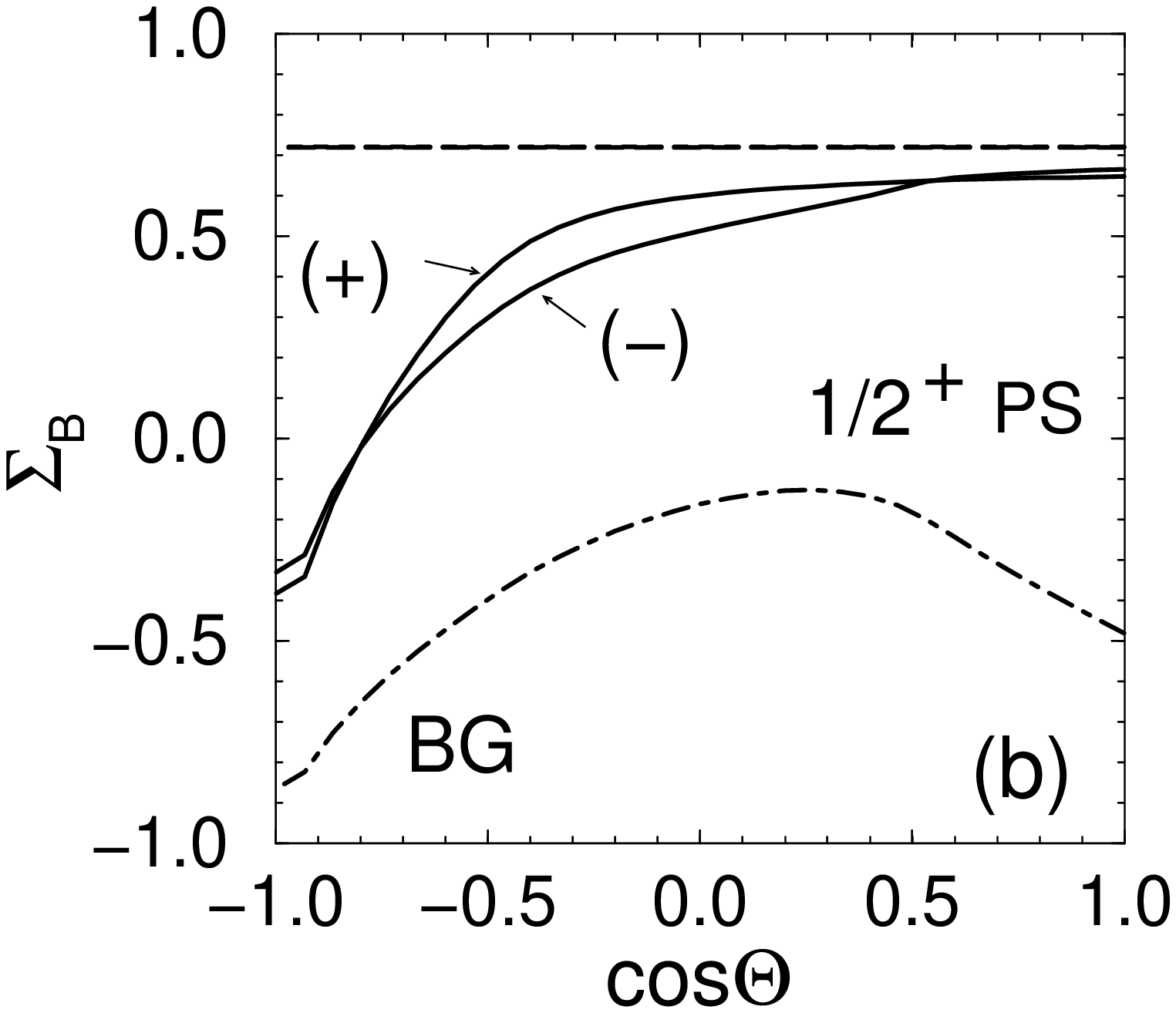}\hfill
 \includegraphics[width=.23\textwidth]{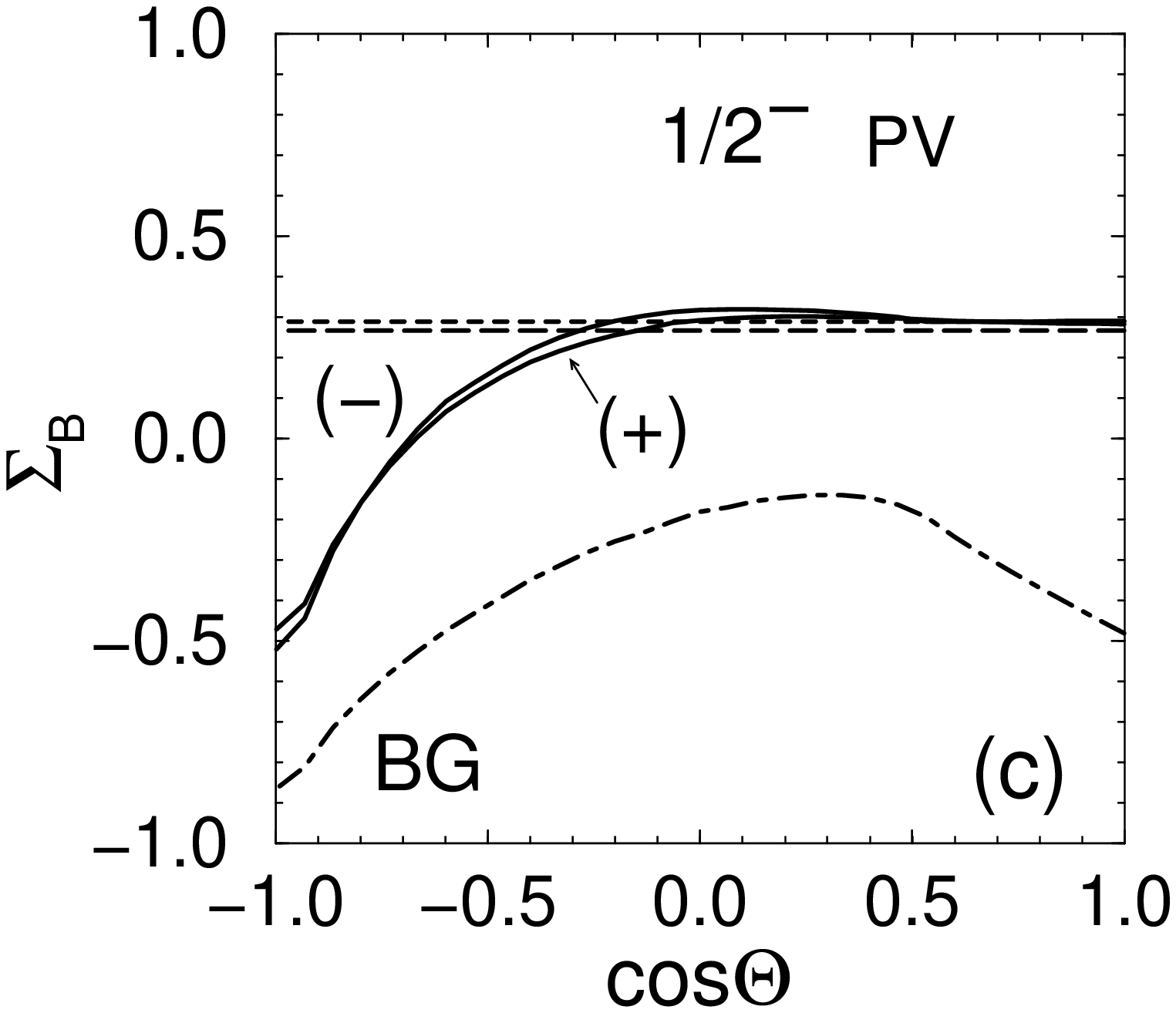}\hfill
 \includegraphics[width=.23\textwidth]{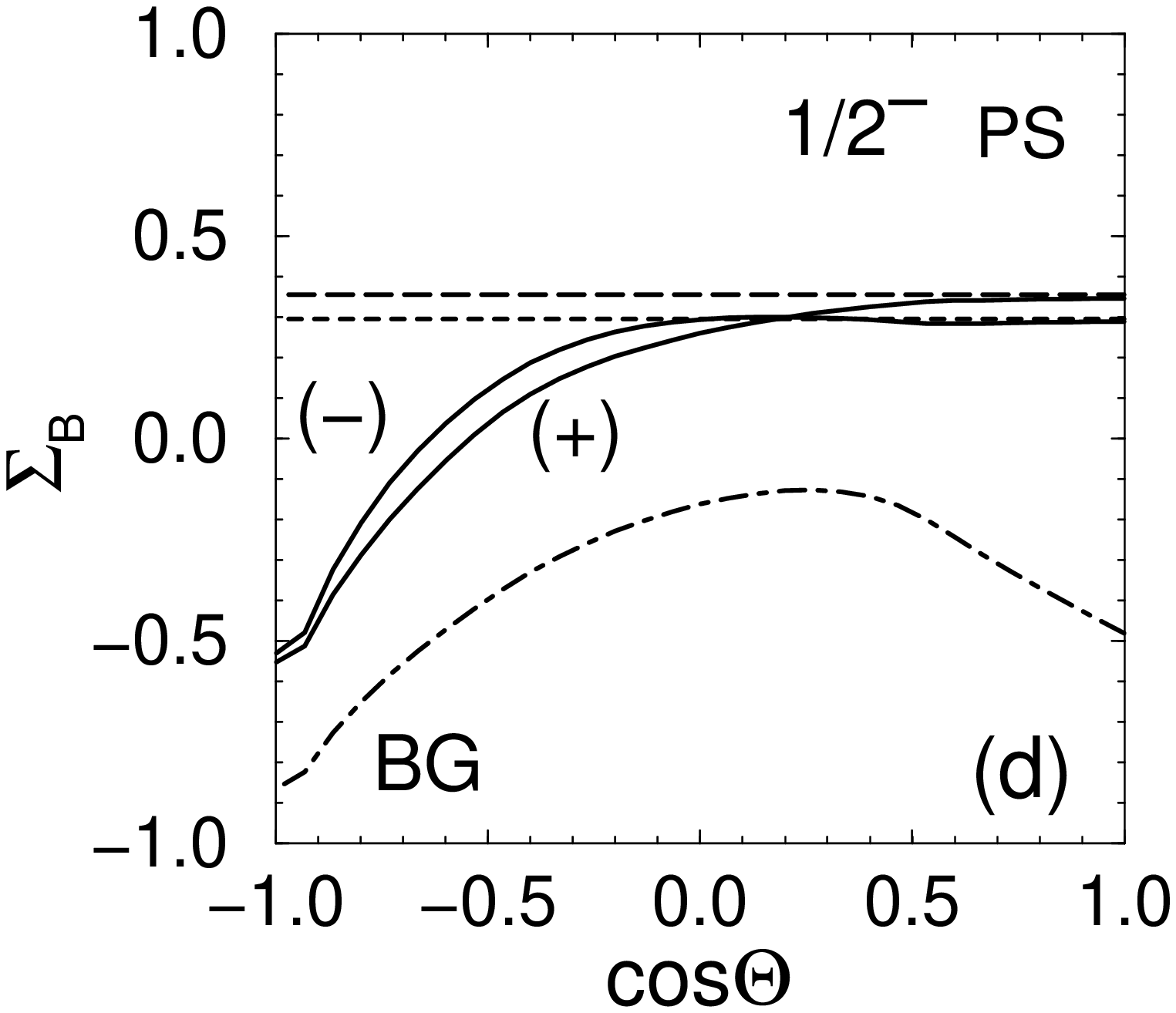}
 \caption{\label{fig:14}\tcaps%
 The beam asymmetry in
 $\gamma n\to nK^+\bar{K}^-$ as a function of the $K$ decay angle.
 Notation is the same as in Fig.~\protect\ref{fig:12}.}

\mbox{}\\

 \includegraphics[width=.23\textwidth]{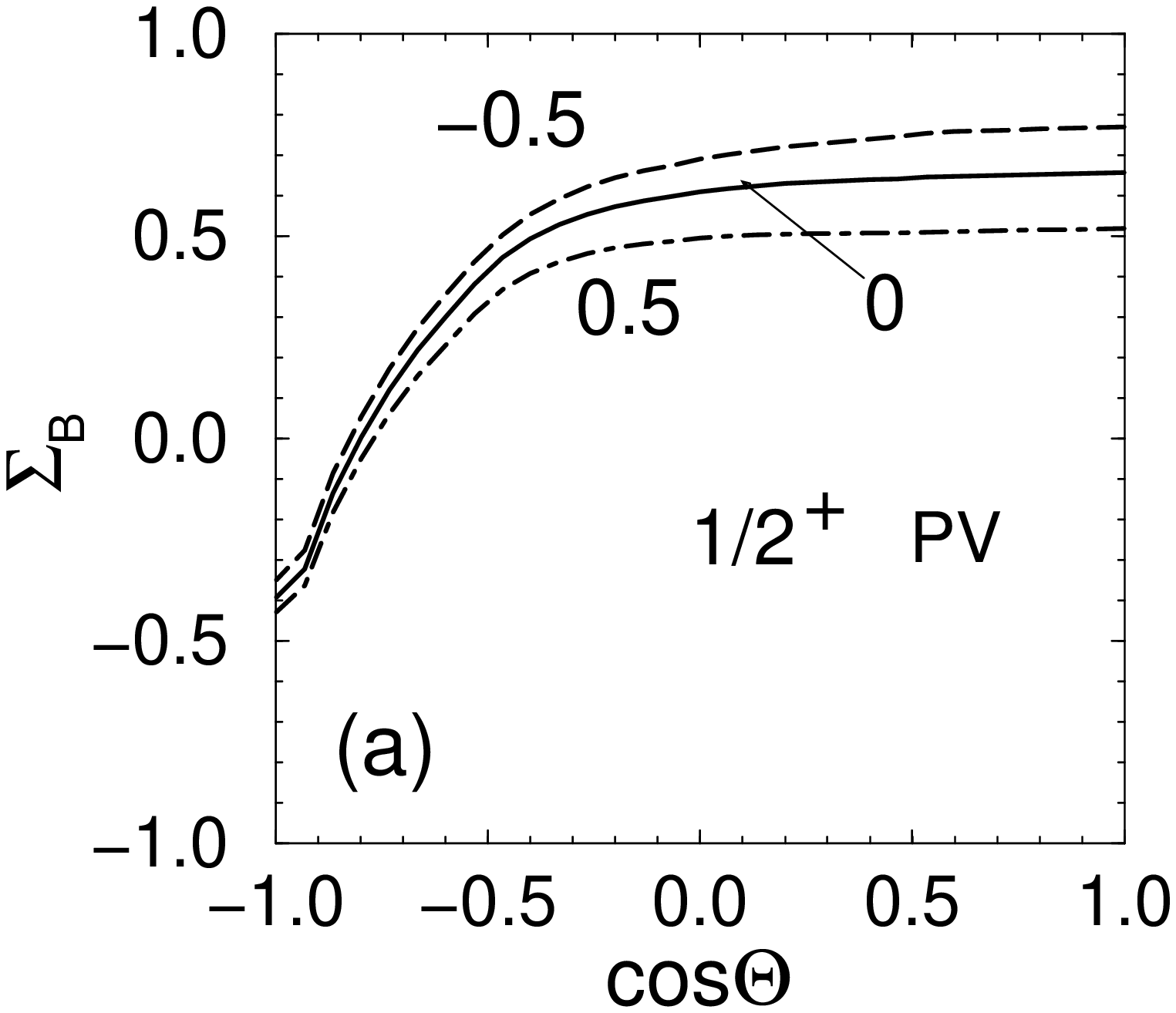}\hfill
 \includegraphics[width=.23\textwidth]{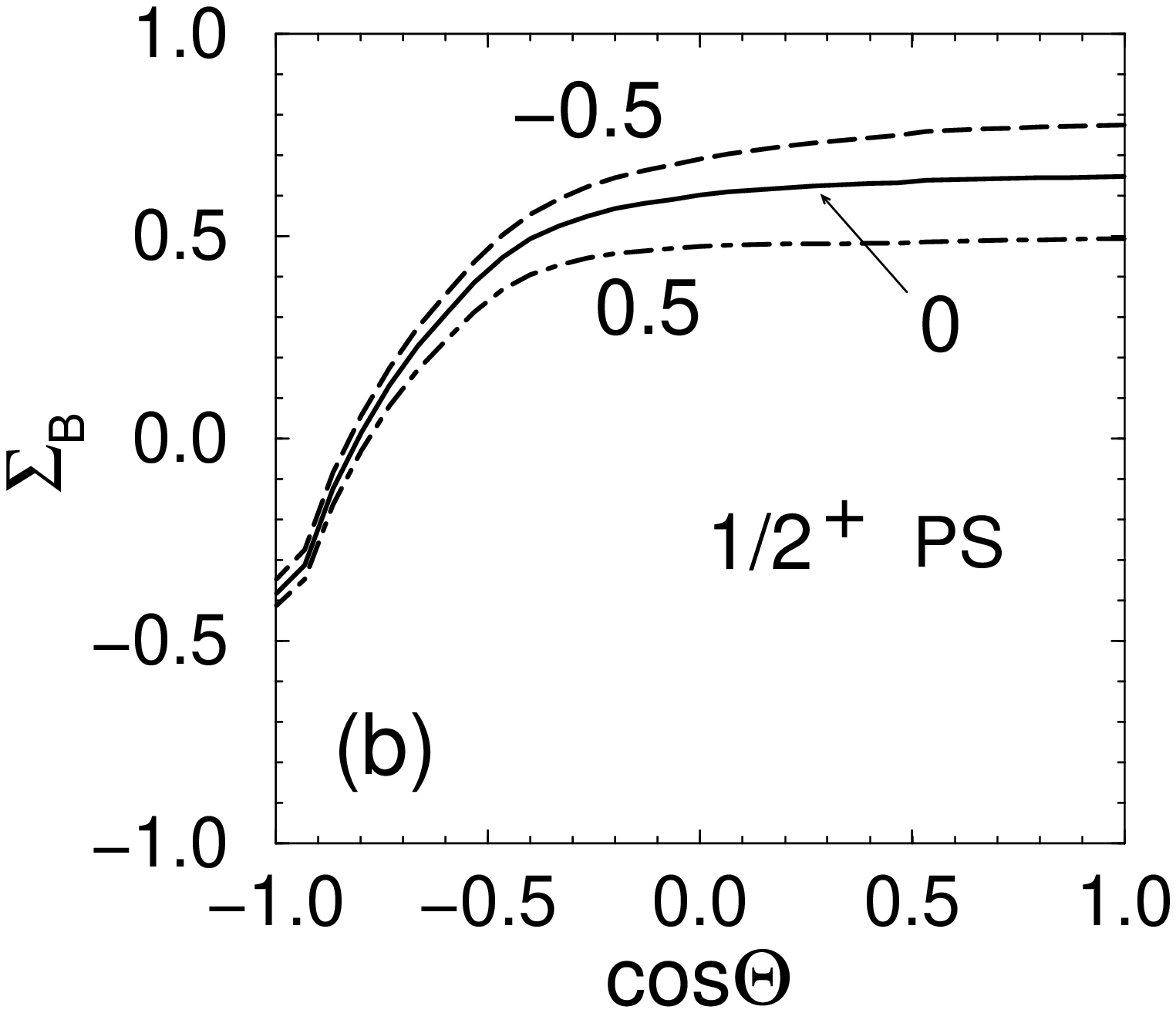}\hfill
 \includegraphics[width=.23\textwidth]{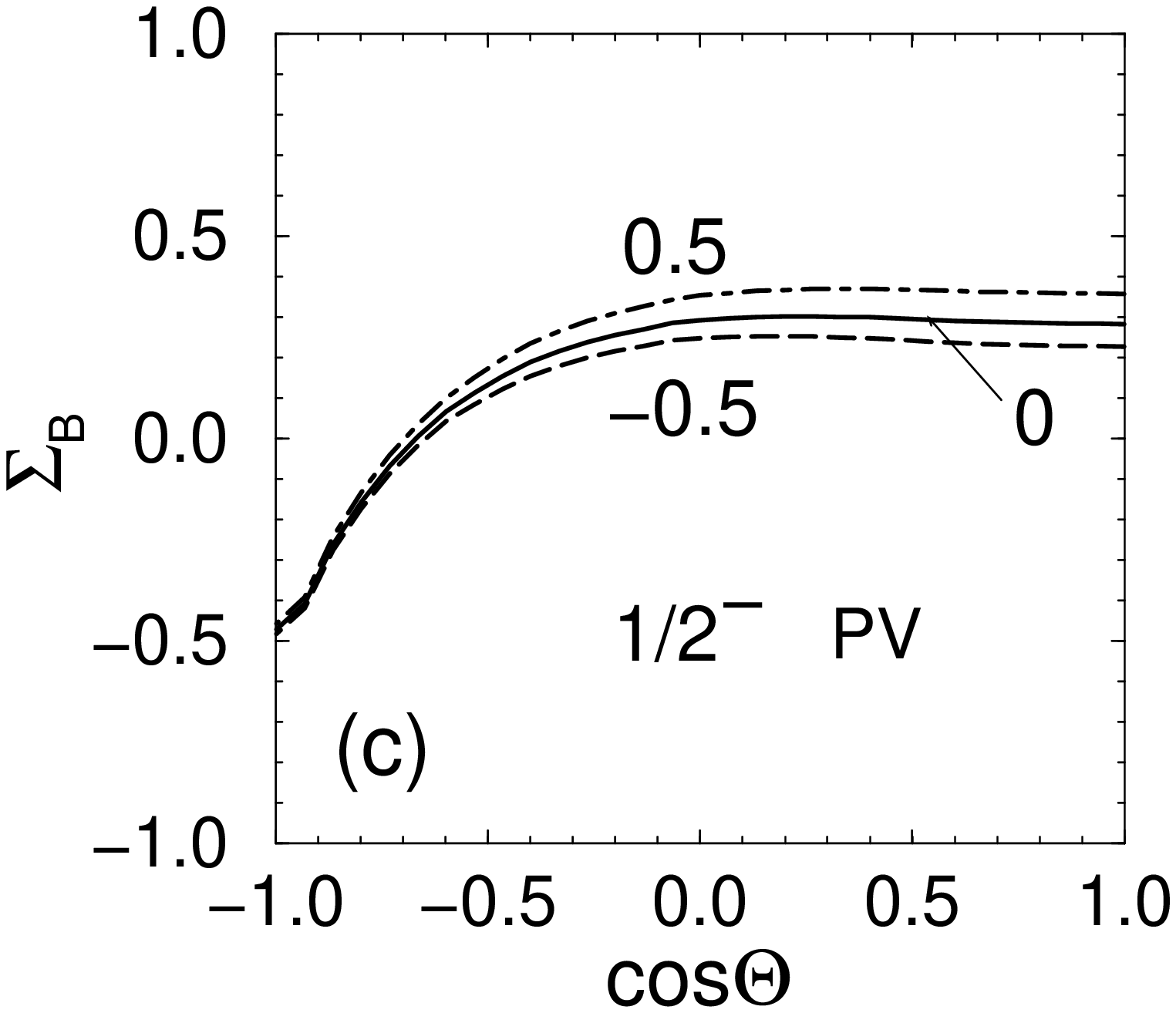}\hfill
 \includegraphics[width=.23\textwidth]{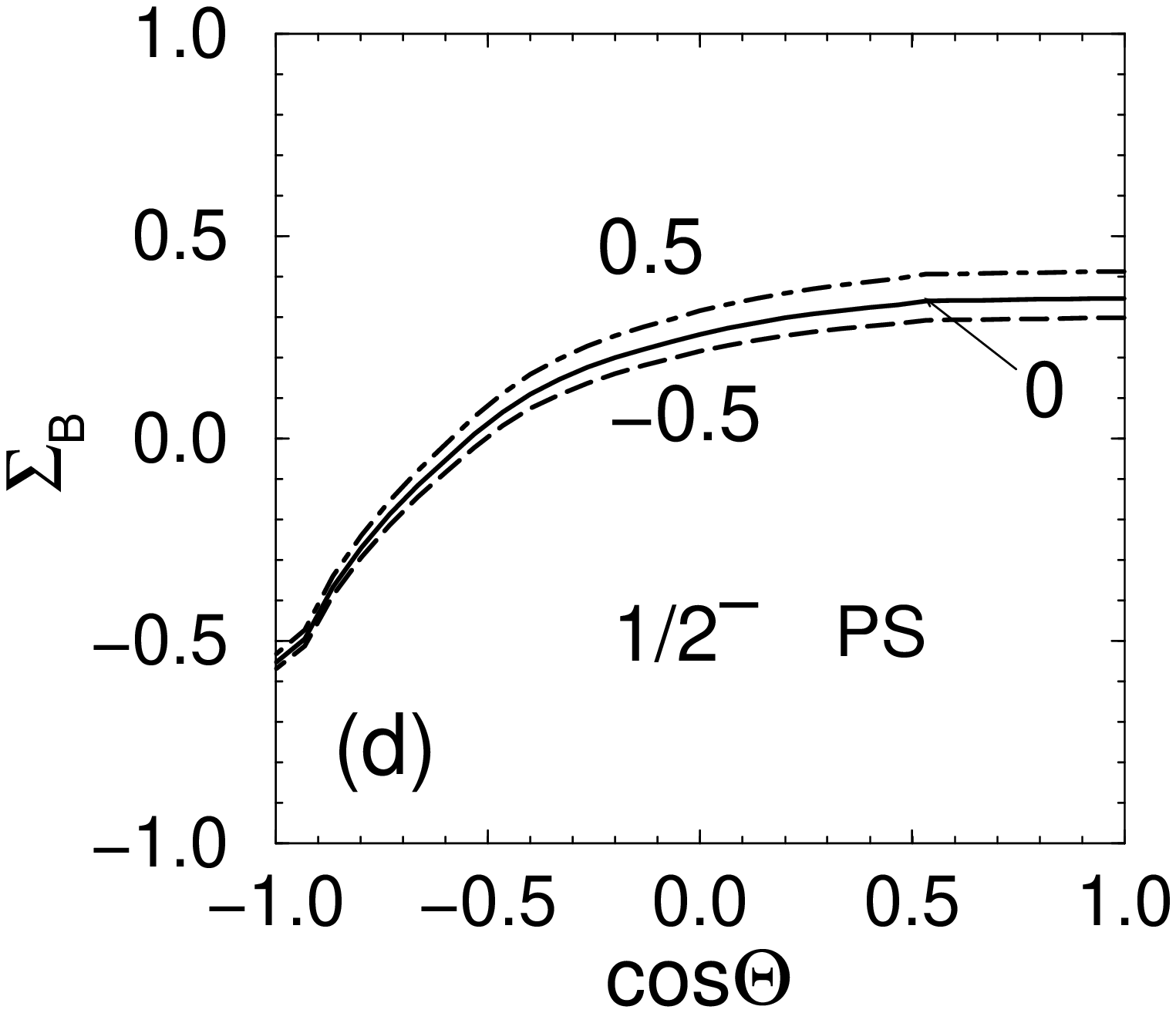}
 \caption{\label{fig:15}\tcaps%
 The beam asymmetry
 $\gamma n\to nK^+\bar{K}^-$ for different values of $\kappa^*$
as a function of the $K$ decay angle.
 Notation is the same as in Fig.~\protect\ref{fig:13}.}
\end{figure}

Figures~\ref{fig:12} and \ref{fig:13}  and Figs.~\ref{fig:14} and \ref{fig:15} show the
results of our calculation for $\gamma p\to p K^0\bar{K}^0$ and $\gamma n\to n K^+ K^-$,
respectively,  for different parity $\pi_\Theta$, different coupling schemes, different
signs of $\alpha$, and different $\kappa^*$. The results for positive and negative
$\pi_\Theta$  are shown in Figs.~\ref{fig:12}-\ref{fig:15} (ab) and (cd), respectively.
The results for the pseudovector (PV) and pseudoscalar (PS) couplings are displayed in
Figs.~\ref{fig:12}-\ref{fig:15} (ac) and (bd), respectively. The asymmetries shown in
Figs.~\ref{fig:12} and \ref{fig:14} are calculated with $\kappa^*=0$; the dependence on
$\kappa^*$ is shown in Figs.~\ref{fig:13} and \ref{fig:15}. In Figs.~\ref{fig:12} and
\ref{fig:14} the asymmetries due to the pure resonant channel are shown by the long
dashed ($\alpha>0$) and dashed ($\alpha<0$) lines. The asymmetry from the background is
shown by the dot-dashed curves.

One can see that the resonant channel contribution gives rise to a positive and constant
beam asymmetry. For the $\gamma p\to p K^0\bar{K}^0$ reaction its  value depends on
$\pi_\Theta$: $\Sigma_B^{+}\approx 2 \Sigma_B^{-}\approx 0.6 - 0.8$. For the $\gamma n\to
n K^+ K^-$ reaction this dependence is rather weak. The asymmetry due to the background
channels is negative with relatively large absolute value. The interplay of the resonant
and background processes results in a strong deviation of $\Sigma_B$ from the constant
(``resonant") values; in all the cases the asymmetry decreases down to negative values
when $\Theta\to \pi$. The variation of $\kappa^*$ modifies the asymmetry at
$\cos\Theta\approx 1$ leaving, however, its shape almost unchanged. In
Refs.~\cite{NT03,Zhao03} the single beam asymmetry for the reaction $\gamma N\to \Theta^+
\bar{K}$ was suggested for the determination of $\pi_\Theta$ by measuring the angular
distribution of the $\bar{K}$ mesons. In our analysis of the two-step process $\gamma N\to
\Theta^+ \bar{K}\to N K\bar{K}$, however, we find that the dependence of $\Sigma_B$ on the
parity $\pi_\Theta$ is not pronounced enough to allow the extraction of the parity of the
$\Theta^+$ state.

 To summarize this subsection, we can conclude that
the beam asymmetry cannot be used as a tool for determining
$\pi_\Theta$. The shapes of $\Sigma_B$ for positive and negative
$\pi_\Theta$ are almost the same. The difference in $\Sigma_B$ in
$\gamma p\to p K^0\bar{K}^0$  mentioned above gets small if we
compare the result for positive $\pi_\Theta$ with $\kappa^*=0.5$
and the result for negative $\pi_\Theta$ with $\kappa^*=-0.5$.  A
similar conclusion is valid for other single spin observables.

\subsubsection{Double spin observables}

 As we have seen, the $\Theta^+$ decay amplitudes in Eq.~(\ref{Dpm}) are related directly
to $\pi_\Theta$. Therefore, observables sensitive to the $\Theta^+$ parity must involve
the spin dependence of the outgoing nucleon. Let us consider one of them, the
target-recoil spin asymmetry, where the spin variables are related to the spin
projections of incoming (target) and outgoing (recoil) nucleons~,
\begin{eqnarray}
 {\cal
A}_{tr}=\frac{\sigma(\uparrow\uparrow)-\sigma({\uparrow\downarrow})}
{\sigma(\uparrow\uparrow)+\sigma({\uparrow\downarrow})}~.
\label{Atr}
 \end{eqnarray}
Here, $\sigma(\uparrow\uparrow)$ and
$\sigma({\uparrow\downarrow})$ are the cross sections without and
with the spin flip transition from the incoming to outgoing nucleon,
respectively. Using Eqs.~(\ref{Dpm}) one can get the asymmetries due to the
pure resonance channel,
\begin{subequations}
\label{AtrT}
\begin{align}
 {\cal A}^+_{tr}(\Theta)&={\cal A}^+_0\cos2\Theta~,
 \qquad
 {\cal A}^-_{tr}(\Theta)={\cal A}^-_0~,\\[1ex]
 &{\cal A}^{\pm}_0=
 \frac{d\sigma^{\pm}_R(\uparrow\uparrow)-d\sigma^{\pm}_R({\uparrow\downarrow})}
 {d\sigma^{\pm}_R(\uparrow\uparrow)+d\sigma^{\pm}_R({\uparrow\downarrow})}~,
 \end{align}
 \end{subequations}
where $\sigma^{+}_R$ and $\sigma^{-}_R$ denote the cross section
for $\Theta^+$ photoproduction with the positive and negative
$\pi_\Theta$ values, respectively. One can see that the spin
asymmetries for positive and negative parities are qualitatively
different to each other. In the case of a negative parity, the
asymmetry is a constant, independent of $\Theta$. For positive
parity, the corresponding asymmetry exhibits a $\cos2\Theta$
dependence which leads to a minimum or a maximum value at
$\Theta=\pi/2$ and a null value at $\Theta=\pi/4$ and $3\pi/4$.

The next observation is related to the production mechanism.
The dominance of the $K^*$-exchange channels leads to a relative
suppression of the spin-flip transitions for positive $\pi_\Theta$.
Therefore, we have
${\cal A}^+_{0}>0$, which results in a minimum of the asymmetry ${\cal A}^+_{tr}
= {{\cal A}_{tr}}_\text{min}$  at $\Theta=\pi/2$.
For  negative $\pi_\Theta$, the dominance of the $K^*$-exchange channels results
in an enhancement of these transitions, so that ${\cal A}^-_{0}<0$.

But this ideal picture is modified when we include the background contribution. The
background processes are dominated by the spin-conserving transition which results in a
positive asymmetry. Therefore, in the case of a negative $\pi_\Theta$, the total
asymmetry may be either positive or negative. In the case of a positive $\pi_\Theta$, the
total asymmetry loses its $\cos2\Theta$ dependence.

\begin{figure}[b]\centering
 \includegraphics[width=.23\textwidth]{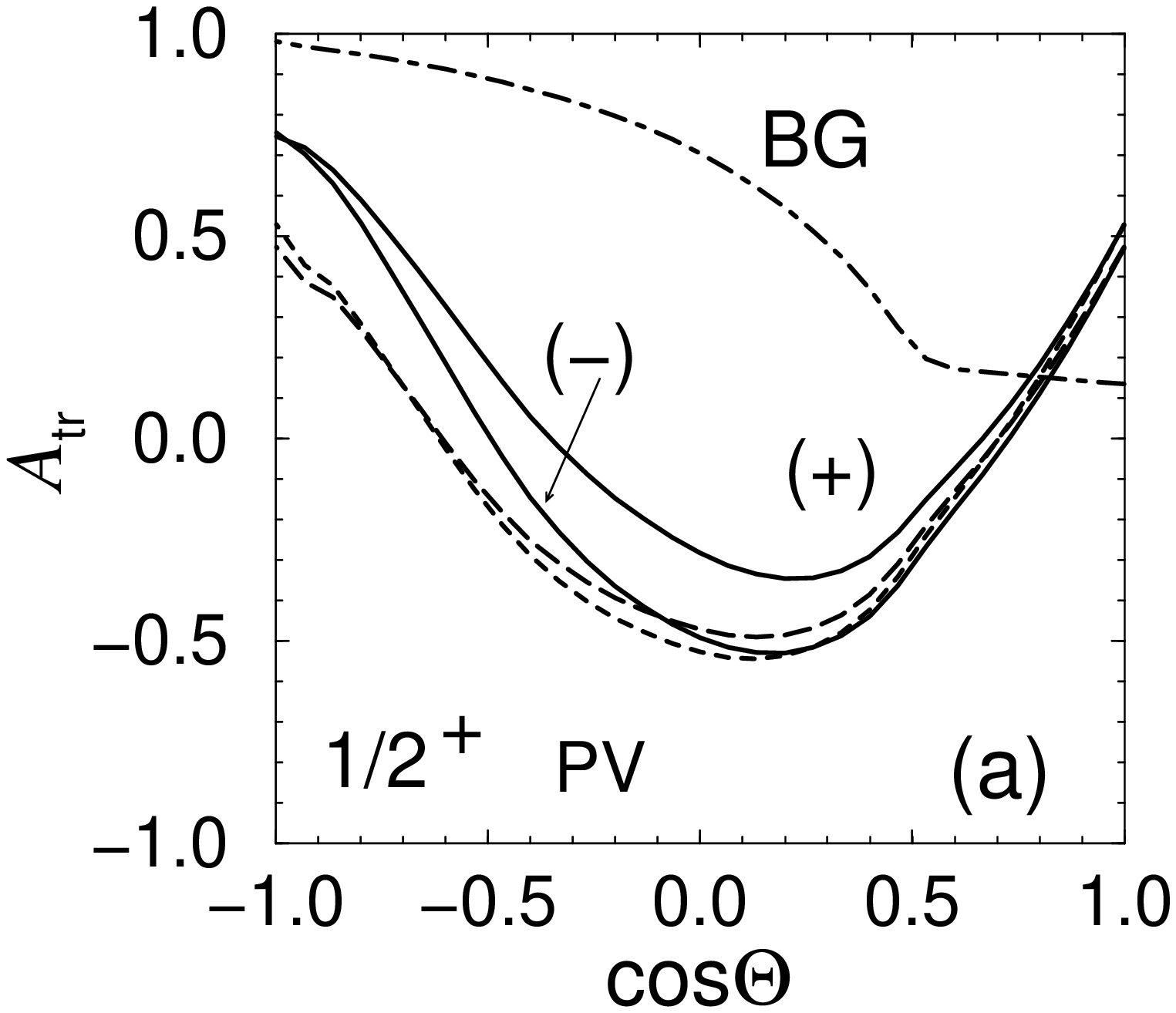}\hfill
 \includegraphics[width=.23\textwidth]{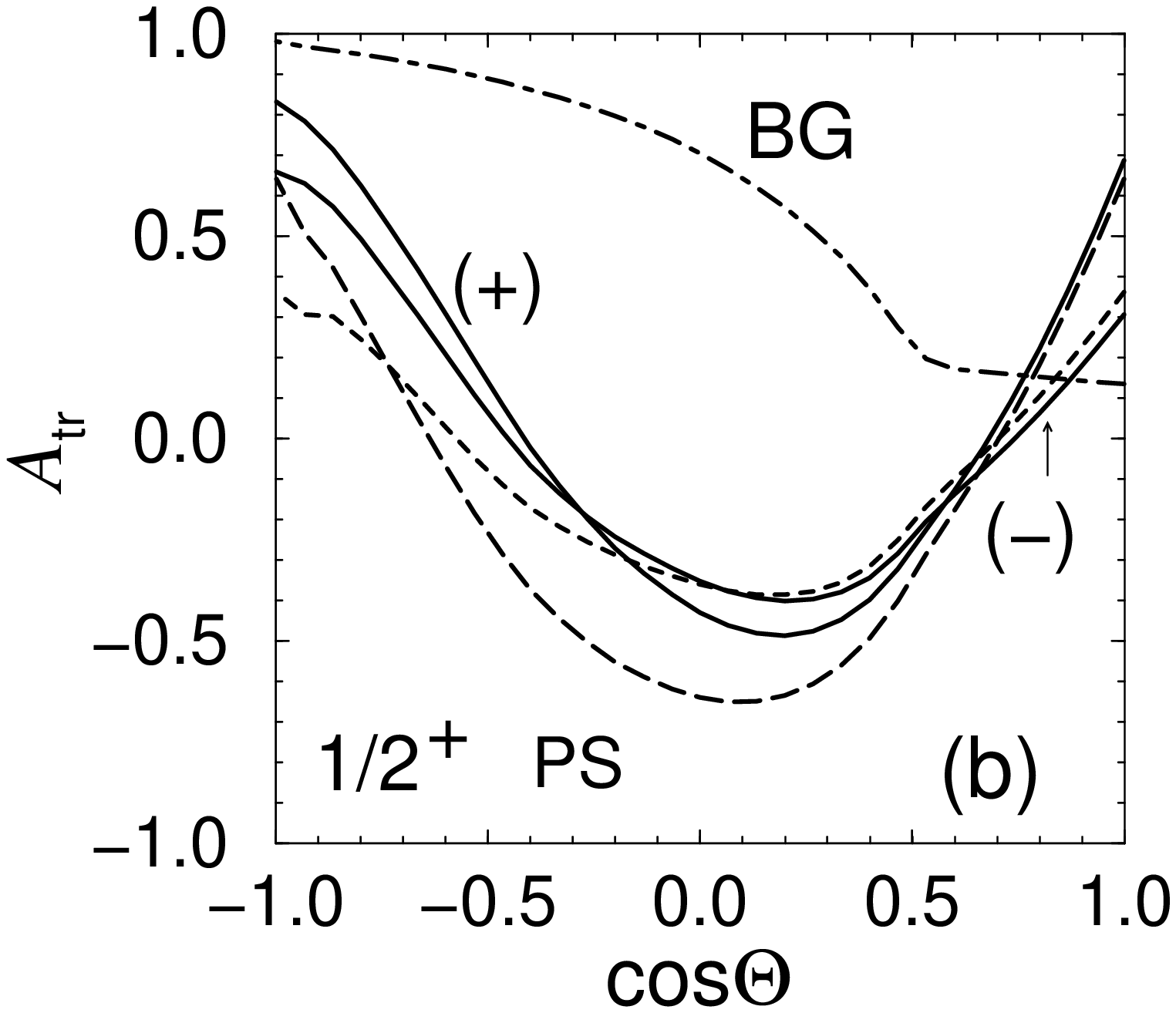}\hfill
 \includegraphics[width=.23\textwidth]{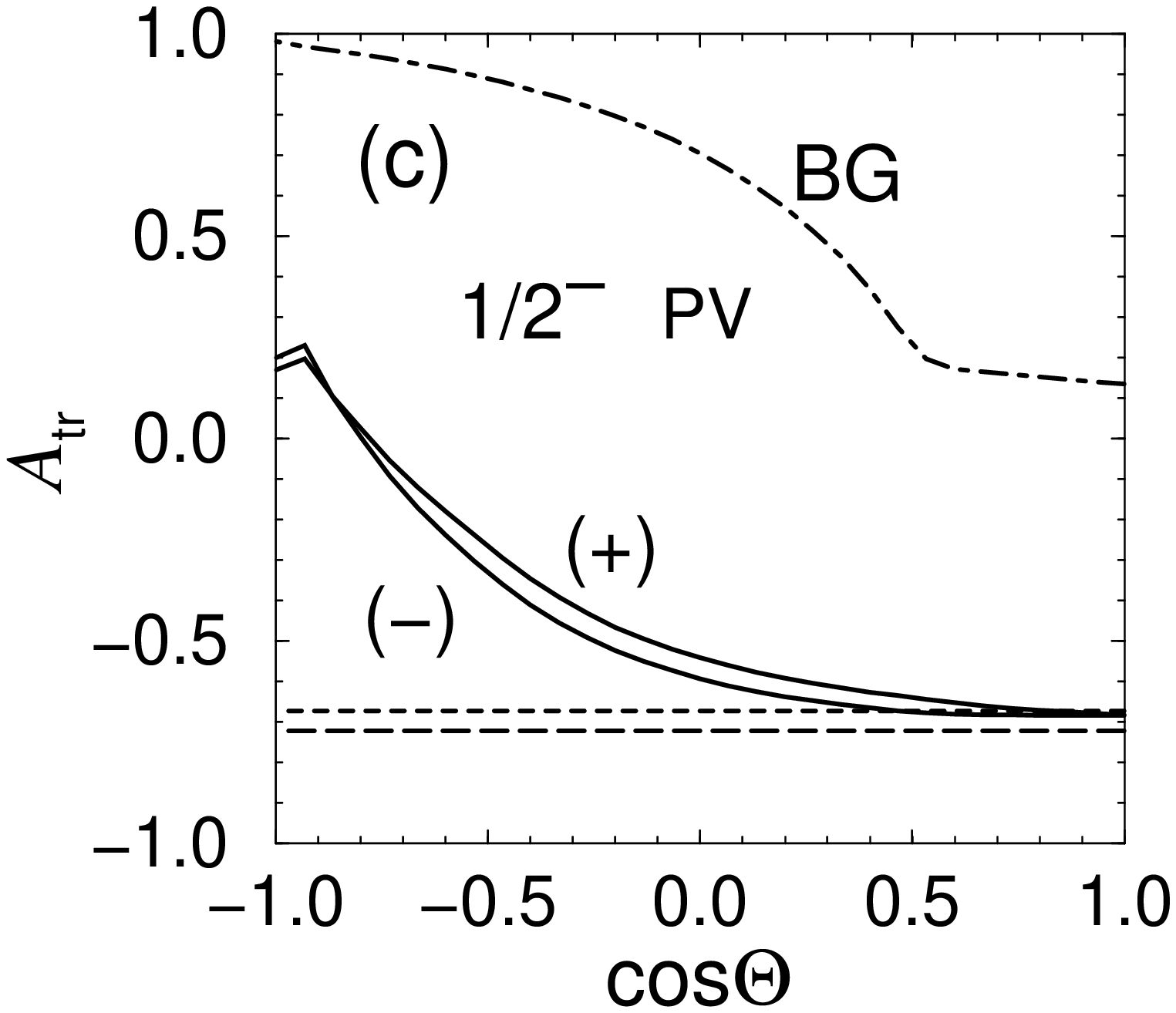}\hfill
 \includegraphics[width=.23\textwidth]{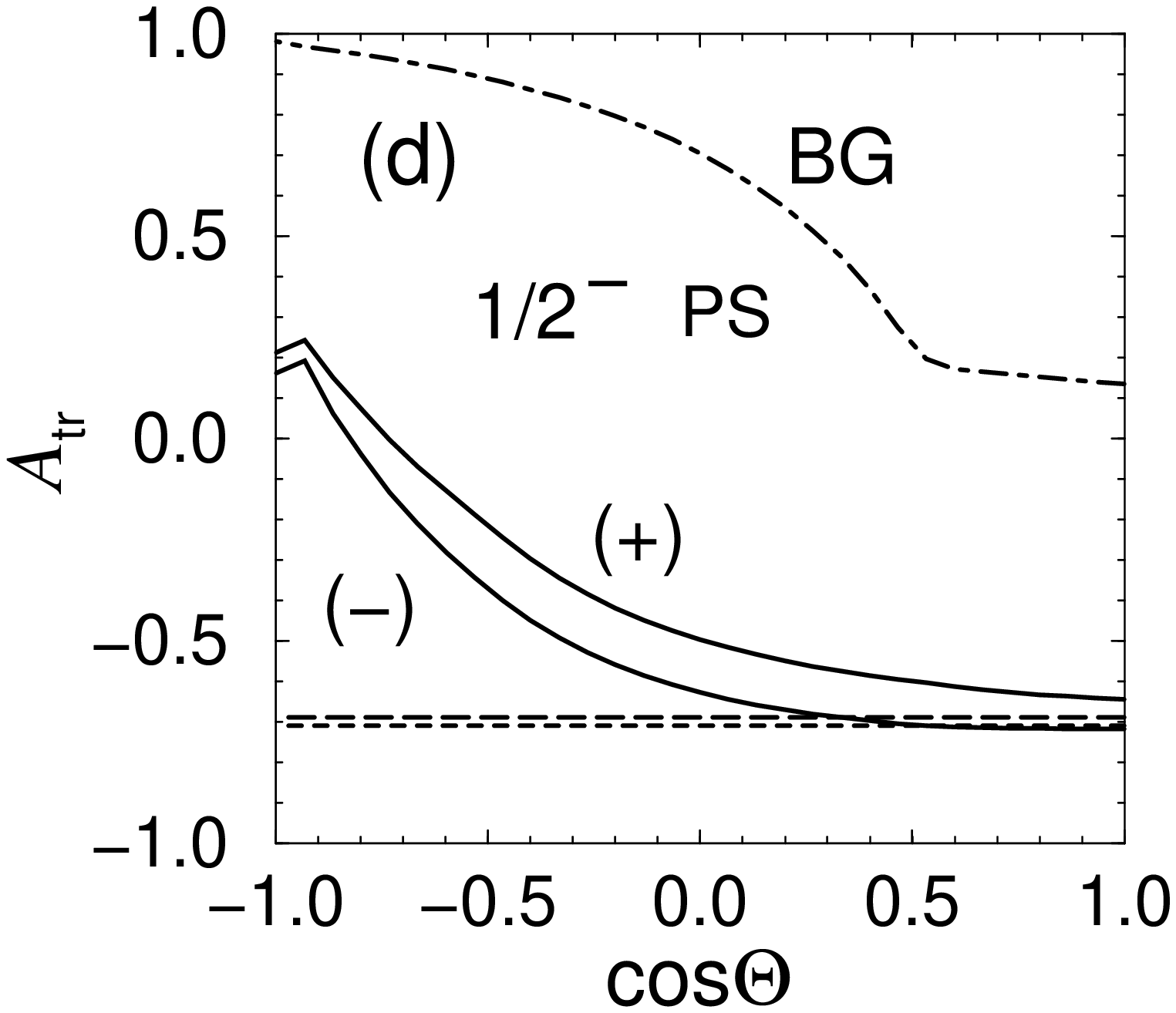}
 \caption{\label{fig:16}\tcaps%
 The double target-recoil  spin asymmetry  $ {\cal A}_{tr}$ in
 $\gamma p\to p K^0\bar{K}^0$ as a function of the $K$ decay angle.
  Notation is the same as in Fig.~\protect\ref{fig:12}.}

\mbox{}\\

 \includegraphics[width=.23\textwidth]{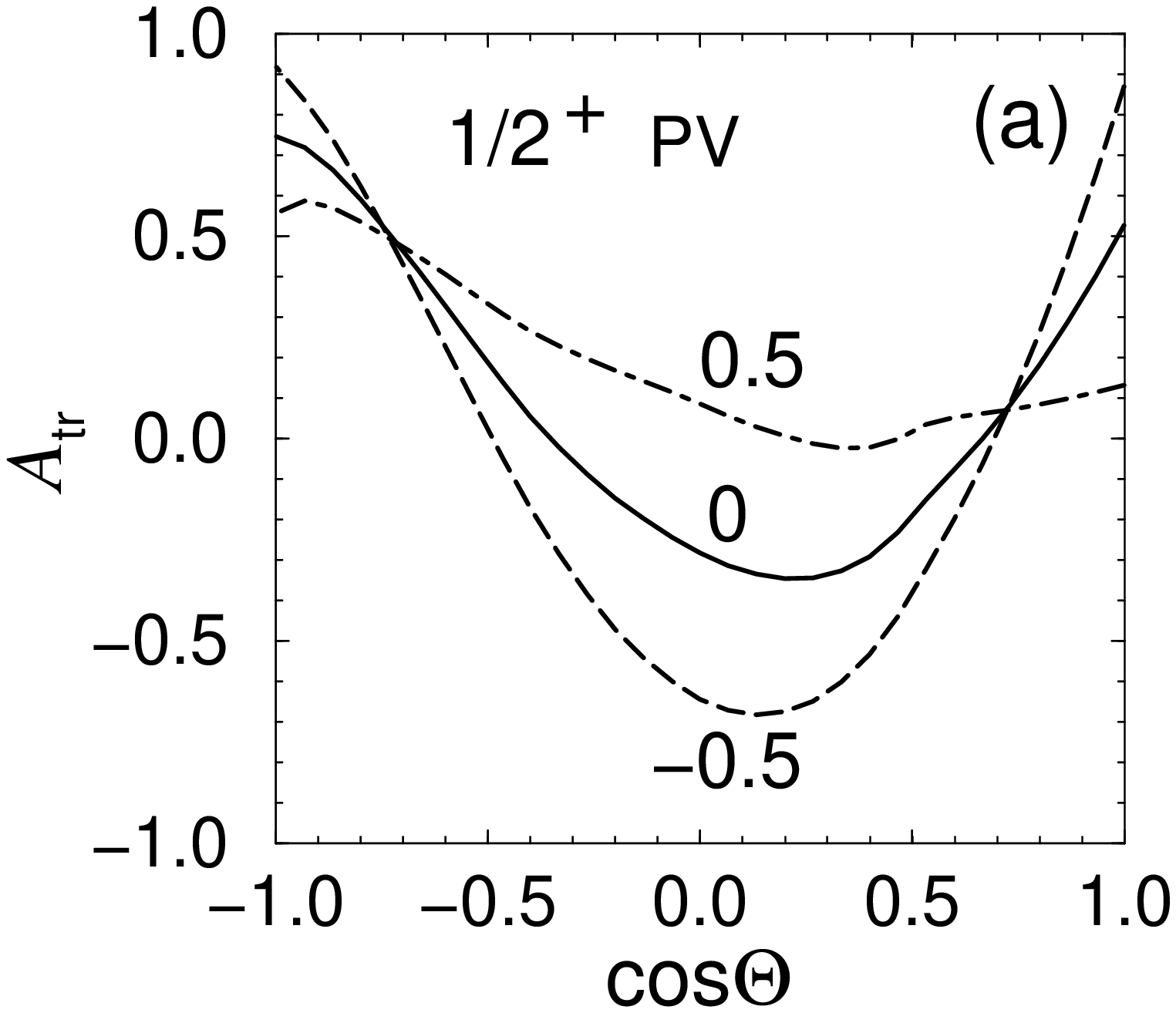}\hfill
 \includegraphics[width=.23\textwidth]{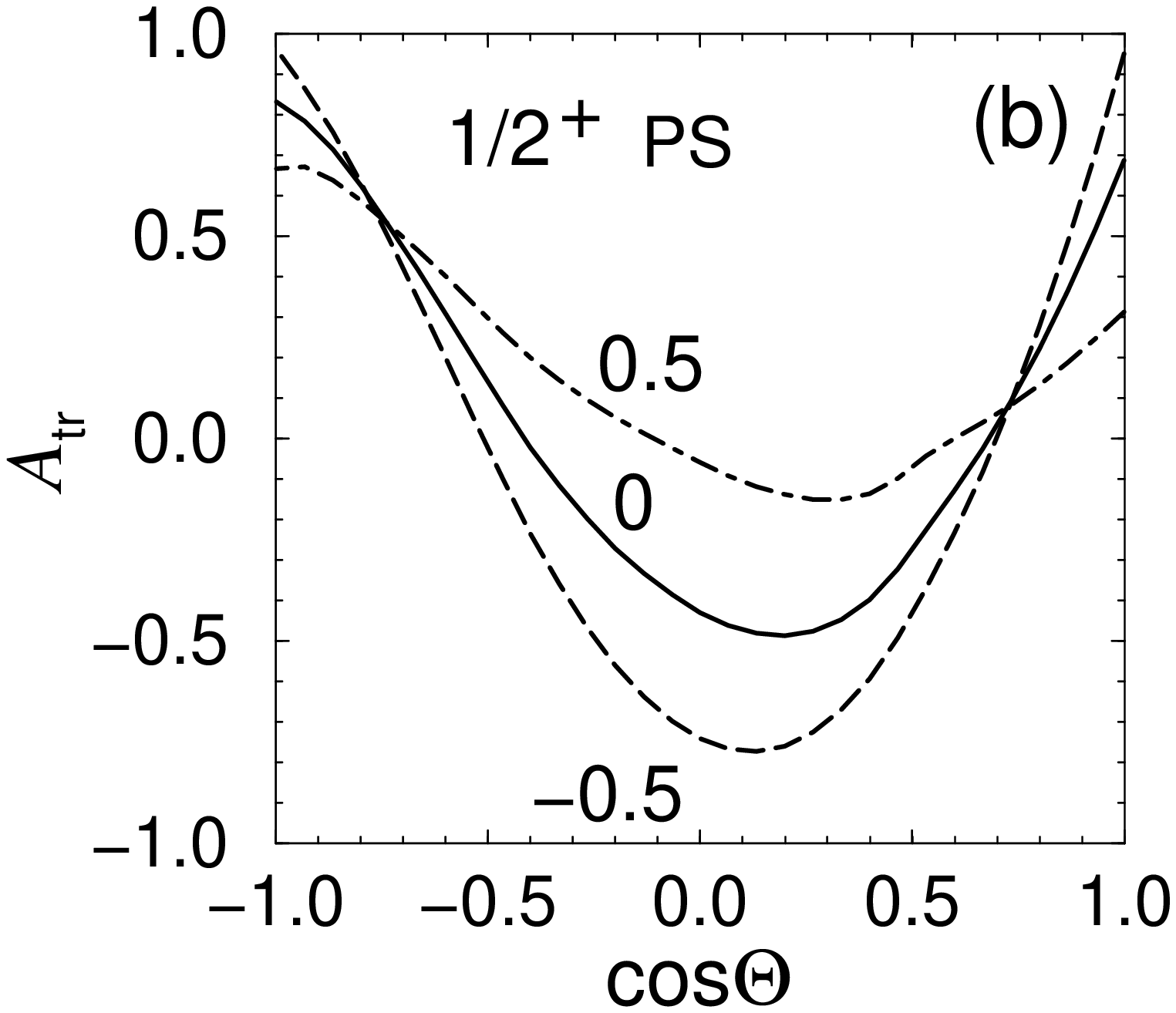}\hfill
 \includegraphics[width=.23\textwidth]{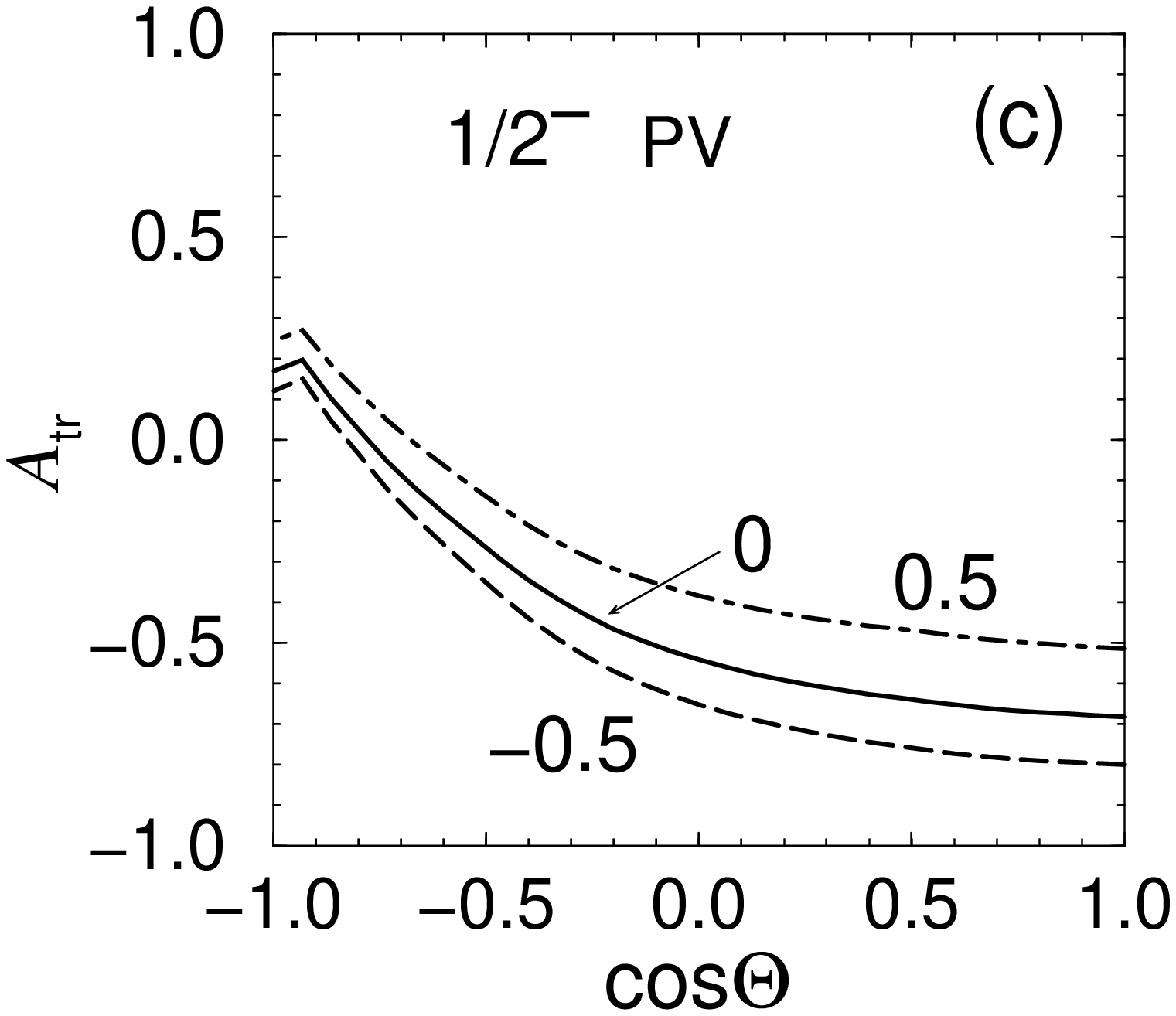}\hfill
 \includegraphics[width=.23\textwidth]{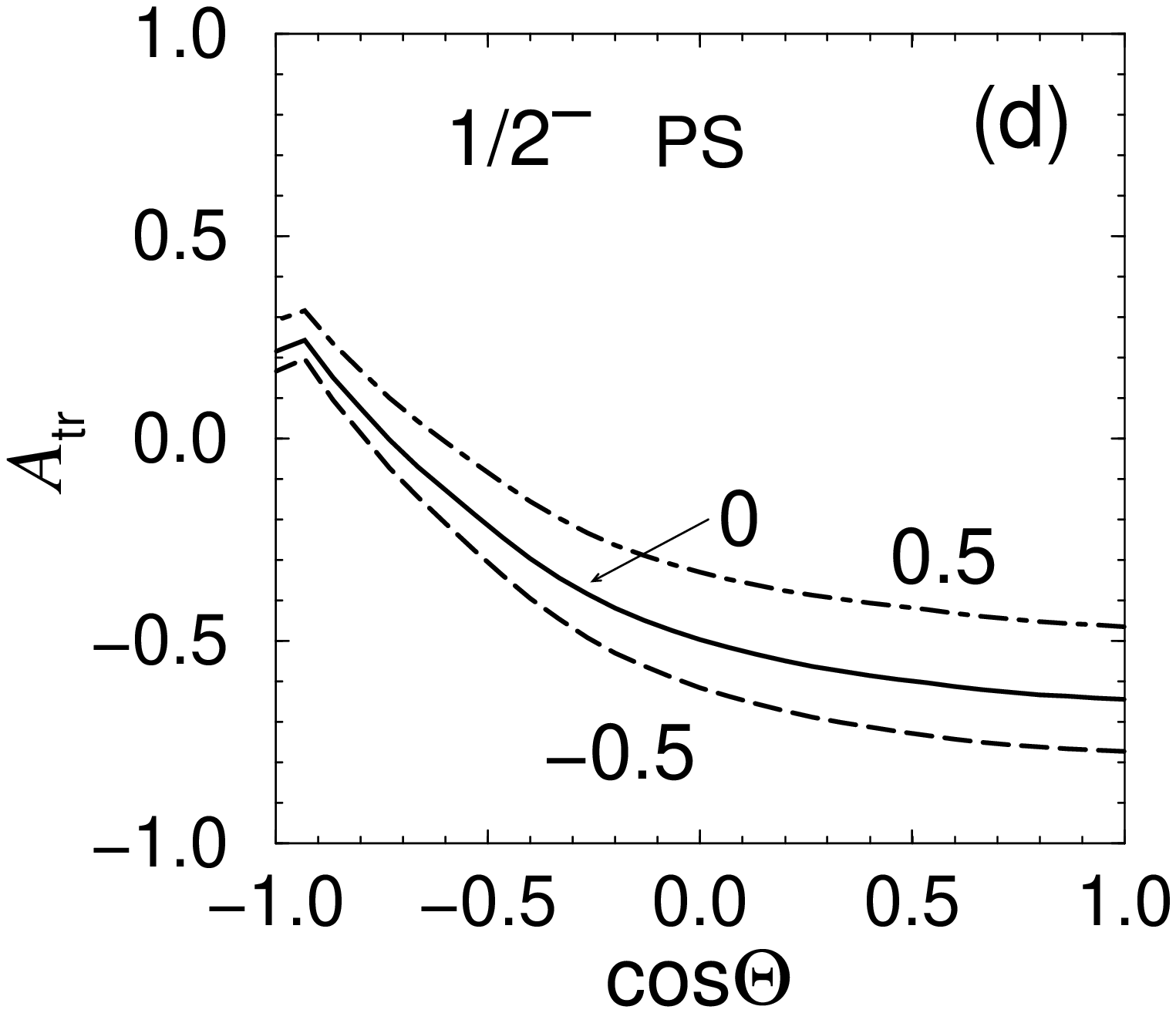}
 \caption{\label{fig:17}\tcaps%
 The double target-recoil spin asymmetry  $ {\cal A}_{tr}$ in
 $\gamma p\to pK^0\bar{K}^0$ as a function of the $K$ decay angle
 for different values of $\kappa^*$.
  Notation is the same as in Fig.~\protect\ref{fig:13}.}
\end{figure}


\begin{figure}[t]\centering
 \includegraphics[width=.23\textwidth]{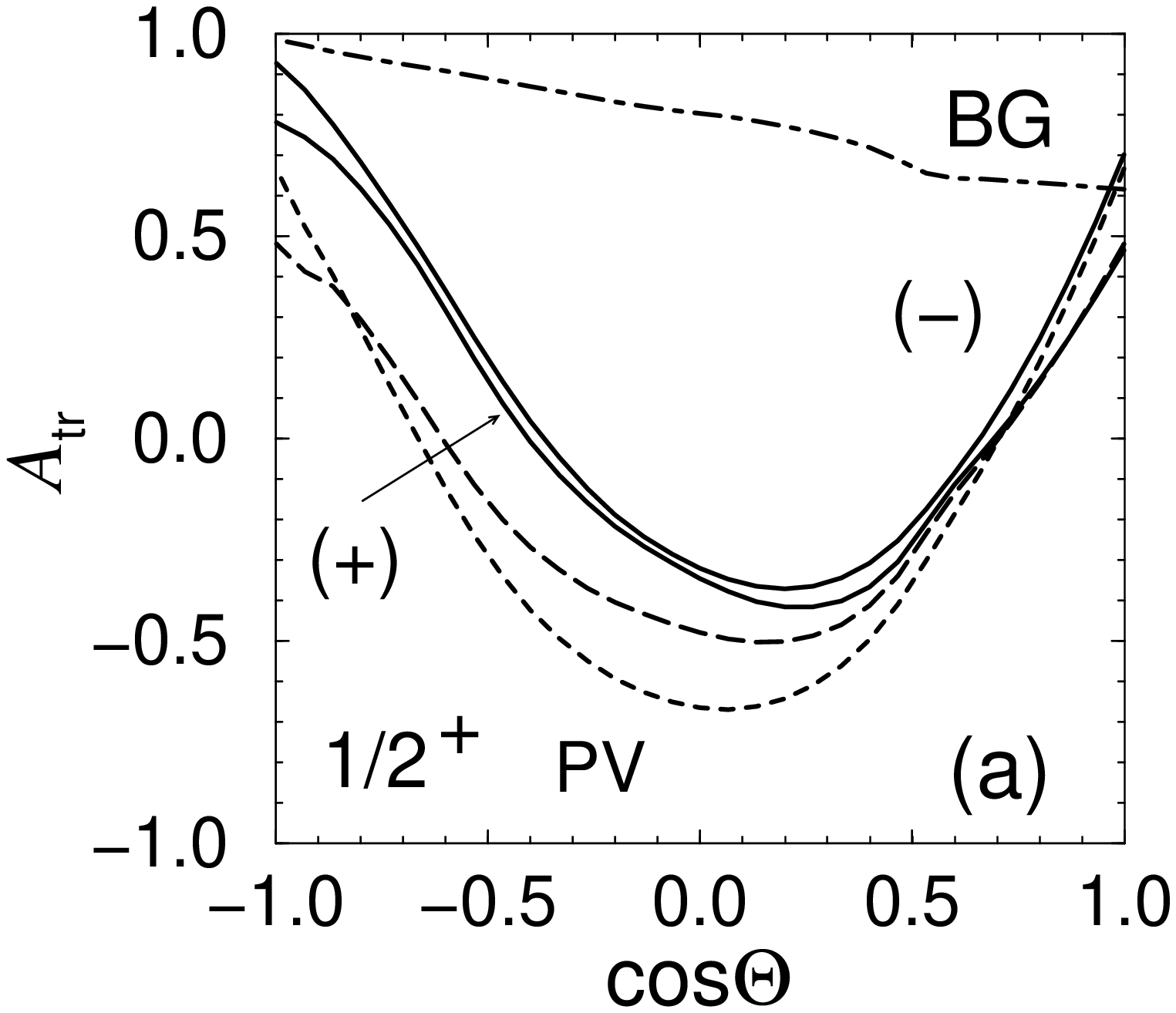}\hfill
 \includegraphics[width=.23\textwidth]{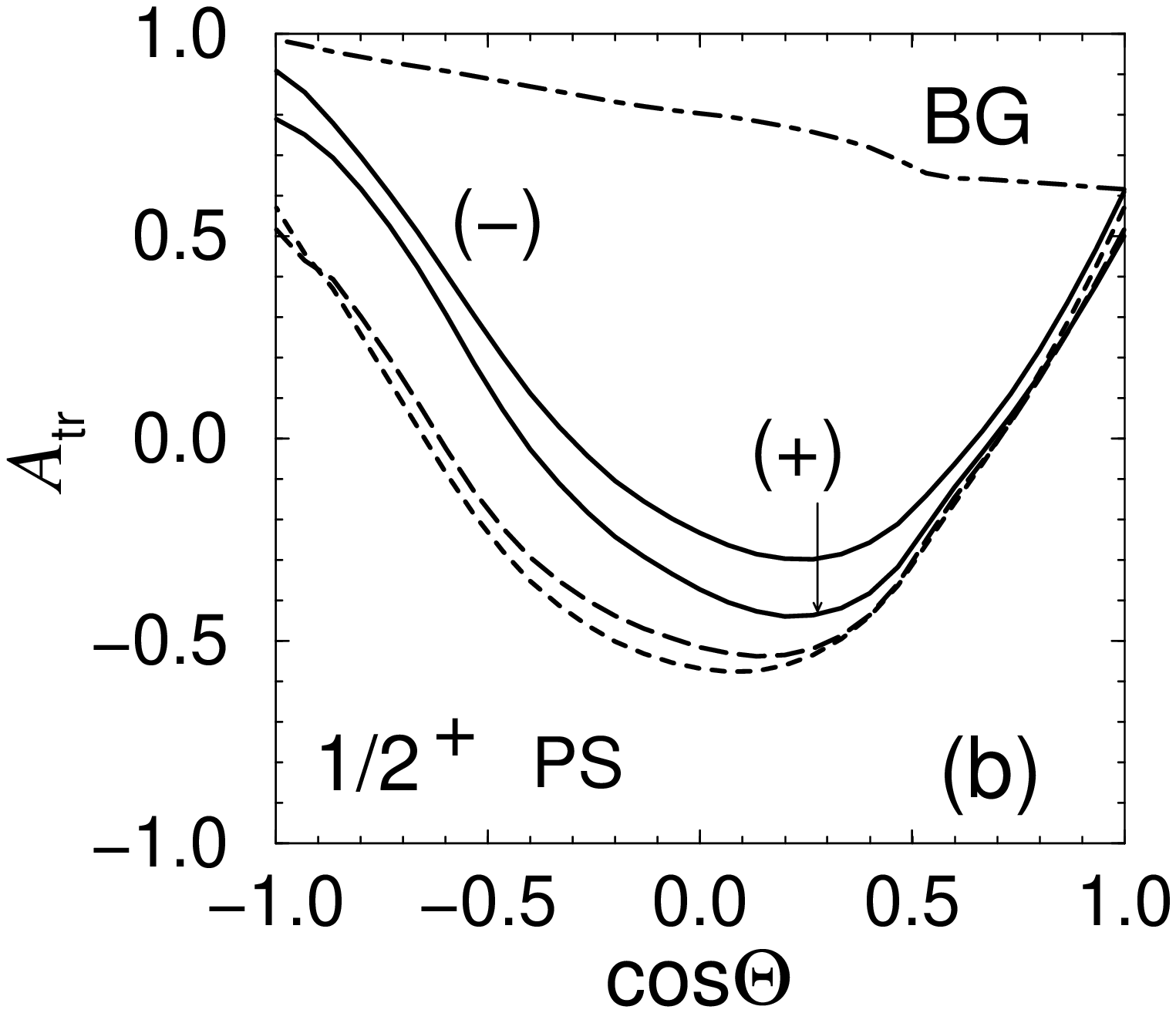}\hfill
 \includegraphics[width=.23\textwidth]{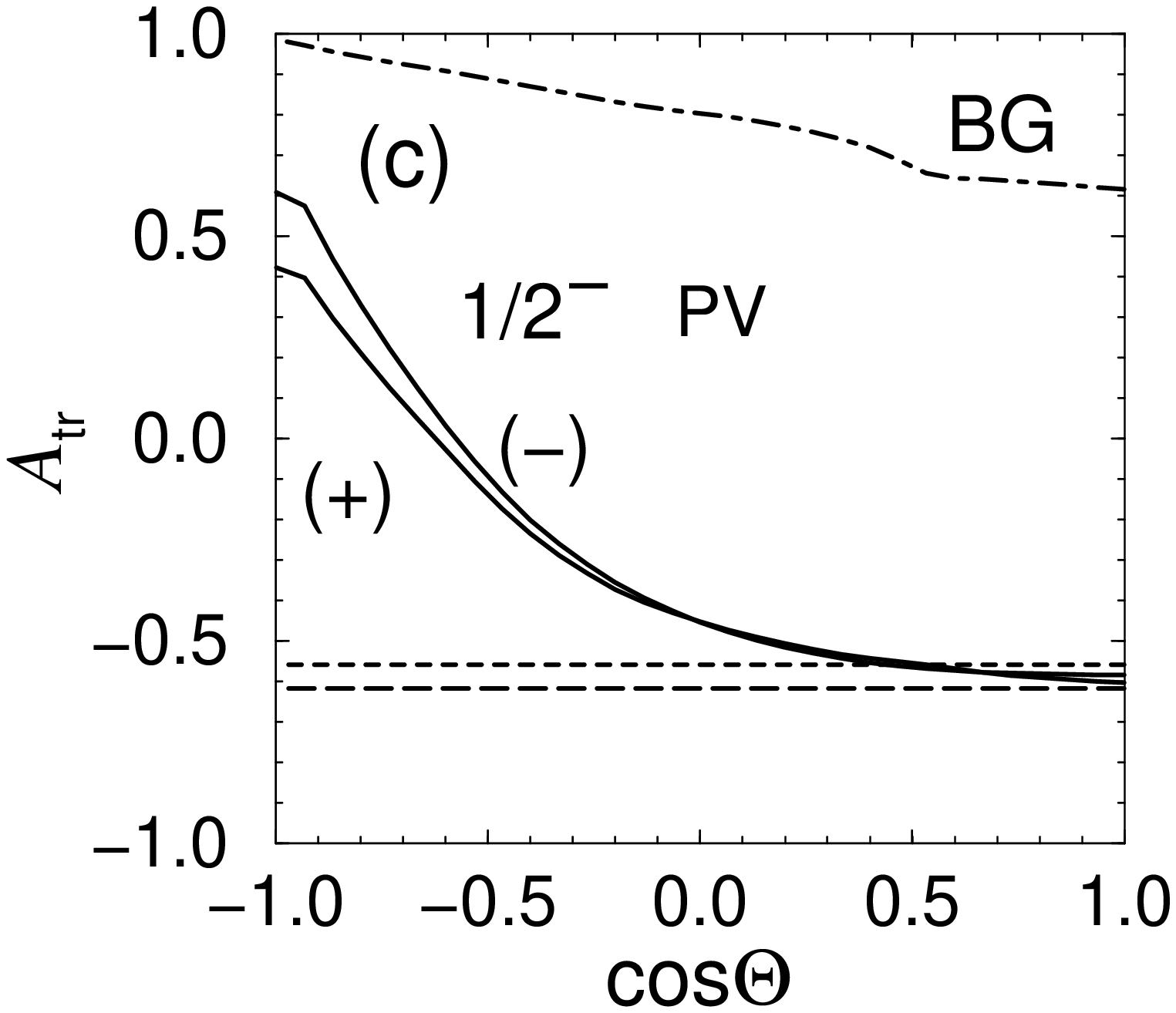}\hfill
 \includegraphics[width=.23\textwidth]{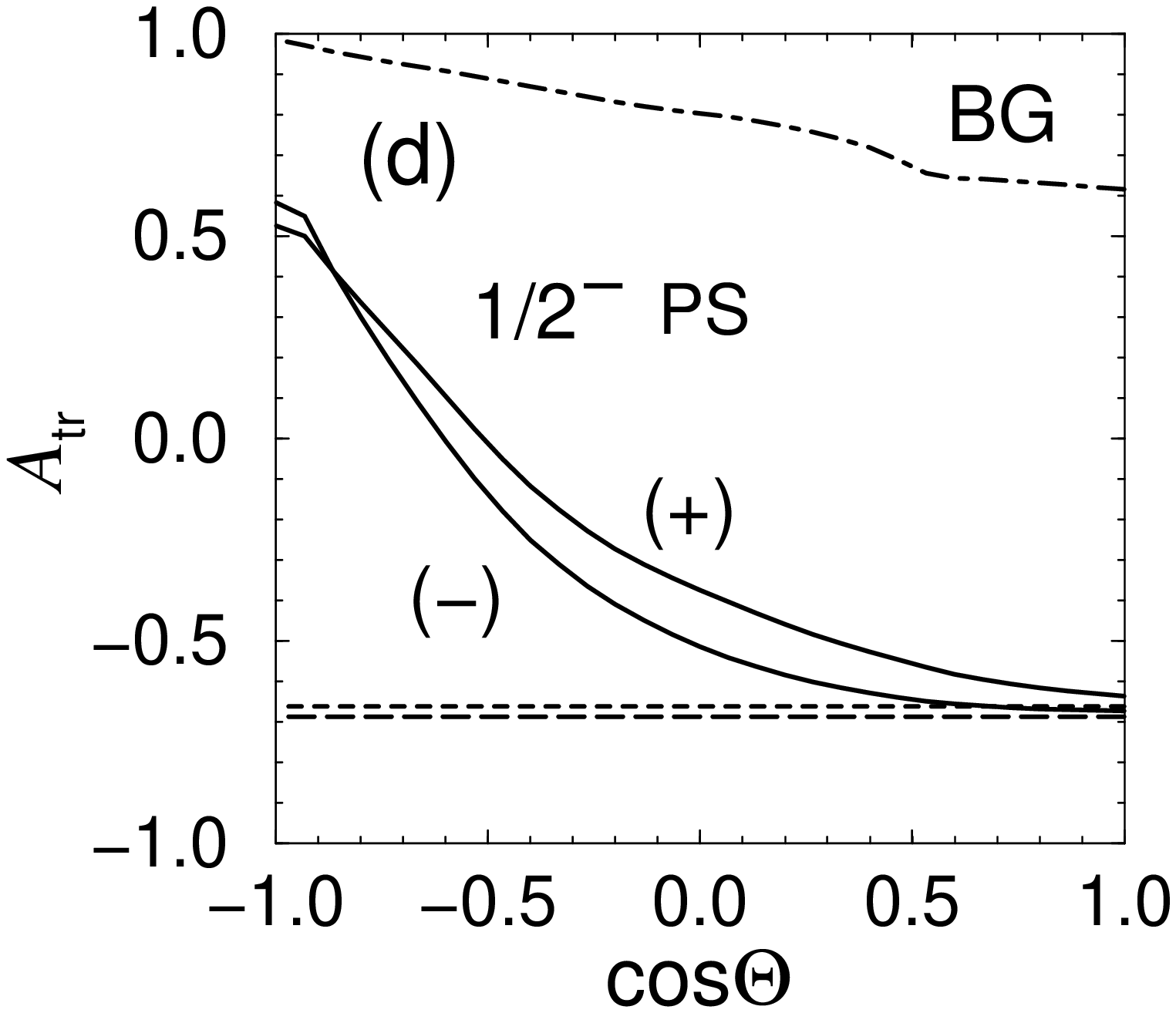}
 \caption{\label{fig:18}\tcaps%
 Same as in Fig.~\protect\ref{fig:16}, for
 $\gamma n\to nK^+ K^-$.}

\mbox{}\\

 \includegraphics[width=.23\textwidth]{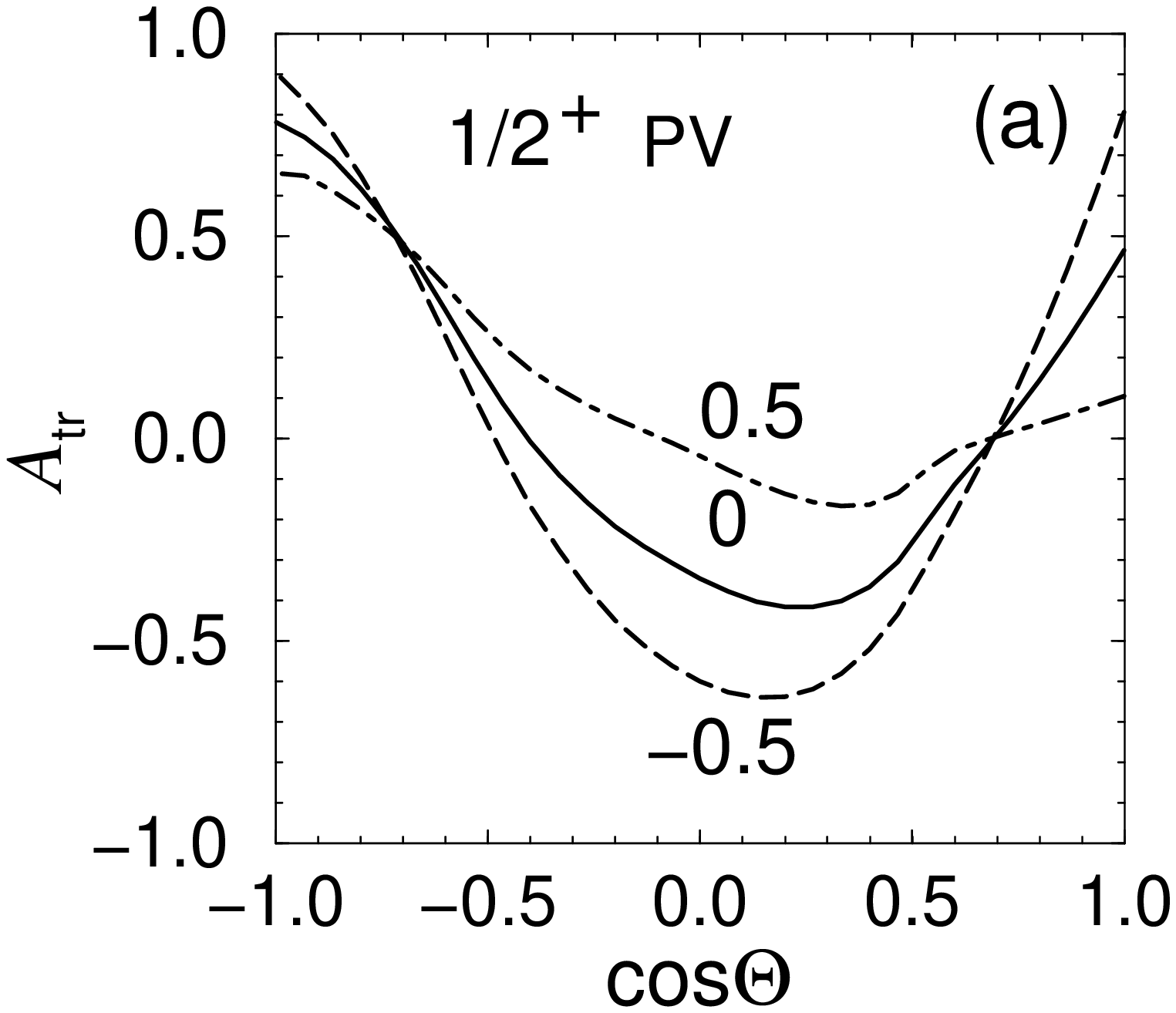}\hfill
 \includegraphics[width=.23\textwidth]{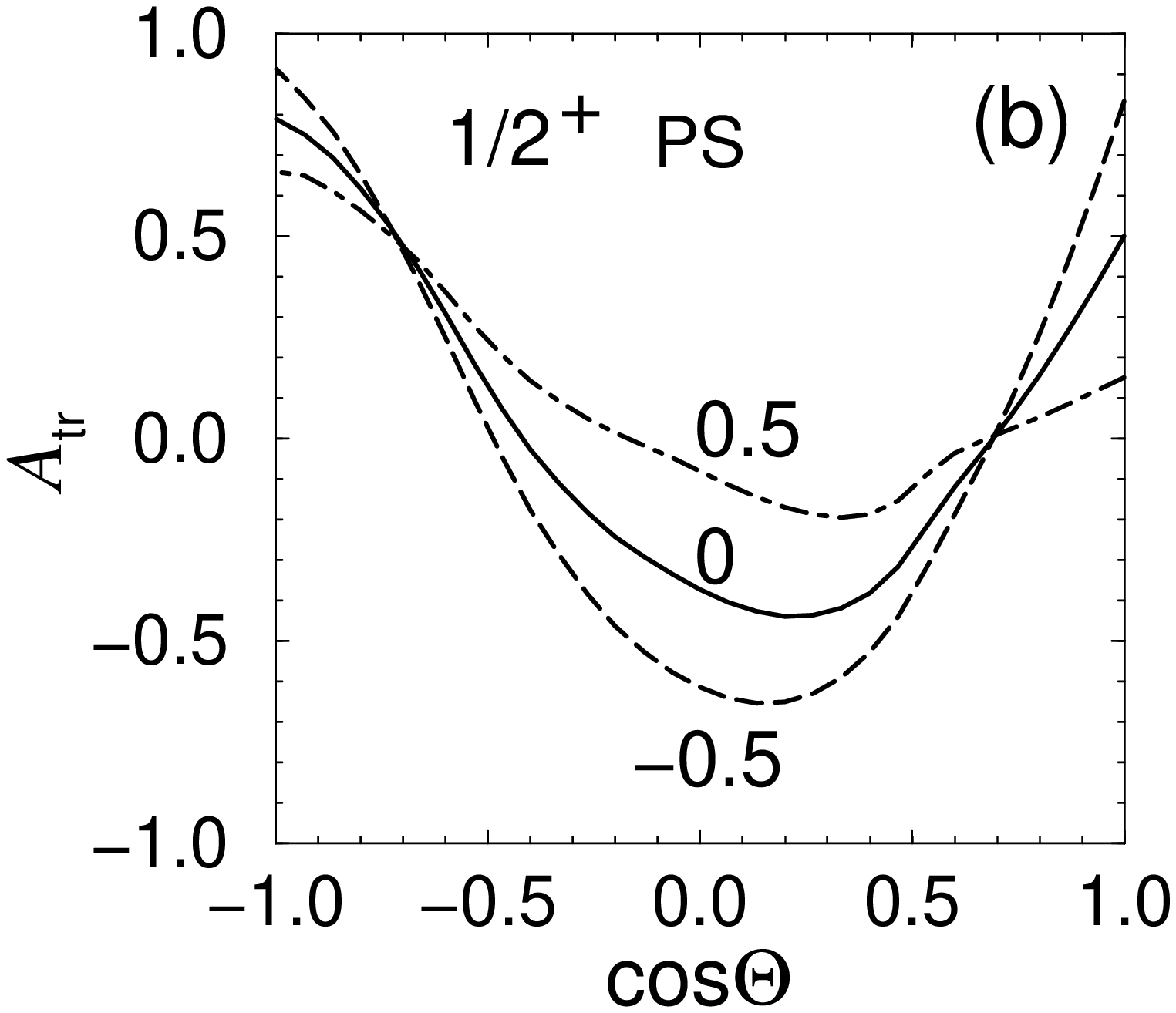}\hfill
 \includegraphics[width=.23\textwidth]{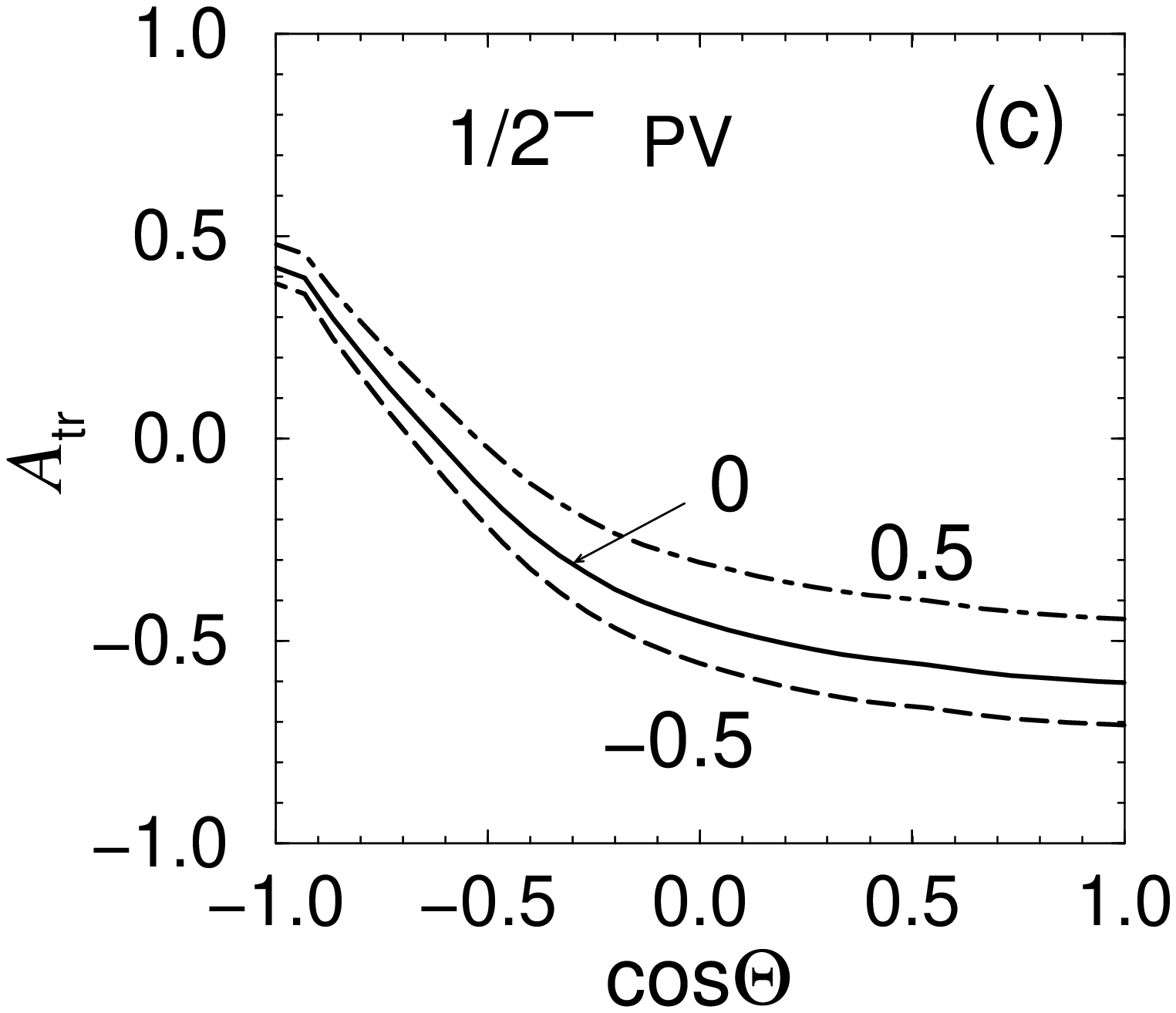}\hfill
 \includegraphics[width=.23\textwidth]{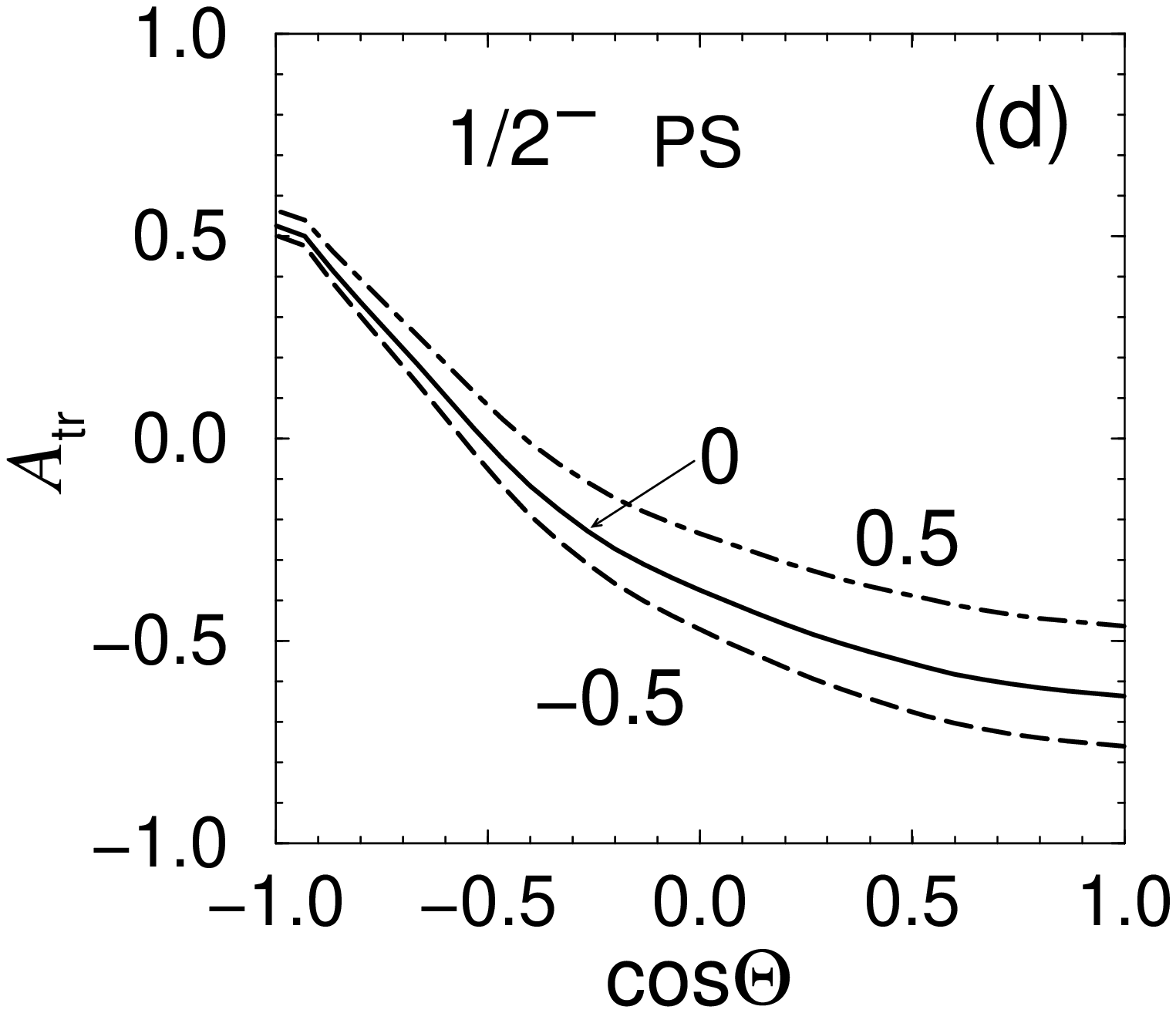}
 \caption{\label{fig:19}\tcaps%
 Same as Fig.~\protect\ref{fig:17}, for $\gamma n\to nK^+ K^-$.}
\end{figure}

Our results for double target-recoil asymmetry are presented in Figs.~\ref{fig:16} and
\ref{fig:17} for the $\gamma p\to p K^0\bar{K}^0$ reaction and in Figs.~\ref{fig:18} and
\ref{fig:19} for the $\gamma n\to n K^+ K^-$ reaction. All the notations are the same as
in Figs.~\ref{fig:12}-\ref{fig:15} for the beam asymmetry. For the pure resonance channel
contribution one can see the $\cos2\Theta$ dependence of ${\cal A}_{tr}$ for the positive
$\pi_\Theta$ and a constant for the negative $\pi_\Theta$. The  ``modulation" ${\cal
A}^+_{0}$ for positive $\pi_\Theta$ depends on the tensor coupling $\kappa^*$ and
decreases when $\kappa^*$ increases. The background contribution modifies the
asymmetries. In the case of a negative $\pi_\Theta$, the asymmetries increases when
$\Theta\to \pi$. In the case of a positive $\pi_\Theta$, one can see a strong
modification of the $\cos2\Theta$ dependence. This modification is especially strong for
larger values of $\kappa^*$. In this case we see almost a monotonic decrease of ${\cal
A}^+_{tr}$ as $\Theta$ varies from $\pi$ to 0 (see Figs.~\ref{fig:17}a
and~\ref{fig:19}a), similarly to the case of the negative $\pi_\Theta$ for large and
negative $\kappa^*$ (see Figs.~\ref{fig:17}d and~\ref{fig:19}d ). Therefore we can
conclude that the  background contribution hampers the use of the double spin observables
for the determination of $\pi_\Theta$ because of a strong interplay of the production
mechanism (in our example it is $\kappa^*$) and effects of the $\Theta^+$ parity in the
transition amplitudes.

\subsubsection{Triple spin observables}

Let us consider the beam asymmetry for the linearly polarized photon beam at a fixed
polarization of the target and the recoil nucleons. The nucleon polarizations are chosen
along the normal to the production plane~\cite{Ejiri,NL04},
 \begin{eqnarray}
  \Sigma_{yy}(\uparrow\uparrow)=
 \frac{\sigma^{\perp}(\uparrow\uparrow) -
 \sigma^{\parallel}(\uparrow\uparrow)}
 {\sigma^{\perp}(\uparrow\uparrow) +
 \sigma^{\parallel}(\uparrow\uparrow)}~,
 \qquad
  \Sigma_{yy}(\uparrow\downarrow)=
 \frac{\sigma^{\perp}(\uparrow\downarrow) -
 \sigma^{\parallel}(\uparrow\downarrow)}
 {\sigma^{\perp}(\uparrow\downarrow) +
 \sigma^{\parallel}(\uparrow\downarrow)}~,
 \label{Byy}
\end{eqnarray}
where  $\sigma(\uparrow\uparrow)$ and $\sigma(\uparrow\downarrow)$
correspond to the spin-conserving and spin-flip transitions
between the initial and the final nucleons, respectively. We
choose these asymmetries because for the $2\to 2$
 ($\gamma N\to \Theta^+
\bar{K}$) reaction,  Bohr's theorem~\cite{Bohr} based on reflection symmetry in the
scattering plane results in
 \begin{eqnarray}
 \Sigma^{\gamma N\to \Theta^+\bar{K}}_{yy}(\uparrow\uparrow)
 =+\pi_{\Theta}~,\qquad
\Sigma^{\gamma N\to \Theta^+\bar{K}}_{yy}(\uparrow\downarrow)
 =- \pi_{\Theta}~.
 \label{Byy22}
\end{eqnarray}
 This prediction is very strict, it does not depend on the production mechanism
(in our case PV or PS coupling schemes, $\alpha$, $\kappa^*$ etc.) and therefore it is
extremely attractive. But unfortunately, the realistic case is more complicated. As we
discussed above, the realistic process is the $2\to 3$ reaction ($\gamma N\to N K\bar{K}$)
and we have to take into account the three-body aspects of the final state. Let us
consider the  coplanar reaction when all three outgoing particles are in the production
plane perpendicular to the nucleon polarization. In this case, Bohr's theorem
predicts
\begin{eqnarray}
\Sigma_{yy}(\uparrow\uparrow)
=\pi_{K}=-1~,\qquad \Sigma_{yy}(\uparrow\downarrow) =-\pi_{K}=+1~,
\label{Byy23}
\end{eqnarray}
independently of the intermediate $\Theta^+$ parity. It is conceivable that  other
``model-independent" predictions made for the $2\to 2$ reaction may suffer from a similar
problem. The only way to use $\Sigma_{yy}$ as a tool to determine the parity of the
$\Theta^+$ pentaquark is to find a kinematical region where this asymmetry is sensitive
to $\pi_{\Theta}$ and insensitive to the production mechanism.

\begin{figure}[b]\centering
 \includegraphics[width=.23\textwidth]{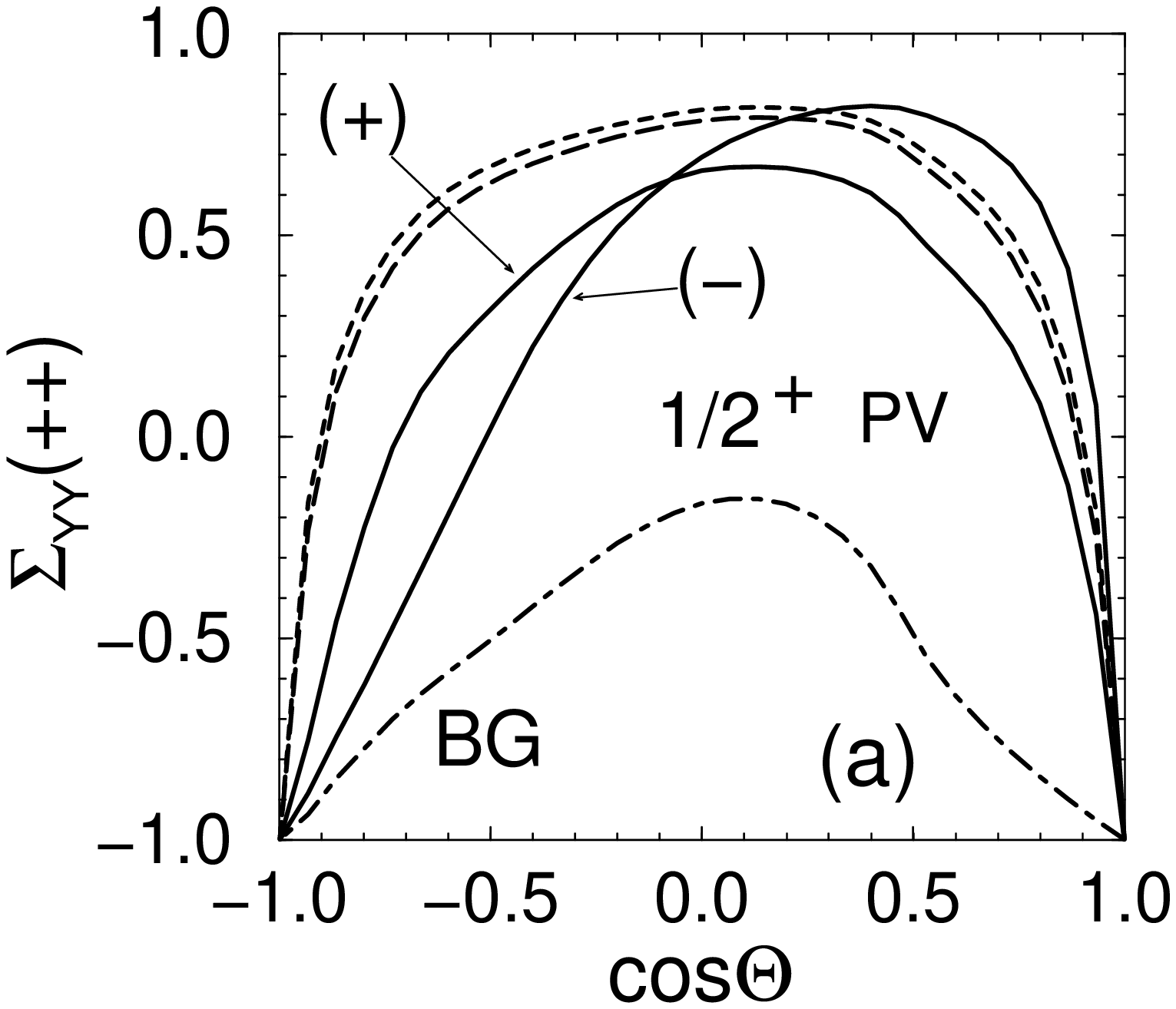}\hfill
 \includegraphics[width=.23\textwidth]{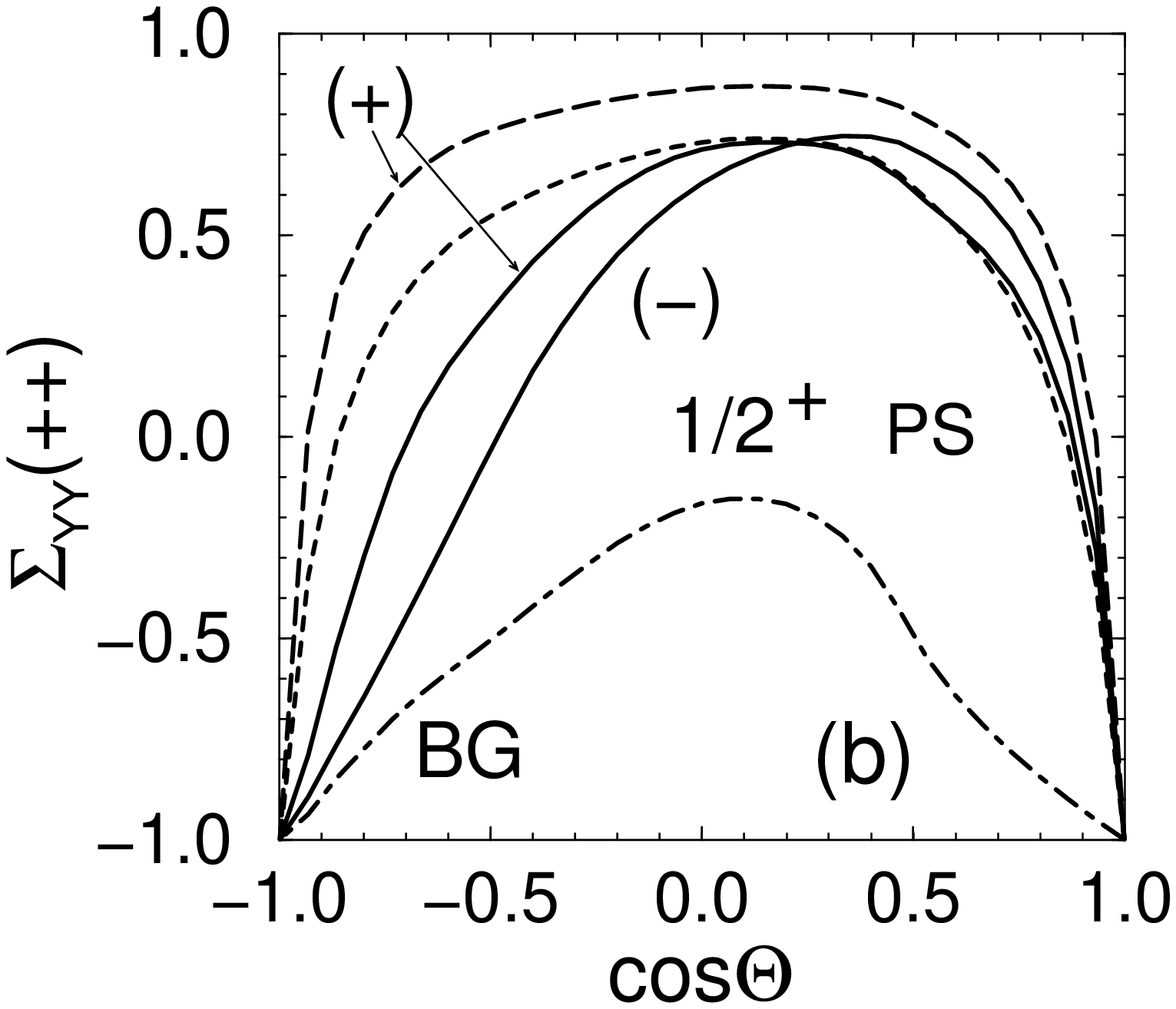}\hfill
 \includegraphics[width=.23\textwidth]{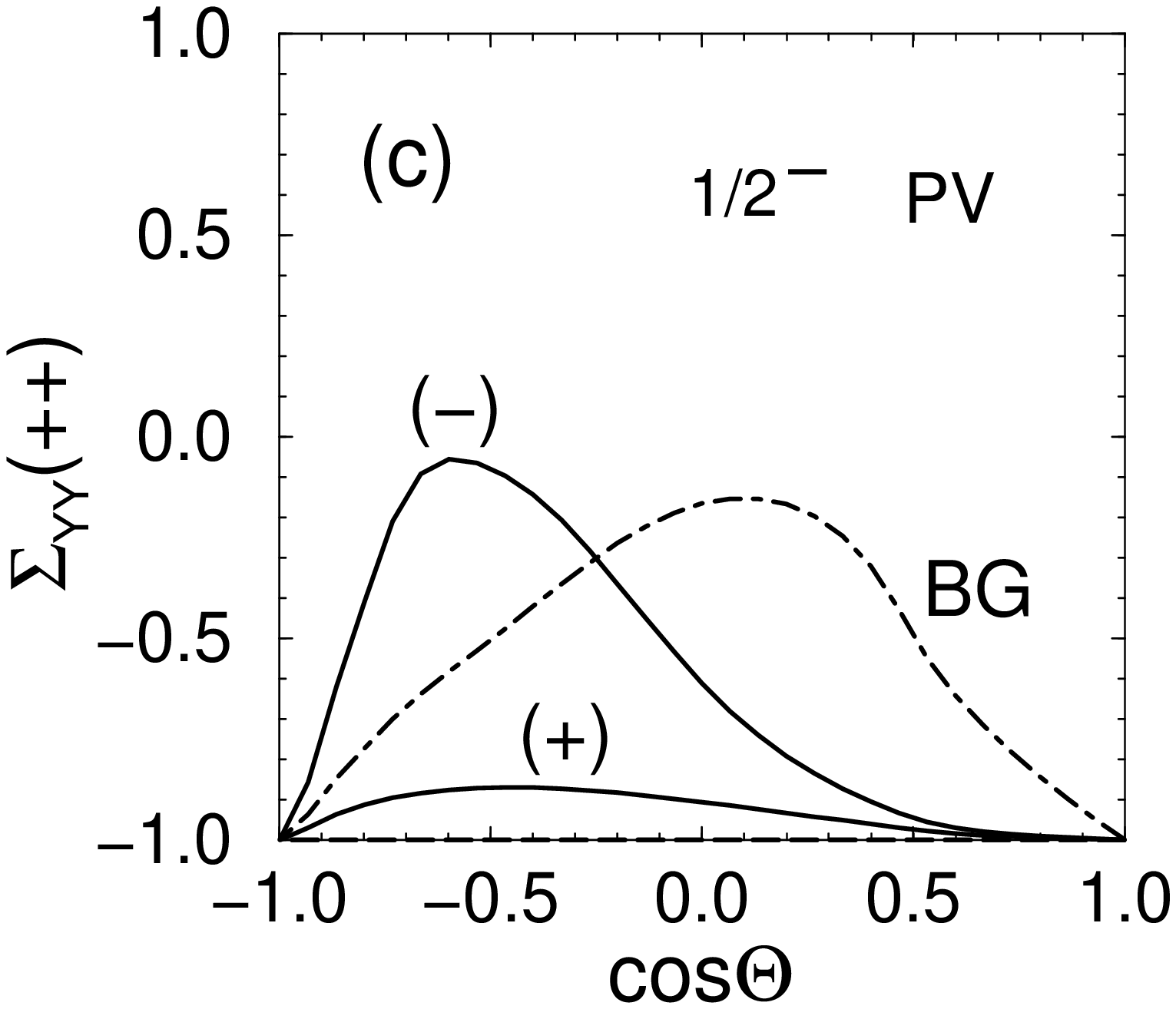}\hfill
 \includegraphics[width=.23\textwidth]{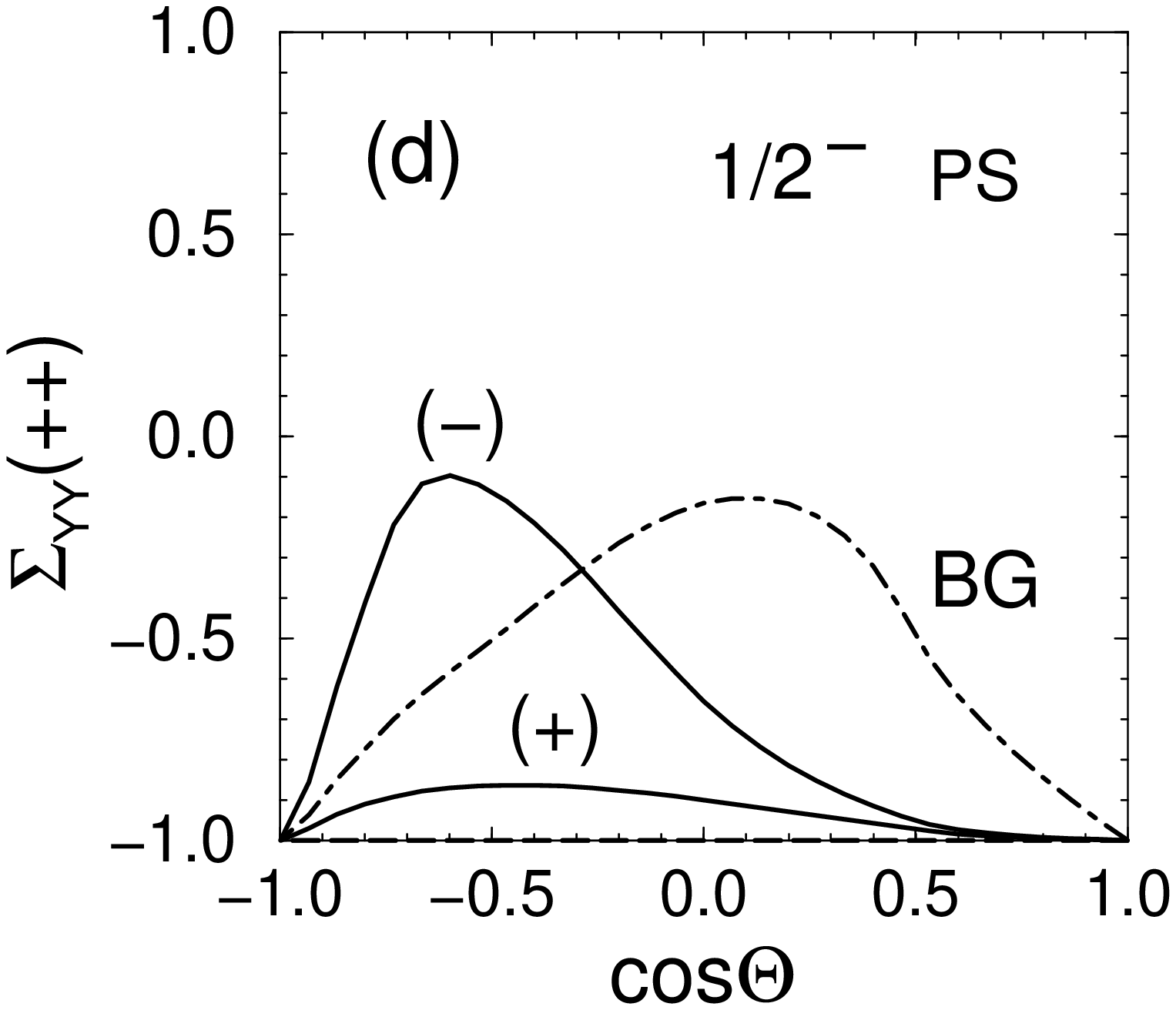}
 \caption{\label{fig:20}\tcaps%
 The triple spin asymmetry $\Sigma_{yy}(\uparrow\uparrow)$
 in $\gamma p\to pK^0\bar{K}^0$ as a function of the $K$ decay angle.
  Notation is the same as in Fig.~\protect\ref{fig:12}.}

\mbox{}\\

\mbox{}\hspace{.23\textwidth}\mbox{}\hfill
 \includegraphics[width=.23\textwidth]{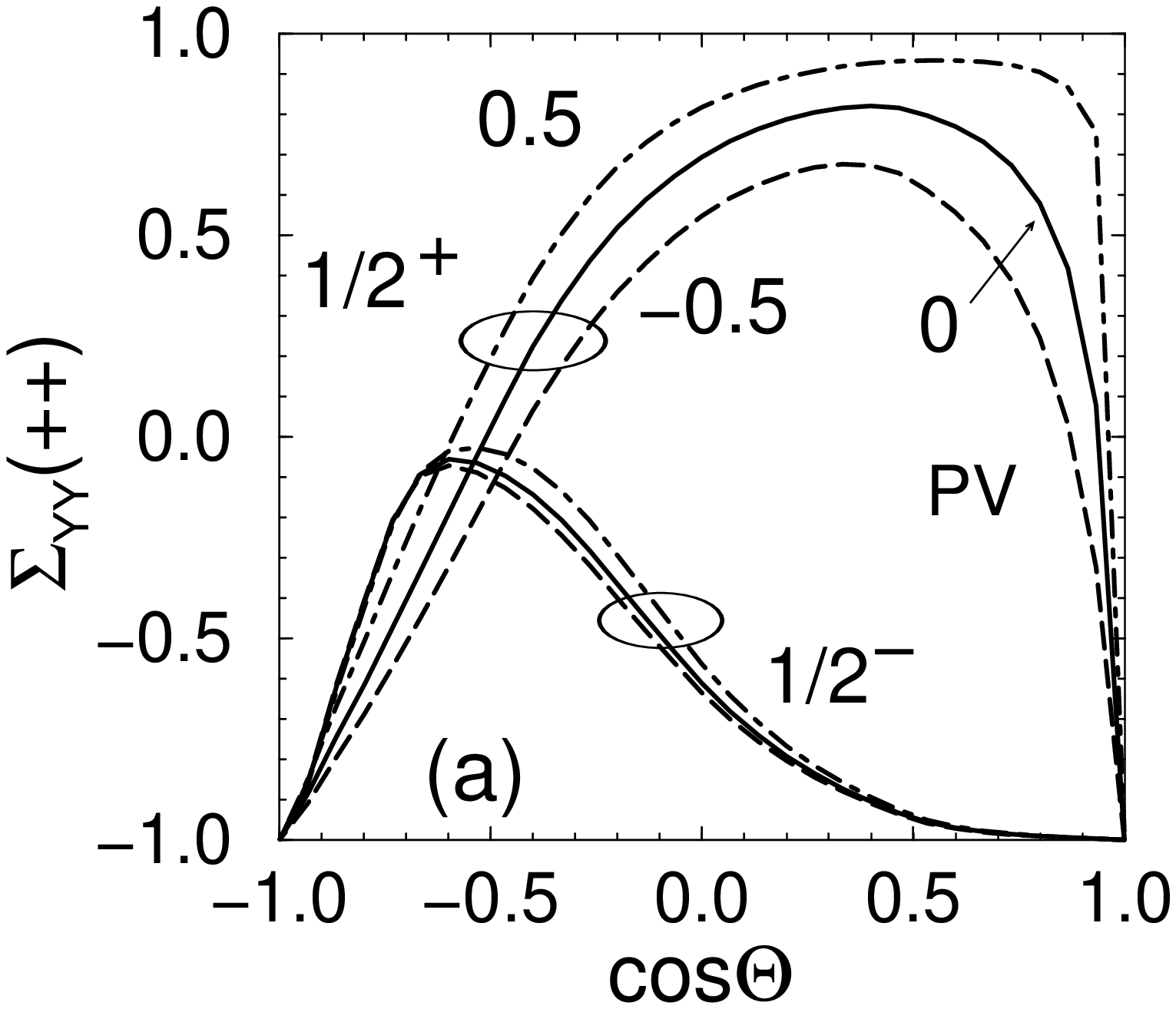}\hfill
 \includegraphics[width=.23\textwidth]{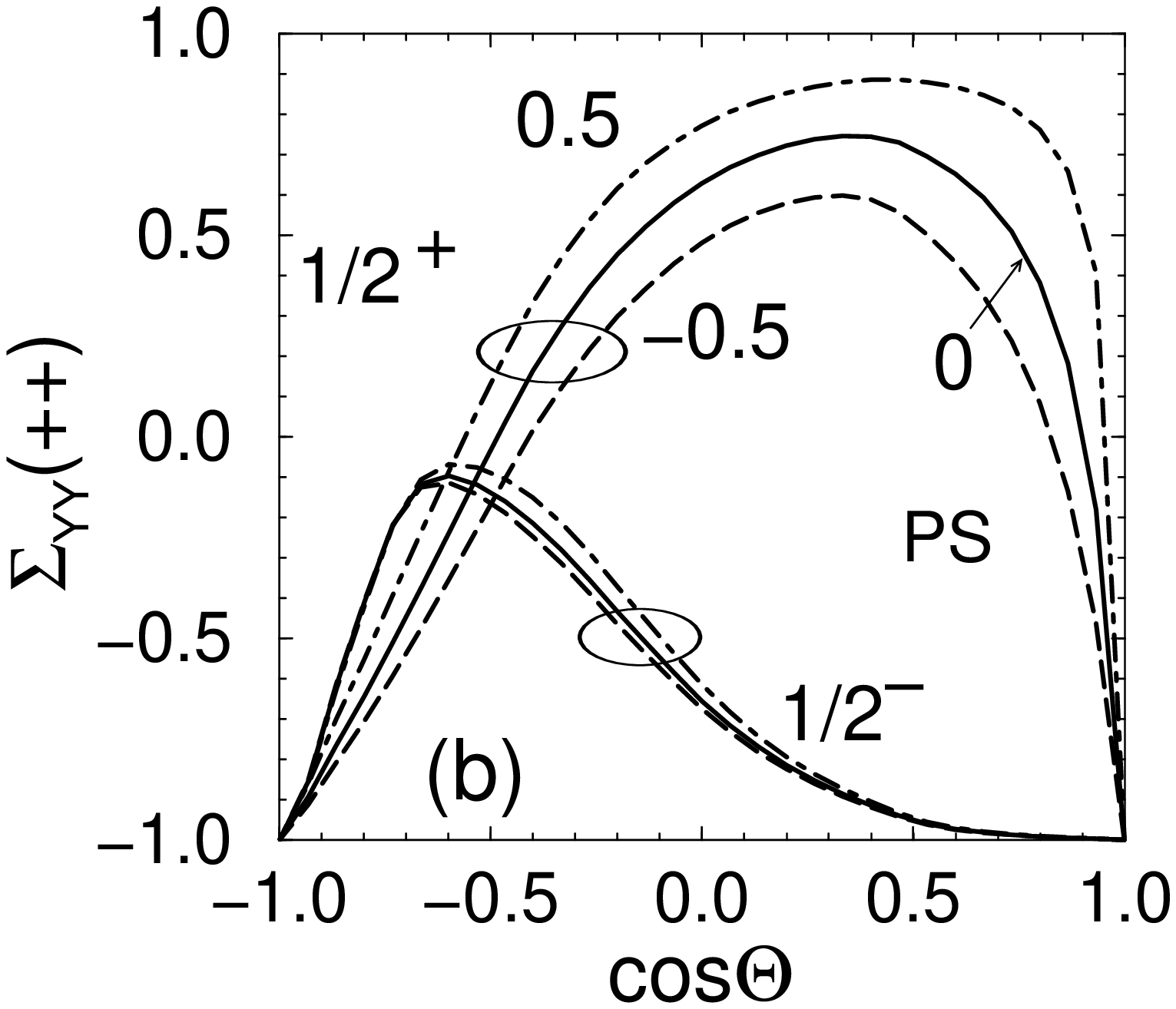}\hfill
 \mbox{}\hspace{.23\textwidth}\mbox{}
 \caption{\label{fig:21}\tcaps%
 The triple spin asymmetry $\Sigma^{}_{yy}(\uparrow\uparrow)$ in
 $\gamma p\to pK^0\bar{K}^0$ as a function of the $K$ decay angle
 for different values of $\kappa^*$ and negative $\alpha$.
 Notation is the same as in Fig.~\protect\ref{fig:13}.}
\end{figure}
\begin{figure}[t]\centering
 \includegraphics[width=.23\textwidth]{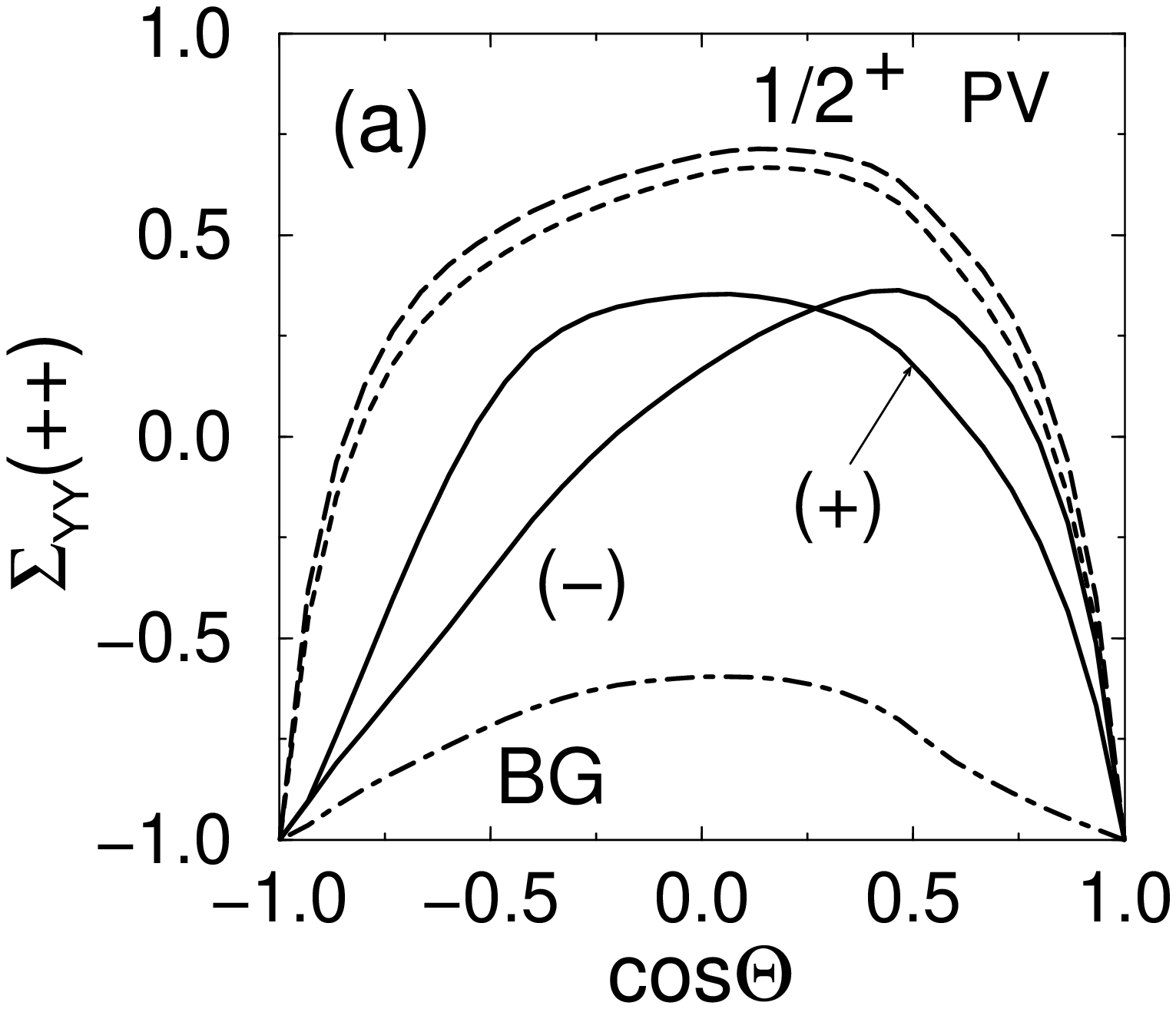}\hfill
 \includegraphics[width=.23\textwidth]{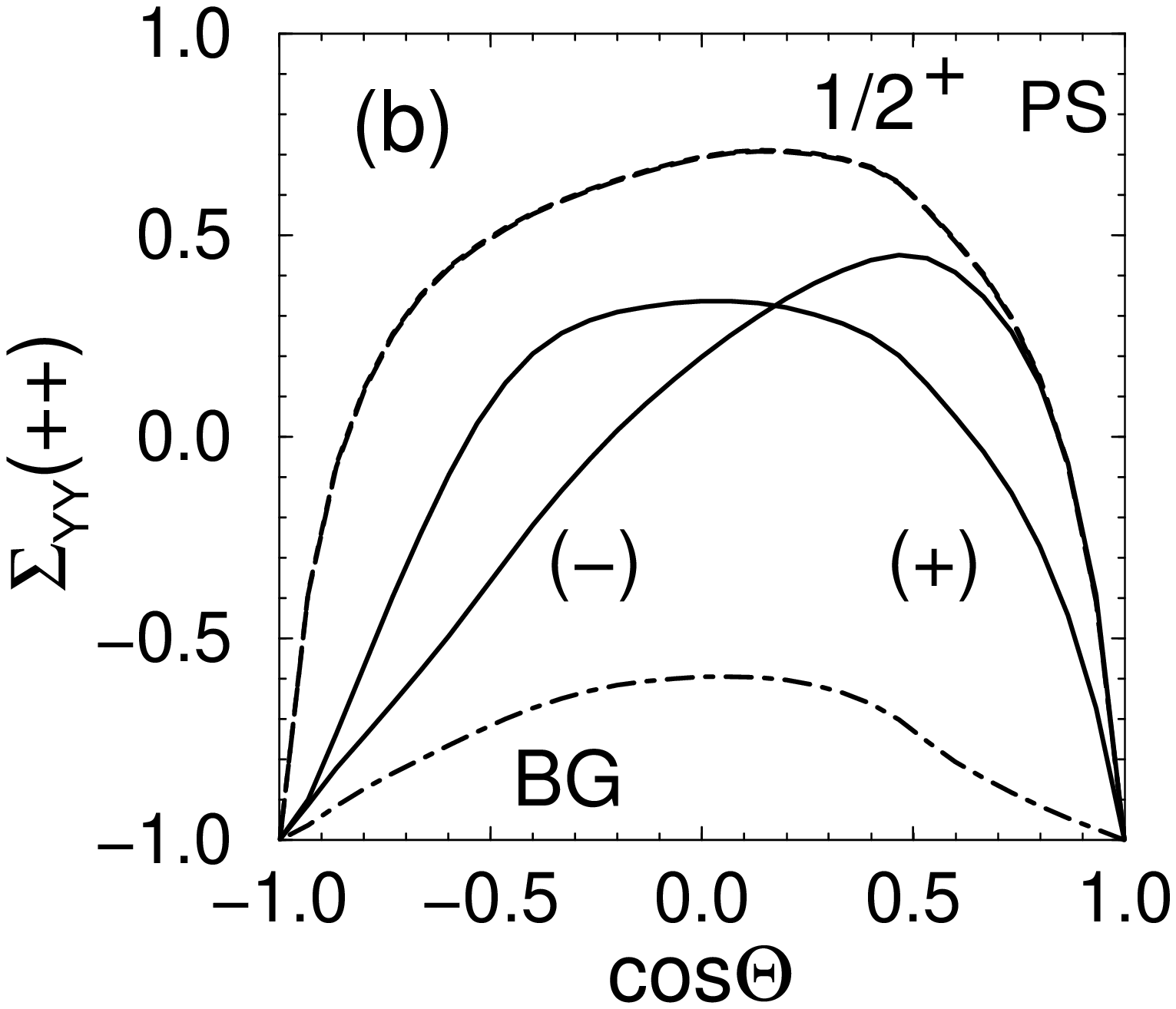}\hfill
 \includegraphics[width=.23\textwidth]{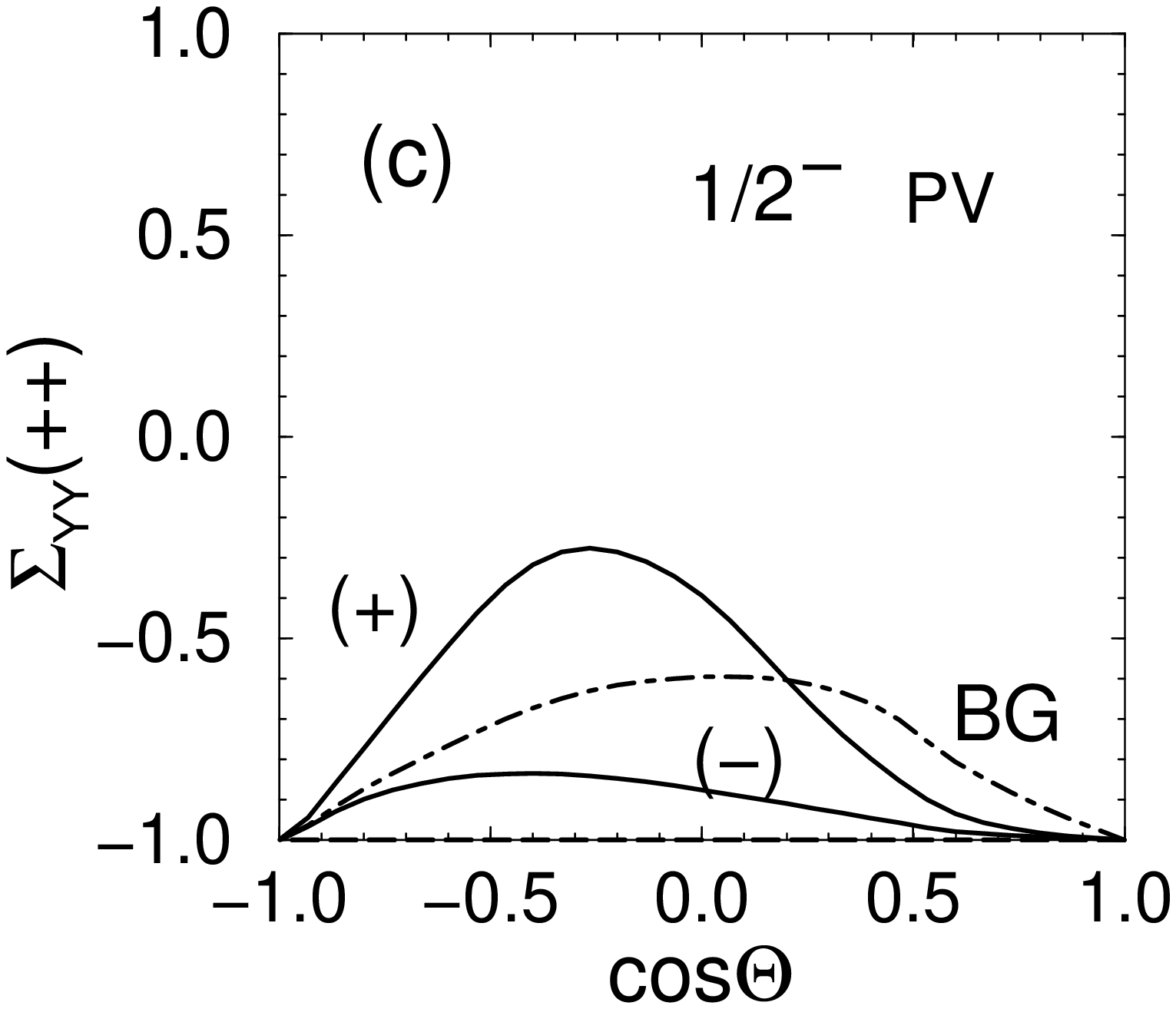}\hfill
 \includegraphics[width=.23\textwidth]{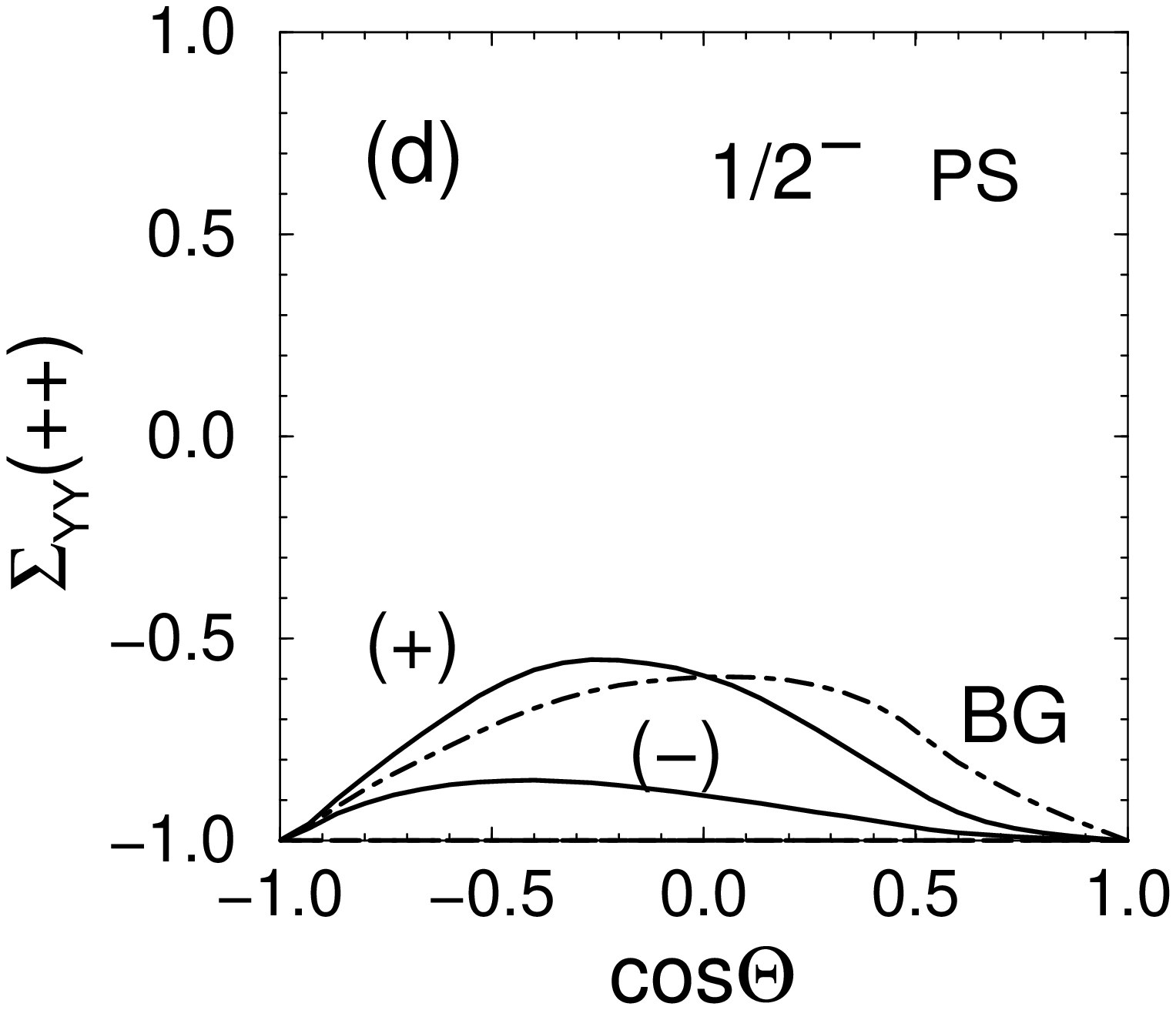}
 \caption{\label{fig:22}\tcaps%
 Same as Fig.~\protect\ref{fig:20}, for $\gamma n\to nK^+ K^-$.}

\mbox{}\\

\mbox{}\hspace{.23\textwidth}\mbox{}\hfill
 \includegraphics[width=.23\textwidth]{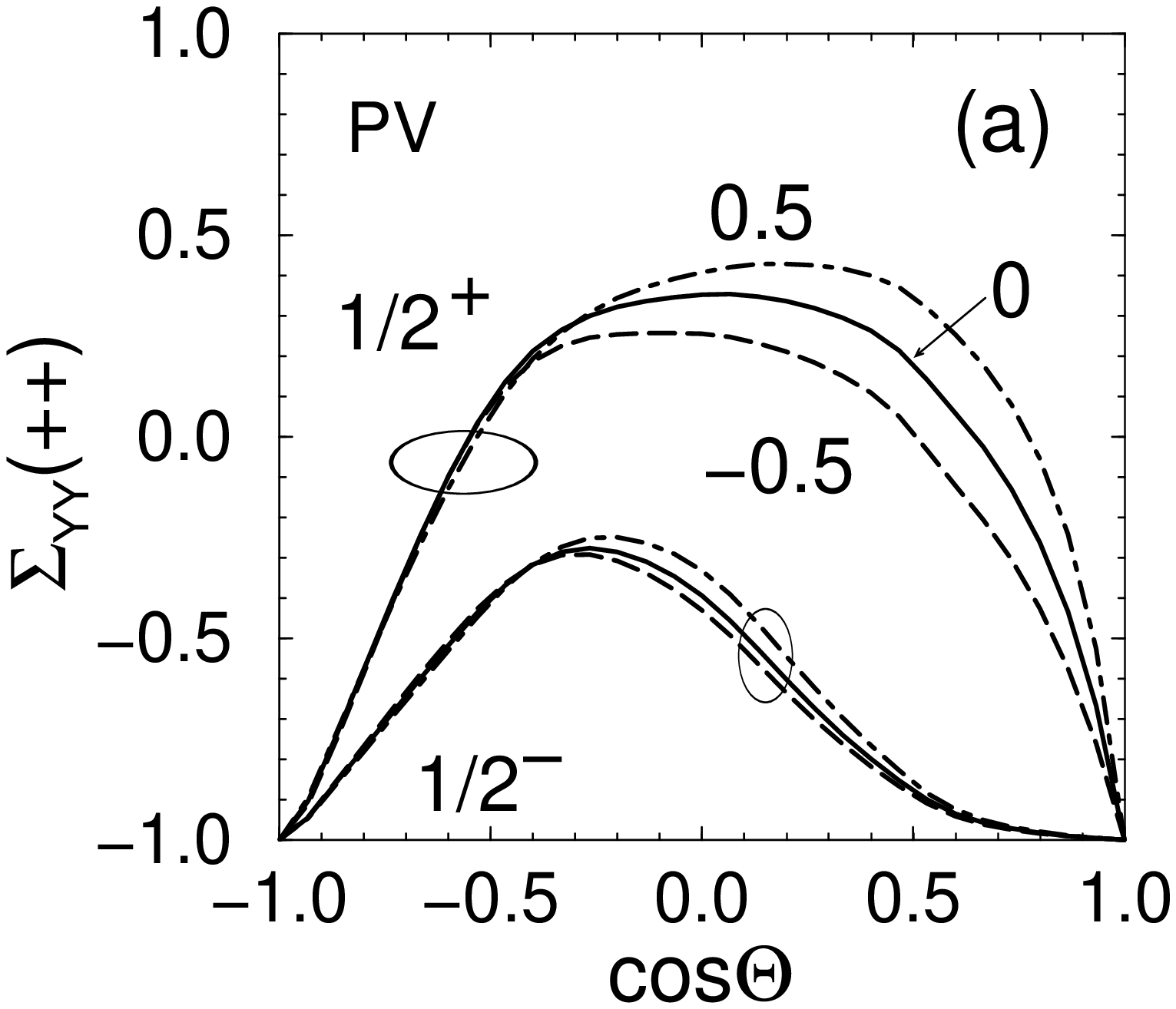}\hfill
 \includegraphics[width=.23\textwidth]{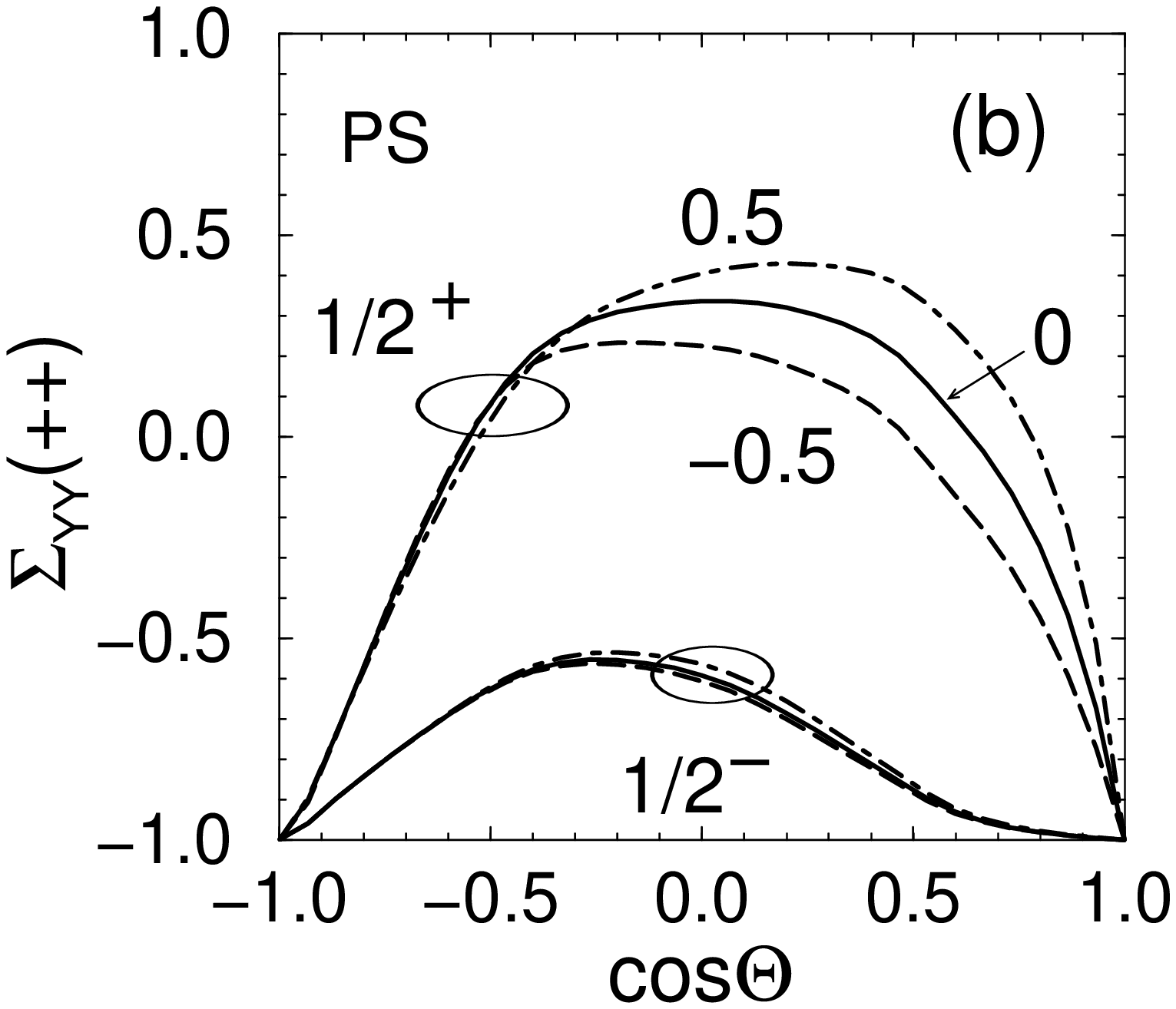}\hfill
 \mbox{}\hspace{.23\textwidth}\mbox{}
 \caption{\label{fig:23}\tcaps%
 Same as Fig.~\protect\ref{fig:21}, for  $\gamma n\to nK^+ K^-$.}
\end{figure}

Consider first the spin-conserving transitions. Figures~\ref{fig:20} and \ref{fig:21}
and Figs.~\ref{fig:22} and \ref{fig:23} show results of our calculation of the triple
spin asymmetry $ \Sigma_{yy}(++)\equiv\Sigma_{yy}(\uparrow\uparrow)$ for $\gamma p\to p
K^0\bar{K}^0$ and $\gamma n\to n K^+ K^-$, respectively, for different $\pi_\Theta$,
different coupling schemes, different signs of $\alpha$, and different $\kappa^*$. The
results for the positive and negative $\pi_\Theta$ are shown in Figs.~\ref{fig:20} and
\ref{fig:22} (ab) and (cd), respectively. In Figs.~\ref{fig:21} and \ref{fig:23} the
results for both parities are displayed simultaneously. The results for pseudovector (PV)
coupling are shown in panels (a) and (c) of Figs.~\ref{fig:20} and \ref{fig:22}, and
panels (a) of Figs.~\ref{fig:21} and \ref{fig:23}; those for pseudoscalar (PS) coupling
are given in panels (b) and (d) of Figs.~\ref{fig:20} and \ref{fig:22}, and panels (b) of
Figs.~\ref{fig:21} and \ref{fig:23}. The asymmetries shown in Figs.~\ref{fig:20} and
\ref{fig:22} are calculated with $\kappa^*=0$; the dependence on $\kappa^*$ is shown in
Figs.~\ref{fig:21} and \ref{fig:23}. In Figs.~\ref{fig:20} and \ref{fig:22} the
asymmetries due to the resonant channel are shown by the long curves ($\alpha>0$) and
dashed ($\alpha<0$) curves. The asymmetry due to the background is shown by the
dot-dashed curves.

The case of $\Theta=0,\pi$ corresponds to the coplanar geometry [Eq.~(\ref{Byy23})],
where $\Sigma_{yy}(\uparrow\uparrow)=-1$ independently of $\pi_\Theta$, reaction
mechanism and input parameters. For the negative $\Theta^+$ parity the asymmetry due to
the resonant channel remains to be $-1$ at all $\cos\Theta$ because of the s-wave
$\Theta^+$ decay, and therefore predictions for $2\to 2$ and $2\to 3$ for this case are
the same.

For positive parity, we have a p-wave decay which leads to a fast increasing asymmetry
from $-1$  up to  positive and large values and results in a specific bell-shape behavior
of $\Sigma_{yy}$. In principle, the shapes of $\Sigma_{yy}$ for different $\pi_\Theta$
are quite different from each other in the region of $-0.8 \lesssim\cos\Theta\lesssim
0.8$ and practically do not depend on the production mechanism. Therefore one could
consider to use this asymmetry for determining $\pi_\Theta$.  Unfortunately, as in the
case of the double target recoil asymmetry, this picture is modified by the interference
between the resonance and background channels. For negative $\pi_\Theta$ and  negative
$\alpha$, one can see a sizeable deviation of $\Sigma_{yy}(\uparrow\uparrow)$ from $-1$
at negative $\cos\Theta$. The effect of the resonance-background interference is large in
the $\gamma n\to n K^+ K^-$ reaction, leading to a decreasing
$\Sigma^+_{yy}(\uparrow\uparrow)$ at $\cos\Theta\approx 0$ and an increasing
$\Sigma^-_{yy}(\uparrow\uparrow)$ at $\cos\Theta\approx 0$--$0.2$ as compared to that in
the $\gamma p\to p K^0 \bar{K}^0$ reaction. Nevertheless, from Figs.~\ref{fig:21}
and~\ref{fig:23} one can  see that in the region of $0.5 \lesssim\cos\Theta \lesssim0.8$,
the asymmetries $\Sigma^{\pm}_{yy}(\uparrow\uparrow)$ for different parities are
qualitatively different from each other. This feature suggests to use this observable for
the determination of $\pi_\Theta$.

\begin{figure}[t] \centering
\mbox{}\hspace{.23\textwidth}\mbox{}\hfill
 \includegraphics[width=.23\textwidth]{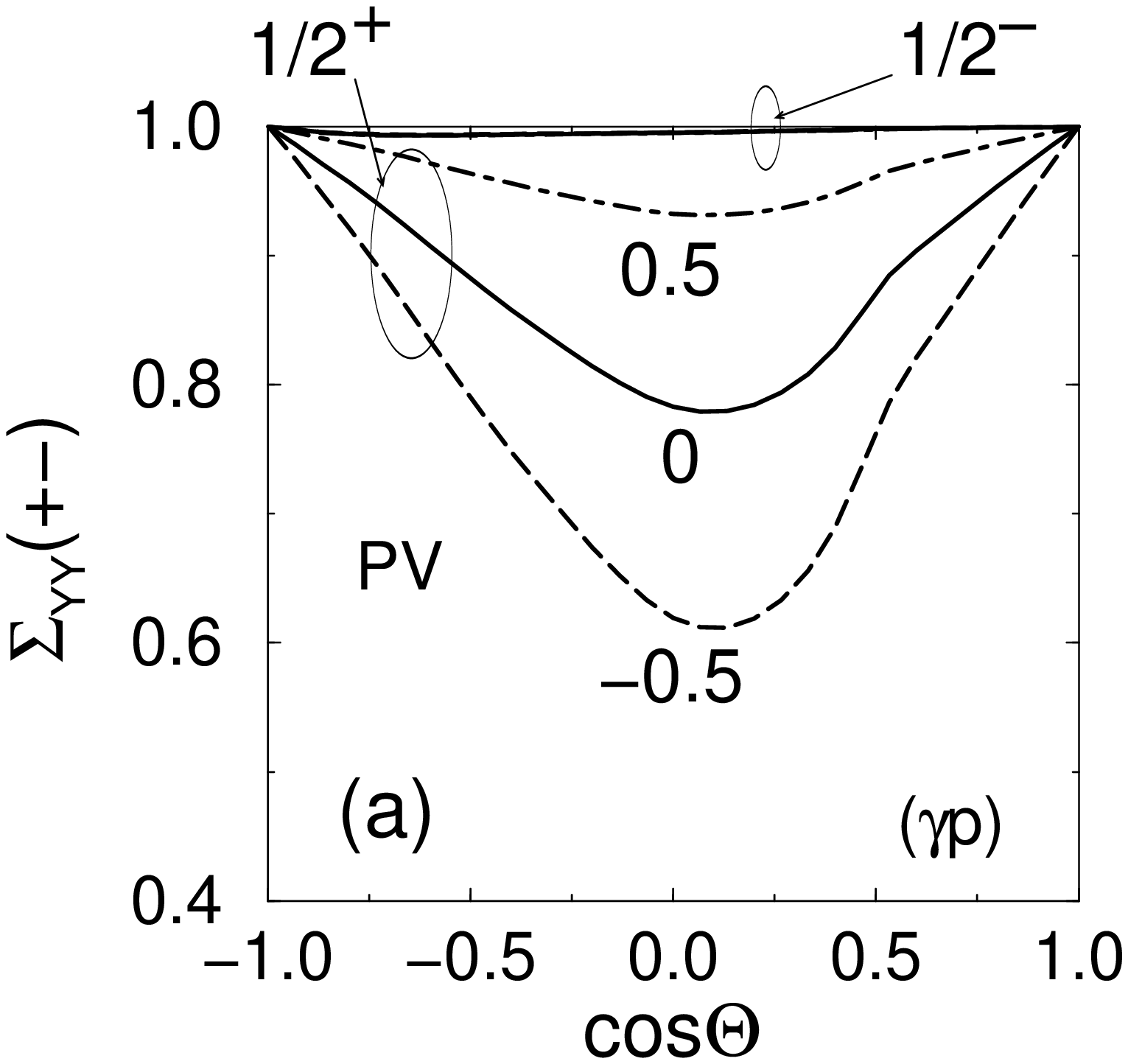}\hfill
 \includegraphics[width=.23\textwidth]{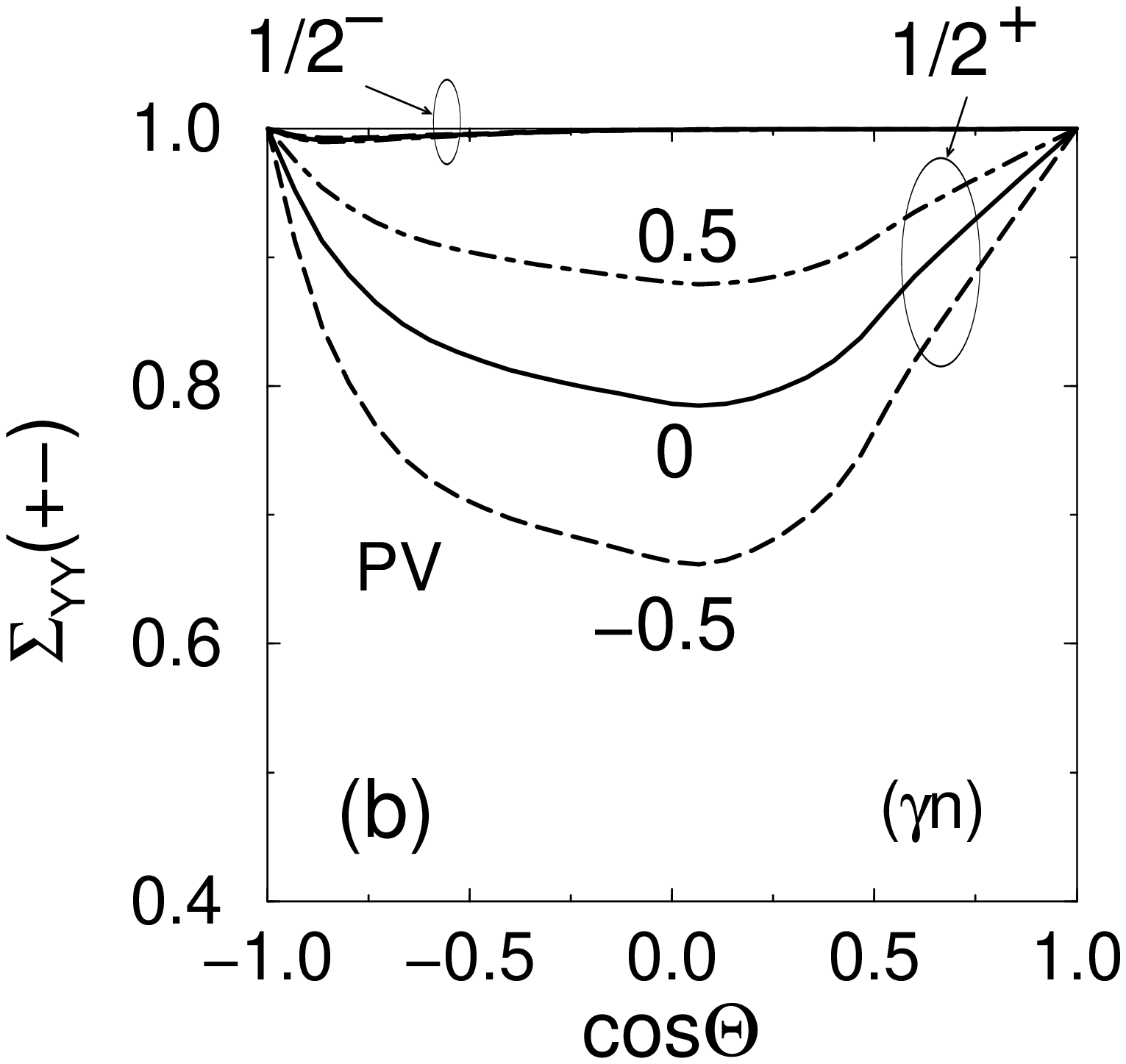}\hfill
\mbox{}\hspace{.23\textwidth}\mbox{}
 \caption{\label{fig:24}\tcaps%
 The triple spin asymmetry $\Sigma^{}_{yy}(\uparrow\downarrow)$
 in  (a) $\gamma p\to pK^0 \bar{K}^0$ and
  (b) $\gamma n\to nK^+ K^-$ as a function of the $K^+$ decay angle
 for different values of $\kappa^*$ and positive  $\bar\alpha$ and the PV-coupling scheme.}
\end{figure}

Consider now the spin-flip transitions. For the vector-meson photoproduction, the
spin-flip transitions are suppressed compared to the spin-conserving ones and, therefore,
the effect of the background in $\Sigma^{\pm}_{yy}(\uparrow\downarrow)$ is much smaller
than in $\Sigma^{\pm}(\uparrow\uparrow)$. Now, $\Sigma^{-}_{yy}(\uparrow\downarrow)$ is
close to its ``model-independent" limit (+1) [cf. Eq.~(\ref{Byy23})]. For the positive
parity we expect $\mbox{$\Sigma^{+}_{yy}(\uparrow\downarrow)$}=+1$ at $\cos\Theta=\pm1$
with some decrease at small $|\cos\Theta|$. The results of the calculations of
$\Sigma_{yy}(+-)\equiv \Sigma_{yy}(\uparrow\downarrow)$ shown in Figs.~\ref{fig:24}a and
b for $\gamma p\to p K^0\bar{K}^0$ and $\gamma n\to n K^+ K^-$, respectively, confirm this
expectation. They are obtained using the PV-coupling scheme and different $\kappa^*$. The
results corresponding to other input parameter values are similar to those shown in
Figs.~\ref{fig:24}a and b, and we do not display them here. One can see a clear
difference between the predictions for positive and negative $\pi_\Theta$. For negative
parity, $\Sigma^{-}_{yy}(\uparrow\downarrow)\approx+1$ at all $\cos\Theta$. For positive
parity, we have  $\Sigma^{+}_{yy}(\uparrow\downarrow)=+1$ at $\cos\Theta=\pm1$ with a
minimum at $\cos\Theta=0$. Unfortunately, the minimum value of
$\Sigma^{+}_{yy}(\uparrow\downarrow)$ depends strongly on the tensor coupling $\kappa^*$;
in particular, at large negative $\kappa^*$, the deviation between
$\Sigma^{+}_{yy}(\uparrow\downarrow)$ and $\Sigma^{-}_{yy}(\uparrow\downarrow)$ may be
rather small. This makes it difficult to use $\Sigma_{yy}(\uparrow\downarrow)$ for the
determination of $\pi_\Theta$.

\section{Summary}

We have analyzed in detail for the first time two essential aspects of the problem
associated with the determination of the parity of the $\Theta^+$ pentaquark from
photoproduction processes.  The first one is the non-resonant background in the reaction
$\gamma N\to NK\bar{K}$. The second one is related to the three-body final states. The
interference between the resonance amplitude and the non-resonant background results in a
modification of all the spin observables from those corresponding to the pure resonance
channels contribution. The effect of the three-body final state means that the
predictions made for the spin ``observables" of the (directly unobservable) two-body
intermediate reaction $\gamma N\to \Theta^+\bar{K}$ are not useful from a practical point
of view. This applies to all ``model-independent" predictions for the $\Theta^+$ parity
determination in $\Theta^+$ photoproduction discussed recently in the
literature~\cite{Ejiri,RT-G04,NL04}.

We have analyzed in detail the non-resonant background. We have found that in the
near-threshold energy region ($E_\gamma\approx 2$ GeV), the non-resonant background is
dominated by the vector ($\phi,\rho$, and $\omega$) meson  photoproduction. The
contributions from the scalar ($\sigma$) and tensor ($a_2,f_2$) mesons are rather small.
However, the later may be important at higher energies. Additional information about the
background structure may be found by studying the relative  angular distribution of the
$K\bar{K}$ pair. In our study, we have shown that, using both the measured ratio of the
resonance to background yields and the calculated non-resonant background, we can find
the strength of the $K^*$ exchange amplitude and reduce the number of unknown parameters.

Finally, we have shown that only in the case of the triple spin asymmetry, one can find a
kinematical ``window" where the predictions are sensitive to the $\Theta^+$ parity and
insensitive to the production mechanism. The present analysis is based of the observed
ratios of the resonant and non-resonant contributions, including the instrument's energy
resolution. The triple spin asymmetries are quite sensitive to the $\Theta^+$ parity, but
not to other parameters. If the instrument's resolution will be improved in future
experiments, the ratio will increase accordingly provided that the intrinsic width $
\Gamma_\Theta$ is smaller than the instrument's resolution. Then, the difference between
the triple spin asymmetries $\Sigma(\uparrow\uparrow)$ for the positive and negative
$\pi_\Theta$ will get quite conspicuous, and thus will provide the $\Theta^+$ parity
almost model-independently.

\acknowledgments

We thank  S.~Dat\'e, K. Hicks, A.~Hosaka, M.~Fujiwara, T.~Mibe, T.~Nakano, Y.~Oh,
Y.~Ohashi, and H.~Toki for fruitful discussion. One of authors (A.I.T.) thanks T.~Tajima,
the director of the Advanced Photon Research Center, Japan Atomic Energy Research
Institute, for his hospitality to stay at SPring-8.

\appendix

\section{Transition operators for resonance channels}

We show here the explicit expressions for the transition operators ${\cal M}_\mu$ in
Eqs.~(\ref{res_ampl}) for the positive and negative $\Theta^+$ parity ($\pi_{\Theta}$)
and for the PV- and PS-coupling schemes.

The specific parameters for the form factor of Eq.~(\ref{FF}) required here are defined by
\begin{equation}
  F_s=F(M_N,s)~,\qquad  F_u=F(M_\Theta,u)~,
  \qquad   \text{and}\qquad
  F_t=F(M_{K^+},t)~.
\label{PVgn-plus-FF}
\end{equation}
In addition, we need the form factor combinations
\begin{equation}
  \widetilde{F}_{tu}=F_t+F_u -F_tF_u
  \qquad   \text{and}\qquad
  \widetilde{F}_{su}=F_s+F_u -F_sF_u
\end{equation}
to construct the contact terms $\mathcal{M}_\mu^c$ given below that make the initial
photoproduction amplitude gauge invariant~\cite{hhgauge,DavWork}. The four-momenta in the
following equations are defined according to the arguments given in the reaction equation
\begin{equation}
  \gamma(k)+N(p) \to \Theta^+(p_\Theta)+\bar{K}(\bar{q})~.
\end{equation}

\subsection{\boldmath $\gamma n\to \Theta^+ K^-$}

\noindent
\underline{$\pi_{\Theta}=+1$;  PV}
\begin{subequations}
\label{PVgn-plus}
\begin{align}
 {\cal M}^t_\mu&=i\frac{eg_{\Theta NK}(k-2\bar q)_\mu\gamma_5}{t-M^2_{K^+}}
 \,F_t~,\label{PVgn-plus-t}\\
 {\cal M}^s_\mu&=i\frac{eg_{\Theta NK}}{M_{\Theta}+M_N}\,
  \gamma_5\fs{\bar{q}}\frac{\fs p+\fs k +M_N}{s-M_N^2}
  \left(i\frac{\kappa_n}{2M_N}\sigma_{\mu\nu}k^\nu\right)\,F_s~,
 \label{PVgn-plus-s}
\displaybreak[1]
 \\
 {\cal M}^u_\mu&=i\frac{eg_{\Theta NK}}{M_{\Theta}+M_N}\,
  \left(\gamma_\mu + i\frac{\kappa_\Theta}{2M_\Theta}\sigma_{\mu\nu}k^\nu\right)
  \frac{\fs p_{\Theta}-\fs k
  + M_\Theta}{u - M_\Theta^2} \gamma_5\fs{\bar{q}}\,F_u~,
 \label{PVgn-plus-u}\\
{\cal M}^c_\mu &=ieg_{\Theta NK}\gamma_5
 \left[\frac{(k-2\bar q)_\mu }{t-M^2_{K^+}}
 \,(\widetilde{F}_{tu}-F_t)
+\frac{(2{p_\Theta} -k)_\mu }{u -
M_\Theta^2}\,(\widetilde{F}_{tu}-F_u) +
\frac{\gamma_\mu}{M_{\Theta}+M_N}\,F_u\right]~.
 \label{PVgn-plus-c}
\end{align}
\end{subequations}

\noindent
\underline{$\pi_{\Theta}=+1$; PS}
\begin{subequations}
\label{PSgn-plus}
\begin{align}
  {\cal M}^t_\mu&=i\frac{eg_{\Theta NK}(k_\mu-2\bar q_\mu)\gamma_5}{t-M^2_{K^+}}
 \,F_t~,\label{PSgn-plus-t}\\
 {\cal M}^s_\mu&=i{eg_{\Theta NK}}
  \gamma_5\frac{\fs p+\fs k +M_N}{s-M_N^2}
  \left(i\frac{\kappa_p}{2M_N}\sigma_{\mu\nu}k^\nu\right)\,F_s~,
 \label{PSgn-plus-s}\\
 {\cal M}^u_\mu&=i{eg_{\Theta NK}}
  \left(\gamma_\mu + i\frac{\kappa_\Theta}{2M_\Theta}\sigma_{\mu\nu}k^\nu\right)
  \frac{\fs p_{\Theta}-\fs k + M_\Theta}{u - M_\Theta^2} \gamma_5\,F_u~,
 \label{PSgn-plus-u}
 \displaybreak[1]
\\
 {\cal M}^c_\mu&=i{eg_{\Theta NK}\gamma_5}
 \left[\frac{(k-2\bar q)_\mu}{t-M^2_{K^+}}
 \,(\widetilde{F}_{tu}-F_t)
 +\frac{(2{p_\Theta}-k)_\mu}{u - M_\Theta^2}\,(\widetilde{F}_{tu}-F_u)\right]~.
 \label{PSgn-plus-c}
\end{align}
\end{subequations}
 For both PS and PV couplings, the positive-parity $t$-channel $K^*$ exchange is given by
\begin{eqnarray}
 {\cal M}^t_\mu(K^*)=\frac{eg_{\gamma KK^*}g_{\Theta NK^*}}{M_{K^*}}\,
  \frac{\varepsilon_{\mu\nu\alpha\beta}k^\alpha \bar q^\beta}{t-M^2_{K^*}}
  \left[\gamma^\nu
  -i\frac{\sigma^{\nu\lambda}(p-p_\Theta)_\lambda}{M_\Theta+M_N} \kappa^*\right]\,F(M_{K^*},t)~.
 \label{gn-plusK*}
\end{eqnarray}

For the negative $\Theta^+$ parity, we have the following amplitudes.

\noindent
\underline{$\pi_{\Theta}=-1$; PV}
\begin{subequations}
\label{PVgn-minus}
\begin{align}
 {\cal M}^t_\mu&= i\frac{eg_{\Theta NK}(k-2\bar q)_\mu}{t-M^2_{K^+}}
 \,F_t~,\label{PVgn-minus-t}\\
{\cal M}^s_\mu&=-i\frac{eg_{\Theta NK}}{M_{\Theta}-M_N}\,
  \fs{\bar{q}}\frac{\fs p+\fs k +M_N}{s-M_N^2}
  \left(i\frac{\kappa_n}{2M_N}\sigma_{\mu\nu}k^\nu\right) F_s~,
 \label{PVgn-minus-s}\\
{\cal M}^u_\mu&=-i\frac{eg_{\Theta NK}}{M_{\Theta}-M_N}\,
  \left(\gamma_\mu + i\frac{\kappa_\Theta}{2M_\Theta}\sigma_{\mu\nu}k^\nu\right)
  \frac{\fs p_{\Theta}-\fs k + M_\Theta}{u - M_\Theta^2}\fs{\bar{q}}\,F_u~,
 \label{PVgn-minus-u}\displaybreak[1]
\\
 {\cal M}^c_\mu&=i{eg_{\Theta NK}}
 \left[\frac{(k-2\bar q)_\mu}{t-M^2_{K^+}}
 \,(\widetilde{F}_{tu}-F_t)
 +\frac{(2{p_\Theta} -k)_\mu}{u - M_\Theta^2}\,(\widetilde{F}_{tu}-F_u)
- \frac{\gamma_\mu}{M_\Theta-M_N}\,{F}_{u}\right]~.
 \label{PVgn-minus-c}
\end{align}
\end{subequations}

\noindent
 \underline{$\pi_{\Theta}=-1$; PS}
\begin{subequations}
\label{PSgn-minus}
\begin{align}
 {\cal M}^t_\mu&=i\frac{eg_{\Theta NK}(k-2\bar q)_\mu}{t-M^2_{K^+}}
 \,F_t~,
 \label{PSgn-minus-t}\\
{\cal M}^s_\mu&=i{eg_{\Theta NK}}
  \frac{\fs p+\fs k +M_N}{s-M_N^2}
  \left(i\frac{\kappa_n}{2M_N}\sigma_{\mu\nu}k^\nu\right)F_s~,
 \label{PSgn-minus-s}\\
{\cal M}^u_\mu&=i{eg_{\Theta NK}}
  \left(\gamma_\mu + i\frac{\kappa_\Theta}{2M_\Theta}\sigma_{\mu\nu}k^\nu\right)
  \frac{\fs p_{\Theta}-\fs k + M_\Theta}{u - M_\Theta^2}\,F_u~,
 \label{PSgn-minus-u}
 \displaybreak[1]
\\
{\cal M}^c_\mu&=i{eg_{\Theta NK}}
 \left[\frac{(k-2\bar q)_\mu}{t-M^2_{K^+}}
 \,(\widetilde{F}_{tu}-F_t)+\frac{(2{p_\Theta}-k)_\mu}
 {u - M_\Theta^2}\,(\widetilde{F}_{tu}-F_u)\right]~.
 \label{PSgn-minus-c}
\end{align}
\end{subequations}
For negative $\Theta^+$ parity, the $K^*$ exchange is given by
\begin{equation}
 {\cal M}^t_\mu(K^*)=\frac{eg_{\gamma KK^*}g_{\Theta NK^*}}{M_{K^*}}\,
   \frac{\varepsilon_{\mu\nu\alpha\beta}k^\alpha \bar q^\beta}{t-M^2_{K^*}}
 \gamma_5 \left[\gamma^\nu
  -i\frac{\sigma^{\nu\lambda}(p-p_\Theta)_\lambda}{M_\Theta+M_N} \kappa^*\right]\,F(M_{K^*},t)~.
 \label{gn-minusK*}
\end{equation}

\subsection{\boldmath $\gamma p\to \Theta^+ \bar{K}^0$}

\noindent
\underline{$\pi_{\Theta}=+1$;  PV}

\begin{subequations}
\label{PVgp-plus}
\begin{align}
{\cal M}^s_\mu&=i\frac{eg_{\Theta NK}}{M_{\Theta}+M_N}\,
  \gamma_5\fs{\bar{q}}\frac{\fs p+\fs k +M_N}{s-M_N^2}
  \left(\gamma_\mu +i\frac{\kappa_p}{2M_N}\sigma_{\mu\nu}k^\nu\right)F_s~,
 \label{PVgp-plus-s}\\
 {\cal M}^u_\mu&=i\frac{eg_{\Theta NK}}{M_{\Theta}+M_N}\,
  \left(\gamma_\mu + i\frac{\kappa_\Theta}{2M_\Theta}\sigma_{\mu\nu}k^\nu\right)
  \frac{\fs p_{\Theta}-\fs k + M_\Theta}{u - M_\Theta^2} \gamma_5\fs{\bar{q}}\,F_u~,
 \label{PVgp-plus-u}\\
 {\cal M}^c_\mu&=
 i\frac{e g_{\Theta NK} }{M_\Theta+M_N} \gamma_5 \fs{\bar{q}}\left[
  \frac{(2{p} +k)_\mu} {s - M_N}\,(\widetilde{F}_{su}-F_s)
  + \frac{(2{p_\Theta} - k)_\mu } {u -
  M_\Theta^2}\,(\widetilde{F}_{su}-F_u)
  \right] ~.
 \label{PVgp-plus-c}
\end{align}
\end{subequations}

\noindent
 \underline{$\pi_{\Theta}=+1$;  PS}
\begin{subequations}
\label{PSgp-plus}
\begin{align}
 {\cal M}^s_\mu&=i{eg_{\Theta NK}}
  \gamma_5\frac{\fs p+\fs k +M_N}{s-M_N^2}
  \left(\gamma_\mu +i\frac{\kappa_p}{2M_N}\sigma_{\mu\nu}k^\nu\right)\,F_s~,
 \label{PSgp-plus-s}\\
 {\cal M}^u_\mu&=i{eg_{\Theta NK}}\,
  \left(\gamma_\mu + i\frac{\kappa_\Theta}{2M_\Theta}\sigma_{\mu\nu}k^\nu\right)
  \frac{\fs p_{\Theta}-\fs k + M_\Theta}{u - M_\Theta^2} \gamma_5\,F_u~,
 \label{PSgp-plus-u}\\
 {\cal M}^c_\mu&=i{eg_{\Theta NK}}\gamma_5\left[
     \frac{(2{p}+k)_\mu} {s - M_N}\,(\widetilde{F}_{su}-F_s)
  +  \frac{(2{p_\Theta} - k)_\mu} {u -
  M_\Theta^2}\,(\widetilde{F}_{su}-F_u)\right]~.
 \label{PSgp-plus-c}
\end{align}
\end{subequations}

The transition amplitudes for the negative $\Theta^+$ parity may be obtained immediately
from Eqs.~(\ref{PVgp-plus}) and (\ref{PSgp-plus}) by the simple substitutions
\begin{subequations}
 \label{PSgp-minus}
\begin{align}
 \text{PV:}&& \frac{\gamma_5}{M_{\Theta}+ M_N}&\to-\frac{1}{M_{\Theta}- M_N}~,
 \\
 \text{PS:}&& {\gamma_5}&\to 1~.
\end{align}
\end{subequations}

\section{Pomeron exchange amplitude}

The invariant amplitude for the Pomeron-exchange process has
the form
 \begin{equation}
  {A}_{fi}^P = - M^{}_P (s,t) \,\Gamma^P_{fi}~.
 \end{equation}
The scalar function $M_P(s,t)$ is described by the Regge
parameterization,
 \begin{equation}
 M^{}_P (s,t)= C^{}_P \, F^{}_1(t) \, F^{}_2 (t)\,\frac{1}{s}
 \left(\frac{s}{s_P} \right)^{\alpha_P^{} (t)} \exp\left[ -
 \frac{i\pi}{2}\alpha_P^{}(t) \right]~,
 \label{Pom:MP}
 \end{equation}
where $F_1^{}(t)$ is the isoscalar electromagnetic form factor of the nucleon and
$F_2^{}(t)$ is the form factor given in Appendix C for the vector-meson--photon--Pomeron
coupling. We also follow Ref.~\cite{DL84-92} to write
 \begin{eqnarray}
  F_1^{} (t) = \frac{ 4 M_N^2-2.8t }{ (4M_N^2-t) (1-t/t_0)^2  }~,\qquad
  F_2^{} (t) = \frac{2\mu_0^2}{(1 - t/M_V^2)(2 \mu_0^2 + M_V^2 - t)}~,
 \end{eqnarray}
where $t_0 = 0.7$ GeV$^2$. The Pomeron trajectory is known to be $\alpha_P^{} (t) = 1.08
+ 0.25\,t$. The strength factor $C_P^{}$ is given by
\begin{eqnarray}
  C_P^{} = \frac{6eg^2}{\gamma_V^{}}~,
 \label{Cp}
 \end{eqnarray}
where $\gamma_V$ is the vector meson decay constant
($\gamma_\omega^2=72.71$, $\gamma^2_\phi=44.22$, and
$\gamma_\rho^2=6.33$).  The parameter $g$ has a meaning of the
Pomeron-quark coupling and it is taken to be the same for all
vector mesons ($g^2=16.6$). The remaining parameters read $\mu_0^2
= 1.1$ GeV$^2$ and  $s_P = 4$ GeV$^2$. The amplitude
$\Gamma^P_{fi}$ reads
 \begin{eqnarray}
  \Gamma^P_{fi}=
  \bar{u}^{}_f \fs{k} \, u^{}_i
 (\varepsilon^*_{\lambda_V}\cdot \varepsilon_{\lambda_\gamma})
  -\bar{u}^{}_f \fs{\varepsilon}_{\lambda_\gamma} u^{}_i
  (\varepsilon^*_{\lambda_V}\cdot k)
 -\bar{u}^{}_f \fs{\varepsilon}^*_{\lambda_V} u^{}_i
   \left[\varepsilon_{\lambda_\gamma}\cdot q
   -\frac{(\varepsilon_{\lambda_\gamma}\cdot \bar p)(k\cdot q)}{\bar p
 \cdot k}\right],
  \label{G-fi}
 \end{eqnarray}
with $\bar p=(p+p')/2$.
$\varepsilon^{}_{\mu} (V)$ and $\varepsilon^{}_{\nu}
 (\gamma)$ are the polarization vectors of the vector meson
 ($\rho,\phi$) and the photon, respectively, and $u_i$=$u^{}_{m_i}(p)$
 [$u_f$=$u^{}_{m_f}(p')$]  is the
 Dirac spinor of the nucleon with momentum $p$ ($p'$) and spin projection
 $m_i$ ($m_f$).

\section{Parameters for the non-resonant background}

 \begin{table}[b]
  \caption{Coupling constants for the non-resonant background.}
  \label{tab:2}
 \begin{tabular}{l@{\quad}|@{\quad}c@{\quad}|@{\quad}c@{\quad}|@{\quad}c}
 \hline\hline
 $ g$&    value  & source & Ref.   \\ \hline
 $g_{\pi  NN}$ &   13.26    & $\phi,\omega$ photoproduction
 &\protect\cite{FS96,Titov03}\\ \hline
 $g_{\eta NN}$ &   3.54    & SU(3)   &\protect\cite{TLTS99}\\ \hline
 $g_{\sigma  NN}$ &   10.03    & $\rho$ photoproduction  &\protect\cite{FS96} \\ \hline
 $g_{\rho  NN};~\kappa_\rho$ & 3.72; 6.71    & Bonn Model &\protect\cite{BonnMod}  \\ \hline
 $eg_{\rho\sigma\gamma}$ &  0.82    & $\rho$ photoproduction&\protect\cite{FS96} \\ \hline
 $eg_{\rho\pi\gamma}$ &  0.16      & $\rho$ photoproduction &\protect\cite{FS96} \\ \hline
 $eg_{\omega\pi\gamma}$ &  0.55    &$\omega$ photoproduction  &\protect\cite{FS96} \\ \hline
 $eg_{\phi\pi\gamma}$ & $- 0.043$    & $\phi\to\pi\gamma$ decay, SU(3)&\protect\cite{PDG} \\ \hline
 $eg_{\phi\eta\gamma}$ & $- 0.214$    & $\phi\to\eta\gamma$ decay, SU(3) &\protect\cite{PDG} \\ \hline
 $g_0\equiv g_{\phi KK}$ & 4.48    & $\phi\to KK$ decay &\protect\cite{PDG} \\ \hline
 $ g_{\rho K^+K^-}$ & $g_0/\sqrt{2}$    &  SU(3) &\protect\cite{PDG} \\ \hline
 $ g_{\rho K^0\bar{K}^0}$ & $-g_0/\sqrt{2}$    &  SU(3) &\protect\cite{PDG} \\ \hline
 $ g_{\sigma KK}=-g_{\sigma \pi\pi}$ &  1.74  & $\sigma\to \pi \pi$ decay, SU(3)
 &\protect\cite{PDG} \\ \hline\hline
\end{tabular}
\end{table}

The parameters of the model that define the amplitude of  the vector-meson
photoproduction are taken from previous studies ~\cite{FS96,Titov03}. The coupling
constants in Eqs.~(\ref{NNpi})-(\ref{VKK}) are based on empirical knowledge, comparison
with the corresponding decay widths and SU(3)-symmetry considerations. For the
$\sigma$-meson photoproduction, we use the Bonn model as listed in Table B.1 (Model II)
of Ref.~\cite{BonnMod}. All coupling constants are displayed in Table~\ref{tab:2}.

 All the vertex functions are
dressed by monopole form factors. We use the following expressions
for their products:
\begin{subequations}
\label{V-FF}
\begin{align}
F_\pi(t)&=\left(\frac{0.6^2-m_\pi^2}{0.6^2 - t}\right)^2~,
\\
F_\eta(t)&=\left(\frac{1.0^2-m_\eta^2}{1.0^2 - t}\right)
\left(\frac{0.9^2-m_\eta^2}{0.9^2 - t}\right)~,
\\
F_\sigma(t)&=\left(\frac{1.0^2-m_\sigma^2}{1.0^2 - t}\right)
\left(\frac{0.9^2-m_\sigma^2}{0.9^2 - t}\right)~,
\\
F_\rho(t)&=\left(\frac{1.3^2-m_\rho^2}{1.3^2 - t}\right)^2~,
\end{align}
\end{subequations}
where all the masses are in GeV and $t$ is in GeV$^2$.

\end{document}